\documentclass[pre,preprintnumbers,amsmath,amssymb,superscriptaddress]{revtex4}
\usepackage{graphicx,color}
\usepackage{bm}% bold math
\usepackage{epsfig}
\usepackage{hyperref}
\usepackage{float}

\usepackage{enumitem}

\newcommand{\isEquivTo}[1]{\underset{#1}{\sim}}

\begin{document}
\title{Run-and-tumble particle in one-dimensional potentials: mean first-passage time and applications}

\begin{abstract}
 We study a one-dimensional run-and-tumble particle (RTP), which is a prototypical model for active system, moving within an arbitrary external potential. Using backward Fokker-Planck equations, we derive the differential equation satisfied by its mean first-passage time (MFPT) to an absorbing target, which, without any loss of generality, is placed at the origin. {Depending on the shape of the potential, we identify four distinct ``phases'', with a corresponding expression for the MFPT in every case, which we derive explicitly.} To illustrate these general expressions, we derive explicit formulae for two specific cases which we study in detail: a double-well potential and a logarithmic potential. We then present different applications of these general formulae to (i) the generalization of the Kramer's escape law for an RTP in the presence of a potential barrier, (ii) the ``trapping'' time of an RTP moving in a harmonic well and (iii) characterizing the efficiency of the optimal search strategy of an RTP subjected to stochastic resetting. Our results reveal that the MFPT of an RTP in an external potential exhibits a far more complex and, at times, counter-intuitive behavior compared to that of a passive particle (e.g., Brownian) in the same potential. 
\end{abstract}

\date{\today}

\author{Mathis Gu\'eneau}
\affiliation{Sorbonne Universit\'e, Laboratoire de Physique Th\'eorique et Hautes Energies, CNRS UMR 7589, 4 Place Jussieu, 75252 Paris Cedex 05, France}
\author{Satya N. Majumdar}
\affiliation{LPTMS, CNRS, Univ.  Paris-Sud,  Universit\'e Paris-Saclay,  91405 Orsay,  France}
\author{Gr\'egory Schehr}
\affiliation{Sorbonne Universit\'e, Laboratoire de Physique Th\'eorique et Hautes Energies, CNRS UMR 7589, 4 Place Jussieu, 75252 Paris Cedex 05, France}

{
\let\clearpage\relax
\maketitle
\newpage
\tableofcontents
}
%\vspace{-10cm}

\section{Introduction}

%During the last few years, there has been a growing interest in the study of \textit{active particles}. 
%Describing and predicting the movements of living organisms is essential for a deeper understanding of their behavior. One effective approach to modeling these movements is through the use of stochastic processes. Among these processes, a notable class is known as \textit{active particles}. 
%The term ''active" distinguishes these particles from ''passive" ones, such as those undergoing Brownian motion. Active particles are typically subjected to colored noises (time-correlated noise) \cite{colorednoise} as opposed to white noises characteristic of passive particles. They are capable of extracting energy from their environment and converting it into their distinctive out-of-equilibrium self-propelled motion \cite{activeintro1,activeintro2,activeintro3,activeintro4,activeintro5}. Observations of bacterial trajectories reveal that small organisms such as \textit{Escherichia coli} \cite{Berg2004} move in a straight line for a certain period before tumbling. A well-known model of active particles to describe them is the run-and-tumble particle (RTP) \cite{ Tailleur_RTP, Cates2012}, or persistent random walk \cite{Kac1974, Ors1990, PRWWeiss}. 

In recent years, there has been a growing interest in the study of active particles, which differ fundamentally from passive particles like those undergoing Brownian motion. The term ``active'' refers to particles that are driven by colored noise (time-correlated noise) \cite{colorednoise}, as opposed to the white noise typical of passive particles. Active particles can extract energy from their surroundings and convert it into {\it out-of-equilibrium} self-propelled motion \cite{activeintro1,activeintro2,activeintro3,activeintro4,activeintro5}. For instance, bacterial trajectory observations show that microorganisms such as Escherichia coli \cite{Berg2004} move in straight lines for a time before undergoing random changes in direction, called ``tumblings''. A widely used model to describe such behavior is the run-and-tumble particle (RTP) model \cite{ Tailleur_RTP, Cates2012}, also known as the persistent random walk~\cite{Kac1974, Ors1990, PRWWeiss}.
%
%At large time, the position $x(t)$ of a free RTP (say in one-dimension) exhibits a diffusive behavior $x(t) \approx \sqrt{D_{\rm eff}\,t}$ with an effective diffusion coefficient $D_{\rm eff}$ \cite{MalakarRTP} \gs{Not sure that this is the best Ref to cite here}. However, in the presence of an external confining potential, the RTP reaches a stationary state, which is usually non-Boltzman~\cite{RTPSS, Sevilla,Velocity_RTP}, exhibiting in some cases accumulation near the boundaries of the support of their spatial distributions \cite{Lee2013, Yang2014, Uspal2015, Duzgun2018, AngelaniHardWalls, hardWallsJoanny, hardWallsCaprini}. While the spatial properties of RTPs under confinement have been well characterized recently, exact analytical results for their first-passage properties \cite{redner, Metzler_book, EVS1} remain sparse. 
%
At large times, the position $x(t)$ of a free RTP (in one dimension, for example) exhibits diffusive behavior characterized by 
${x(t) \approx \sqrt{2 D_{\rm eff} \,t}}$ where $D_{\rm eff}$ is the effective diffusion coefficient. However, in the presence of an external confining potential, the RTP reaches a stationary state, which is generally non-Boltzmann \cite{RTPSS, Sevilla,Velocity_RTP}. In certain cases, this leads to accumulation near the boundaries of the support of their spatial distribution \cite{Lee2013, Yang2014, Uspal2015, Duzgun2018, AngelaniHardWalls, hardWallsJoanny, hardWallsCaprini,RTPSS, Sevilla}. While the spatial properties of confined RTPs have been well characterized recently, exact analytical results for their first-passage properties remain relatively sparse.

%Important observables for characterizing these first-passage properties include the survival probability (the probability for the particle to stay in a region of space for a fixed duration) and the exit probability (the probability of leaving an interval at a given time), which contains all the information of the first-passage time distribution. For a free RTP these quantities have been computed in one \cite{MalakarRTP, Debruyne, Singh2020, Singh2022, MFPT1DABP} and higher dimensions \cite{RTPsurvivalMori,TVB12,RBV16}. Recent studies have focused on adding partially absorbing boundaries \cite{BressloffStickyBoundaries,EPJEAngelani, AngelaniGenericBC, RTPpartiallyAbsorbingTarget}, and some in the presence of an external potential \cite{ExitProbaShort,letter, RTPSS, Kardar,naftaliSS, SurvivalRPTlinear}, although such situations of RTPs under confinement pose significant challenges, even in one-dimension which is the case we will consider in the rest of this paper. 

Important observables for characterizing these first-passage properties include the survival probability (the probability for the particle to remain within a specific region of space for a fixed duration) and the exit probability (the probability of leaving an interval at a given time), which encodes all the information of the first-passage time distribution \cite{redner, Metzler_book, EVS1}. For a free RTP, these quantities have been computed in one \cite{MalakarRTP, Debruyne, Singh2020, Singh2022, MFPT1DABP} and higher \cite{RTPsurvivalMori,TVB12,RBV16} dimensions. Recent studies have focused on the addition of partially absorbing boundaries \cite{BressloffStickyBoundaries,EPJEAngelani, AngelaniGenericBC, RTPpartiallyAbsorbingTarget}. Further works have also considered the case 
of RTPs in the presence of external potentials \cite{ExitProbaShort,letter, RTPSS, Kardar, naftaliSS, SurvivalRPTlinear}. However, obtaining analytical results in this situation involving RTPs under confinement is quite challenging, even in one dimension, 
which is the case considered in the present paper.

In the presence of a linear `V-shaped' potential, $V(x) = \alpha |x|$, a closed-form expression for the survival probability of an RTP has been derived {\cite{Debruyne, SurvivalRPTlinearMORI, SurvivalRPTlinear}} giving access to all first-passage properties. {Exact results were also obtained for the survival probability of a free RTP in the presence of stochastic resetting \cite{resettingRTP} which, in the large time limit, also amounts, to some extent, to adding an effective linear V-shaped potential \cite{resettingPRL, resettingReview}.} However, in most of the cases, obtaining a closed-form equation for the survival probability is a difficult task. Hence, a first step toward understanding first-passage properties of active particles is the study of the mean first-passage time (MFPT) to an absorbing boundary, i.e., the first moment of the first-passage time distribution. 
%
%For an RTP moving within a harmonic potential $V(x) = \alpha x^2$, the Laplace transform of the survival probability is known in terms of hypergeometric functions \cite{RTPSS}, but extracting the mean first-passage time from it is still a difficult task~\cite{letter}. Therefore, developing a method to compute the mean first-passage time is crucial.
%
In a recent paper \cite{letter}, we derived an explicit expression for the MFPT of an RTP in the presence of a specific class of confining potentials of the form $V(x)= \alpha |x|^p$ with $p\geq 1$. In particular, we demonstrated the existence of an optimal value of the tumbling rate that minimizes the MFPT for $p>1$, which was the main focus of that paper. However, it turns out that the derivation of the MFPT strongly depends on the shape of the potential \cite{letter}. Consequently, the formula derived in \cite{letter} only applies to a restricted class of potentials, e.g., it does not apply to potentials of the form $V(x)= \alpha |x|^p$ with $p<1$. The goal of this paper is to develop a general method to compute analytically the MFPT of an RTP moving inside an (essentially) arbitrary potential, which may not necessarily be confining. More precisely, we identify four different characteristic shapes of potentials (which we call ``phases'' in the following, see Figs. \ref{MFPTPhases12} and \ref{MFPTPhase34}) which lead to different functional forms of the MFPT. For each of these phases, we provide relevant examples of potentials for which the MFPT can be computed explicitly. Interestingly, in all these examples, the MFPT can be expressed in terms of hypergeometric functions. 

We then present several applications of these formulae. In particular, we discuss in detail the application of our exact formula of the MFPT for a double well potential to the calculation of transition rates, often used to characterise activation processes in chemistry and biology \cite{kramers}. The general method developed here allows us to address the active version of the Kramers' law: what is the mean time for a particle to jump from the bottom of a well to the top of a barrier? For a passive particle, 
this rate, in the limit of small temperature $T$, is simply proportional to the exponential of the ratio of the barrier height over $k_B T$, with $k_B$ the Boltzman constant. {The extension of Kramer's law to colored-noise (i.e., with temporal correlations) is a challenging problem, which has already attracted some attention in the past, see e.g.~\cite{RTPSS4,coloredKramers}}. More recent studies have explored this question for active particles both experimentally \cite{delago} and theoretically, via approximations, perturbative approaches or scaling arguments \cite{Woillez, Woillez2, Hanggi_kramers, Sharma, WexlerKramers, kramersCaprini}. Here we present an exact calculation of this transition rate, based on our analytical results for the MFPT, in the case of a purely active noise and establish contact with these previous approaches.

%Note that most of these works consider a combination of both an active and a thermal noise, instead here we only consider the effect of a purely active noise (e.g. as in \cite{naftaliSS}), and we propose exact analytical results.

The rest of the paper is organized as follows. In Section \ref{Mainresultssection}, we first introduce the model of RTP studied here and present the main results of the paper. Depending on the shape of the potential, we provide an explicit expression of the MFPT. In the next two sections we illustrate these formal results by providing explicit analytical expressions for the MFPT for two different types of potentials: a double-well potential in Section \ref{doublewellsection} and a logarithmic potential in Section \ref{logpotsection}. In Section \ref{application}, we discuss three applications of the developed methods: estimating the ``trapping time'' of an RTP (i.e., the time it takes for an RTP to reach the support of its stationary distribution), deriving Kramers' law for an RTP, and identifying the optimal search strategy between a potential-driven RTP and a resetting RTP (in the absence of any potential). Finally, we conclude in Section \ref{conclusionsection}.

%\blue{GS: some Refs are certainly still missing. Please check!}

\section{Main results}\label{Mainresultssection}

\subsection{Definition of the model and differential equations for the MFPT}
We consider a run-and-tumble particle (RTP) whose position is denoted by $x(t) \in [0,+\infty[$. The RTP starts from $x(t=0)=x_0\geq0$ and its equation of motion reads
\begin{equation}
    \frac{dx(t)}{dt} = \dot{x}(t)=f(x) + v_0\, \sigma(t)\, ,
\label{langeRTP}
\end{equation}
where $v_0$ represents a constant speed, and $f(x)$ denotes the force acting on the RTP. 
%This force eventually derives from a potential $V(x)$, i.e., $f(x) = -V'(x)$. The stochastic part of the dynamics is described by a telegraphic noise $\sigma(t)$ which alternates values between $\pm 1$, which thus mimics the tumblings, i.e., the change of direction, of the particle.   
%Here, we consider Poissonian tumblings, i.e., the time intervals $\tau$ between two tumbling events are exponentially distributed with rate $\gamma >0$, such that $p(\tau)=\gamma\, e^{-\gamma \tau}$. Alternatively, the dynamics of the telegraphic noise $\sigma(t)$ can be described as follows. 

This force derives from a potential \( V(x) \), i.e., \( f(x) = -V'(x) \). The stochastic part of the dynamics is driven by a telegraphic noise \( \sigma(t) \), which alternates between the values \( \pm 1 \). This noise mimics the tumblings of the particle, i.e., the random changes in direction. Here, we consider Poissonian tumblings, where the time intervals \( \tau \) between successive tumbling events are exponentially distributed with rate \( \gamma > 0 \), such that \( p(\tau) = \gamma\, e^{-\gamma \tau} \). Alternatively, the dynamics of the telegraphic noise \( \sigma(t) \) can be described as follows. Within the infinitesimal time interval $[t,t+dt]$, the noise $\sigma(t)$ evolves via the rule
\begin{eqnarray}
\sigma(t+dt) = \begin{cases}
\sigma(t) &\text{, with probability } (1 - \gamma \, dt)\\
-\sigma(t) &\text{, with probability } \gamma\, dt 
\end{cases}\, .
\end{eqnarray}
Initially, we have $\sigma(0) = + 1$ or  $\sigma(0) = - 1$, with equal probability $1/2$. Here we are interested in the calculation of $\tau_\gamma(x_0)$ which is 
the mean first-passage time (MFPT) at the origin $x=0$, i.e., the average time it takes for an RTP to reach the origin for the first time, starting from $x_0 \geq 0$. Of course $\tau_\gamma(x_0)$ is also a function of the speed $v_0$ and of the force $f(x)$. Note that the MFPT to a target located at $X>0$ can be easily obtained from the expression of the MFPT to the origin by shifting the potential $V(x) \to V(x+X)$.

To compute the MFPT, it is useful to first consider the survival probabilities $Q^\pm(x_0,t)$. They are the probabilities that the particle remains in the positive region of the real line up to time $t$, with initially $\sigma(0) = \pm 1$. 
One can show~\cite{RTPSS} that these probabilities obey the following coupled differential equations (for completeness, a derivation is provided in Appendix~\ref{survivalappendix})
\begin{eqnarray}
        \partial_t Q^+(x_0,t) = \left[f(x_0) + v_0\right]\partial_{x_0} Q^+(x_0,t) -\gamma\, Q^+(x_0,t) + \gamma\, Q^-(x_0,t)\, ,\label{survivalequation1} \\
        \partial_t Q^-(x_0,t) = \left[f(x_0) - v_0\right]\partial_{x_0} Q^-(x_0,t) -\gamma\, Q^-(x_0,t) + \gamma\, Q^+(x_0,t)\, .
\label{survivalequation2}
\end{eqnarray}
%\begin{figure}[t]
%    \centering
%    \includegraphics[width=0.8\linewidth]{MFPTdef.pdf}
%  \caption{Consider a run-and-tumble particle (represented here by a bacteria) moving in a potential $V(x)$, that starts its motion on the positive region of the line (left plot) at time $t=0$. The first passage time to the origin $T$ is the first time the particle reaches the position $x=0$. In this paper, we compute the mean first passage time to the origin, i.e., the stochastic average of the quantity $T$.}
%  \label{MFPTdef} 
%\end{figure}
If the initial state of the particle is chosen with equal probability, the ``average'' survival probability of an RTP is $Q(x_0,t)=(Q^+(x_0,t)+Q^-(x_0,t))/2$.
From Eqs.~(\ref{survivalequation1}) and (\ref{survivalequation2}), we will now derive coupled differential equations for the MFPTs $\tau_\gamma^\pm(x_0)$, i.e., the MFPT starting from $x_0 \geq 0$ with $\sigma(0) = \pm 1$. To proceed, it is useful to define $Q(x_0,t)$ as the probability that the first-passage time to the origin $T$, {starting from $x_0$,} is larger than $t$, i.e., 
\begin{equation}
   Q(x_0,t) = {\rm Prob.} (T>t) = 1 - {\rm Prob.} (T<t)\, . 
\end{equation}
The survival probability is therefore directly related to the cumulative distribution of the random variable $T$, such that we have the important relation $F_{\rm fp}(x_0,t)=-\partial_t Q(x_0,t)$, where $F_{\rm fp}(x_0,t)$ is the probability density function (PDF) of the first-passage time $T$. This means that $F_{\rm fp}(x_0,t)\, dt$ is the probability that an RTP initially located at $x_0$ reaches the origin for the first time in the interval $[t,t+dt]$. The MFPT $\tau_\gamma(x_0)$ is thus given by
\begin{eqnarray}
    &\tau_\gamma(x_0)&  = \int_0^{+\infty}dt\, t\, F_{\rm fp}(x_0,t) = -\int_0^{+\infty}dt\, t\, \partial_t Q(x_0,t)\,, 
\end{eqnarray}
and similarly 
\begin{eqnarray}    
   & \tau_\gamma^\pm(x_0) &=  -\int_0^{+\infty}dt\, t\, \partial_t Q^\pm(x_0,t)
    \, .
\label{defMFTP}
\end{eqnarray}
To find the coupled differential equations for $\tau_\gamma^\pm(x_0)$ (assuming that both MFPTs are finite), we differentiate Eqs.~(\ref{survivalequation1}) and (\ref{survivalequation2}) with respect to $t$, multiply by $t$, and integrate them over $t$ from $0$ to $+\infty$. Using the initial conditions $Q^\pm(x_0>0,t=0)=1$ and assuming that $Q^\pm(x_0,t=+\infty) = 0$ (otherwise the MFPT is infinite) one obtains~\cite{letter, RTPPSG, Kardar}
\begin{eqnarray}\label{coupledtaupm1}
    && \left[f(x_0)+ v_0\right] \partial_{x_0}\tau_\gamma^+(x_0) - \gamma \tau_\gamma^+(x_0) + \gamma \tau_\gamma^-(x_0)= -1 \, ,\\
&&\left[f(x_0) - v_0\right] \partial_{x_0}\tau_\gamma^-(x_0) + \gamma \tau_\gamma^+(x_0)  -\gamma \tau_\gamma^-(x_0)  = -1\, .
\label{coupledtaupm2}
\end{eqnarray}
From the coupled equations~(\ref{coupledtaupm1}) and (\ref{coupledtaupm2}) one can show that the MFPT $\tau_\gamma(x_0) = \left(\tau_\gamma^+(x_0) + \tau_\gamma^-(x_0)\right)/2$ obeys a closed ordinary differential equation \cite{letter}
\begin{equation}
\left[v_0^2 -f(x_0)^2\right] \partial^2_{x_0} \tau_\gamma(x_0) + 2f(x_0)\left[\gamma-f'(x_0) \right]\partial_{x_0} \tau_\gamma(x_0) = f'(x_0)- 2\gamma\, .
\label{ODE2ndTau}
\end{equation}
The computation of $\tau_\gamma(x_0)$ gives us also the full solution for $\tau_\gamma^\pm(x_0)$. Indeed, by manipulating Eqs. (\ref{coupledtaupm1}) and (\ref{coupledtaupm2}), one can express $\tau_\gamma^\pm(x_0)$ in terms of $\tau_\gamma(x_0)$ and its first derivative as \cite{letter}
\begin{eqnarray}
    &&\tau_\gamma^-(x_0) = \frac{f(x_0)}{2\gamma\, v_0}- \frac{1}{2\gamma v_0}\left[v_0^2-f(x_0)^2\right]\partial_{x_0} \tau_\gamma(x_0) + \tau_\gamma(x_0) \, ,\label{ODEtauminus}\\
&&\tau_\gamma^+(x_0) = - \frac{f(x_0)}{2\gamma\, v_0}+  \frac{1}{2\gamma v_0}\left[v_0^2-f(x_0)^2\right]\partial_{x_0} \tau_\gamma(x_0) + \tau_\gamma(x_0) \, . \label{ODEtauplus}
\end{eqnarray}
Interestingly, the differential equation satisfied by $\tau_\gamma(x_0)$ is a second order differential equation but it is a first order differential equation for $\partial_{x_0} \tau_\gamma(x_0)$. It can thus be solved explicitly for any $f(x)$ up to two integration constants, which have to be fixed by appropriate boundary conditions. {We will see below that these conditions depend in a rather subtle way on the force $f(x)$, which eventually lead to the different ``phases'' discussed below (see Figs. \ref{MFPTPhases12} and~\ref{MFPTPhase34}).}

%A first condition can easily be obtained by noting that if an RTP that starts its motion at the origin with a negative speed, it crosses the origin instantly, thus $\tau_\gamma^-(x_0=0) = 0$ (supposing $f(0)<v_0$). The second condition is in general more subtle to derive and it will greatly depend on the force $f(x)$ as we demonstrate below. 

We conclude this subsection by a comment on the higher moments of the PDF $F_{\rm fp}(x_0,t)$. Here we focus on the MFPT, which is the first moment of this PDF. In fact, starting from Eqs.~(\ref{survivalequation1}) and (\ref{survivalequation2}) and performing manipulations very similar to the ones presented here and leading to 
Eq. (\ref{ODE2ndTau}) -- see Appendix \ref{momentsappendix} -- one can in principle compute recursively the higher moments $\langle T^n \rangle$ of this PDF with $n \geq 2$. These higher moments can be obtained from the following differential equation
\begin{eqnarray}
    && \left(v_0^2-f(x_0)^2\right)\partial_{x_0}^2\langle T^{n} \rangle  + 2f(x_0)\left[\gamma-f'(x_0) \right] \partial_{x_0}\langle T^{n} \rangle \nonumber \\
    &&= n\, (f'(x_0)-2\gamma)\langle T^{n-1} \rangle+2n f(x_0)\partial_{x_0}\langle T^{n-1} \rangle+ n^2 \langle T^{n-2} \rangle \quad, \quad n \geq 2 \;,
\label{highermoments}
\end{eqnarray}
which generalizes the recursion relations found in the passive case \cite{kramers, diffusivemoments} -- see also Appendix \ref{momentsappendix}. Interestingly, the differential operator acting on $\langle T^n \rangle$ on the left hand side of (\ref{highermoments}) does not depend on $n$ -- and is thus the same as in the case $n=1$ in Eq.~(\ref{ODE2ndTau}). However the right hand side depends explicitly on $n$ and involves the lower moments of~$T$. Thus, in principle, these differential equations can be solved recursively. The analysis of this equation for $n \geq 2$ for an arbitrary force field $f(x)$ seems however quite challenging and is deferred to future studies. 

%\gs{Use $\langle T^n \rangle$ also in the Appendix B.}

\subsection{Explicit solutions in different phases}

Having established the differential equations \eqref{ODE2ndTau}-\eqref{ODEtauplus} satisfied by $\tau_\gamma^{\pm}(x_0)$ and $\tau_\gamma(x_0) = (\tau_\gamma^+(x_0) + \tau_\gamma^-(x_0))/2$, we now need to fix the appropriate boundary conditions for these functions. Here we will restrict our analysis to the case where $f(x)$ is continuous. A first boundary condition can be easily found from the following argument. First, we notice that if $f(0) \geq v_0$, the origin can never be crossed, starting from $x_0 \geq 0$ since, from Eq. (\ref{langeRTP}), the force felt by the particle at $0$ is always positive, namely $f(0) \pm v_0 \geq 0$ if $f(0) \geq v_0$. Therefore the computation of the MFPT for $f(0) \geq v_0$ is trivial since $\tau^{\pm}_\gamma(x_0)=\tau_\gamma(x_0) = + \infty$. Hence we only need to consider situations where $f(0) < v_0$. 
Let us then analyse the MFPT starting at the origin $x_0=0$. In this case, from Eq. (\ref{langeRTP}), we see that $\dot{x}(0) = f(0) + \sigma(0) v_0$. Therefore, if $\sigma(0) = -1$, the initial velocity is negative: namely $\dot{x}(0) = f(0) - v_0 < 0$ since $-v_0<f(0)<v_0$ (while the initial velocity $\dot{x}(0) > 0$ if $\sigma(0) = +1$). Thus if $\sigma(0)= -1$ and $x_0=0$ the position of the RTP $x(t)<0$ for $t=0^+$ and we thus obtain the first boundary condition
    \begin{eqnarray} \label{first_cond}
    \tau_\gamma^-(0) = 0 \;,
    \end{eqnarray}
while a priori $\tau_\gamma^+(0) >0$. 

The second boundary condition, which still needs to be fixed, depends crucially on the existence or not of fixed points (or turning points) of the dynamics, i.e., values of $x$ such that $f(x) \pm v_0 = 0$. In general, we need to distinguish different types of turning points~\cite{Velocity_RTP}: negative (respectively positive) turning points when $f(x_-)=-v_0$ (respectively $f(x_+)=v_0$), that may be stable and denoted by $x_{\pm}^s$ (respectively unstable and denoted by $x_{\pm}^u$) if $f'(x_\pm^{s})<0$ (respectively $f'(x_\pm^{u})>0$). The stability of the turning points $x_\pm$ has a strong influence on the behavior of $\tau_\gamma(x_0)$ around $x_0=x_\pm$ (see below and also in Appendix \ref{behavior_xneg}). Specifically, depending on the starting position $x_0$, the particle may become permanently trapped in certain regions of space, making the origin inaccessible from $x_0$. For instance, in the presence of a positive turning point $x_+$, if $x_0>x_+$ the RTP is unable to reach the origin and therefore in this case $\tau_\gamma(x_0) = + \infty$ (see e.g., the right panel of Fig. \ref{MFPTPhase34} when $x_0 > x_+^u$).

To cover most of the different force landscapes $f(x)$, it turns out that it is sufficient to compute the MFPT in four different situations 
where there are either no turning point -- ``Phase I'' and ``Phase II'' -- or only one turning point -- ``Phase III'' and ``Phase IV''. Note that when there is one turning point, since $f(0)<v_0$, this turning point can be either negative-stable (``Phase III'') or positive-unstable (``Phase IV'') -- corresponding to $f(0)>-v_0$ -- as well as negative-unstable -- corresponding to $f(0)<-v_0$. Note however that this turning point can not be positive-stable (because $f(0)<v_0$). In fact, it turns out that the MFPT in the case of a negative-unstable turning point can actually be obtained by combining the results of Phase II and Phase I, as discussed below (see also the right panel of Fig. \ref{MFPTPhase5}). Hence there are only four distinct phases to consider. The formula for the MFPT in more complicated situations, i.e., with a higher number of fixed points can then be obtained by ``gluing'' together the results obtained from these four ``building blocks'', as discussed and illustrated below in some specific examples. 

%
%origin is not accessible to the particle such that $Q^\pm(x_0,t=+\infty) = 1$.  \\
%The Langevin equation (\ref{langeRTP}) has fixed points of the dynamics. There are indeed values of $x$ such that the velocity of the RTP is $\dot{x}(t)=0$, i.e. turning points of the force $f(x) = \pm v_0$ when $\sigma(t) = \mp 1$. 
%
%
%
We now present our results for these four different phases, which lead to different expressions of the MFPT.

\begin{figure}[t]
    \centering
    \includegraphics[width=1\linewidth]{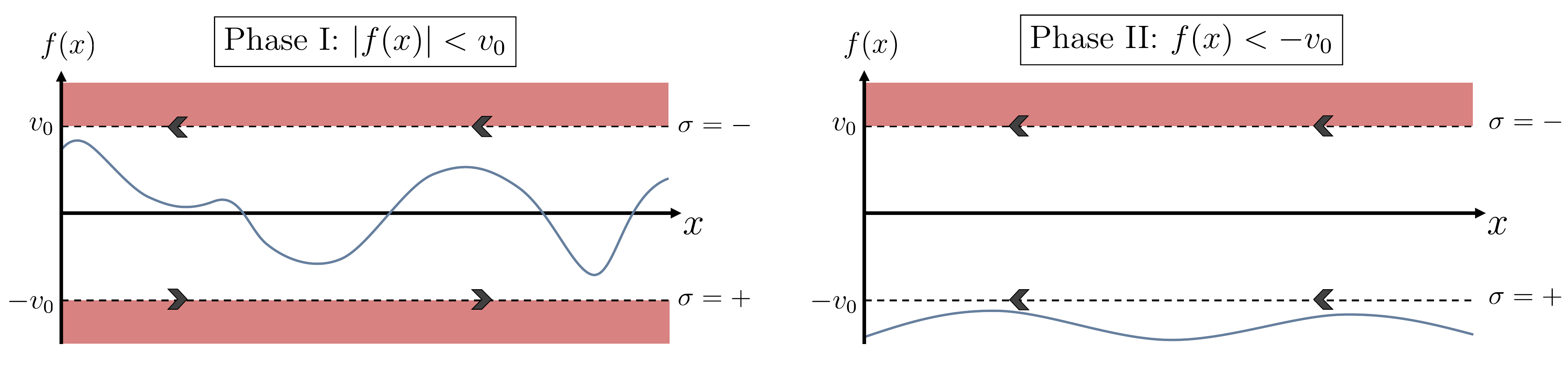}
  \caption{\textbf{Left panel}: We show a force such that $|f(x)|<v_0$ which correspond to phase I. \textbf{Right panel}:  in phase II, the force is bounded such that $f(x)<-v_0$. The arrows on the dotted lines show the direction of the velocity of the RTP in state $\sigma=\pm$. If the arrow is directed to the right (left), the velocity is positive (negative) in this region.}
  \label{MFPTPhases12} 
\end{figure}

\begin{itemize}
    
    \item \textit{Phase I: $|f(x)|<v_0$}. In this case there is no turning point -- see the left panel of Fig. \ref{MFPTPhases12} for a schematic description of this situation. To fix the second boundary condition (in addition to the first one in Eq. \eqref{first_cond}), we impose a reflecting barrier at position $L>x_0$, and then take the limit $L\to \infty$.  One can show that this leads to the following boundary condition (see \cite{letter, RTPPSG})
\begin{equation}
    \partial_{x_0}\tau_\gamma^+(x_0)\Big|_{x_0=L}=0\,.
    \label{condioni}
\end{equation}
One can then solve the Eqs. \eqref{ODE2ndTau}-\eqref{ODEtauplus} with these two boundary conditions (\ref{first_cond}) and (\ref{condioni}) and eventually take the limit $L \to \infty$. This leads to the following expression for the MFPT in this case 
    \begin{equation}
\begin{split}
     \tau_\gamma(x_0) &=   \frac{1}{2\gamma}+\int_{0}^{+\infty} \frac{dy}{v_0-f(y)} \, \text{exp}\left[\int_0^{y}du\, \frac{2\gamma f(u)}{v_0^2-f(u)^2}\right]  \\
     &-\int_0^{x_0}dz\, \frac{1}{v_0^2-f(z)^2} \int_{z}^{+\infty} dy \left(f'(y)- 2\gamma \right) \text{exp}\left[\int_y^{z}du\, \frac{-2\gamma f(u)}{v_0^2-f(u)^2}\right]\,,
\end{split}
\label{phase1solLinf}
\end{equation}
while $\tau_\gamma^{\pm}(x_0)$ can be obtained from Eqs. \eqref{ODEtauminus} and \eqref{ODEtauplus}. 
    
A simple example belonging to this phase I is the linear potential $V(x) = \alpha |x|$, with $-v_0<-\alpha<v_0$. For positive values of $x \geq 0$, the force is $f(x) = -\alpha$. Hence substituting this expression for $f(x)$ in Eq. (\ref{phase1solLinf}), one obtains (see also \cite{letter,Debruyne, SurvivalRPTlinearMORI})  
    \begin{eqnarray}
  \tau_\gamma(x_0) =  \frac{x_0}{\alpha} + \frac{v_0}{2 \alpha \gamma}\, .
  \label{taulinearphase2aINTRO}
\end{eqnarray}

 % \item \textit{Phase I.b.} Suppose first that $-v_0<f(x)<v_0$. Now, assume $\exists X>0$ such that $\forall x \geq X$ we have $f(x) = 0$. In these conditions, for all possible starting positions of the RTP, there is a non-zero probability that the particle reach the region above $X$ where it become diffusive at large time, and thus can go to infinity. Therefore, some contributions to the MFPT are infinite and we conclude  $\tau_\gamma(x_0) = +\infty$ - see the right panel of Figure \ref{MFPTPhases12} . The simplest example one could think of for this phase is the linear potential $V(x) = \alpha |x|$ when $\alpha = 0$, leading to a null force.

    \item \textit{Phase II: $f(x) < -v_0$}. Clearly here there is no turning point -- this phase is illustrated on the right panel of Fig.~\ref{MFPTPhases12}. Since $f(0) < -v_0$, if the particle starts at the origin $x_0 = 0$, the initial velocity $\dot{x}(0) = f(0) + \sigma(0) v_0 < 0$ in any of the two states $\sigma(0)= \pm 1$. In this case, the boundary conditions are thus   
\begin{eqnarray}
     &&\tau_\gamma^+(x_0=0) = 0\, , \label{conditiondtau2b}\\
     &&\tau_\gamma^-(x_0=0) = 0\, . \label{tauminuscondition2b}
\end{eqnarray}
By solving Eq. (\ref{ODE2ndTau}) with these two boundary conditions \eqref{conditiondtau2b} and \eqref{tauminuscondition2b}, we show in 
Appendix \ref{phase2appendix}, that the MFPT is given by
\begin{equation}
    \tau_\gamma(x_0) =   \int_{0}^{x_0}dz\, \frac{1}{v_0^2-f(z)^2}\left[\int_0^{z} dy \left(f'(y)- 2\gamma\right)\text{exp}\left(\int_y^{z}du\, \frac{-2\gamma f(u)}{v_0^2-f(u)^2}\right) + f(0)\, \text{exp}\left(\int_0^{z}du\, \frac{-2\gamma f(u)}{v_0^2-f(u)^2}\right)\right]\, .
\label{solutionf(0)leqmainresults}
\end{equation}

Here also, the linear potential $V(x)=\alpha |x|$  with $\alpha > v_0$
 is an illustration for this phase. By substituting $f(x) = - \alpha < -v_0$ in the general expression \eqref{solutionf(0)leqmainresults} one finds 
\begin{eqnarray}
\tau_\gamma(x_0) = 
\frac{x_0}{\alpha}+ \frac{v_0^2}{2 \alpha^2 \gamma}\left(1- e^{-\frac{2\alpha \gamma x_0}{\alpha^2-v_0^2}}\right)\, ,
\end{eqnarray}
thus recovering the result obtained in \cite{letter,Debruyne, SurvivalRPTlinearMORI}. 

    \item \textit{Phase III.} Here, we consider a force $f(x)<v_0$ such that $|f(0)|<v_0$ but, at variance with Phase I, we assume here that there is a unique stable negative turning point $x_-^s$, i.e., $f(x_-^s) = -v_0$ - see Figure \ref{MFPTPhase34}. This phase was studied in Ref. \cite{letter}. The second condition arises when writing Fokker-Plank equations~(\ref{survivalequation1}) and (\ref{survivalequation2}) at $x_-^s$ and it reads~\cite{letter}  
\begin{equation}
    \lim_{x_0 \to x_-^s} \left(f(x_0)+v_0\right)\partial_{x_0}\tau_\gamma^+(x_0)=0\, .
\label{condionii}
\end{equation}
Solving Eq.~(\ref{ODE2ndTau}) with boundary conditions (\ref{first_cond}) and (\ref{condionii}) leads to 
    \begin{equation}
    \begin{split}
     \tau_\gamma(x_0) &=   \frac{1}{2\gamma}+\int_{0}^{x_-^s} \frac{dy}{v_0-f(y)} \, \text{exp}\left[\int_0^{y}du\, \frac{2\gamma f(u)}{v_0^2-f(u)^2}\right]  \\
     &+\int_0^{x_0}dz\, \frac{1}{v_0^2-f(z)^2} \int_{x_-^s}^{z} dy \left(f'(y)- 2\gamma \right) \text{exp}\left[\int_y^{z}du\, \frac{-2\gamma f(u)}{v_0^2-f(u)^2}\right]\, .
    \end{split}
    \label{phase3sol}
    \end{equation}
For a particle inside the interval $[0,x_-^s[$, the turning point acts as a reflective hard wall explaining the similarities in the formulae (\ref{phase3sol}) and (\ref{phase1solLinf}) (see the Supp. mat. of \cite{letterSM}).
\begin{figure}[t]
    \centering
    \includegraphics[width=1\linewidth]{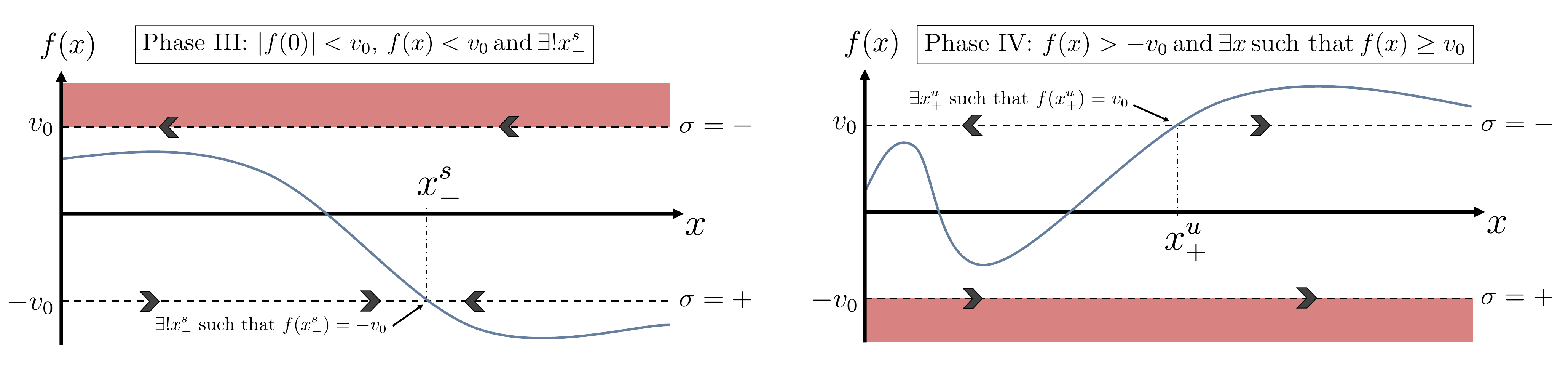}
  \caption{\textbf{Left panel:} we show an example of force from Phase III in which there is a unique stable negative turning point $x_-^s$ and we have $|f(0)|<v_0$ as well as $f(x)<v_0$. \textbf{Right Panel:} in Phase IV, the turning point is positive and unstable. The direction of the velocity of the RTP in state $\sigma=\pm$ is indicated by the arrows on the dotted lines. When the arrow points to the right (left), the velocity is positive (negative) in that region.}
  \label{MFPTPhase34} 
\end{figure}
Phase III includes for instance the harmonic potential $V(x) = \mu\, x^2/2$ with $\mu>0$ leading to a force applied on the RTP given by $f(x) =-\mu\, x$. Since $x>0$ and $\mu >0$, we indeed have  $f(x)< v_0$, and there is a stable negative turning point $x_-^s = v_0/\mu$. This case is studied in details in \cite{letter}. The MFPT is given by
\begin{equation}
    \begin{split}
\tau_\gamma(x_0) &=  \frac{\sqrt{\pi}}{2\gamma} \frac{\Gamma\left(1+\frac{\gamma}{\mu}\right)}{\Gamma\left(\frac{1}{2}+\frac{\gamma}{\mu}\right)}\left[1 + 2\gamma\frac{  x_0}{v_0} \, {}_2 F_1\left(\frac{1}{2},1 +\frac{\gamma}{\mu},\frac{3}{2},\frac{\mu^2x_0^2}{v_0^2}\right)\right]\\
&- (2\gamma + \mu) \frac{x_0^2}{2v_0^2} \, {}_3 F_2\left(\{1,1,\frac{3}{2}+\frac{\gamma}{\mu}\};\{\frac{3}{2},2\};\frac{\mu^2x_0^2}{v_0^2}\right)\, ,
\end{split}
\end{equation}
where $_2F_1(.;z)$ and $_3F_2(.;z)$ are hypergeometric functions \cite{Grad}. Another interesting example is the double-well potential $V(x) = \alpha/2\, (|x|-1)^2$. The force is then $f(x)=-\alpha(x-1)$ and if $0<\alpha<v_0$, this is indeed a phase~III case (see Fig. \ref{doublewellsection}). We study it in section \ref{phase3DWsection}.

\begin{figure}[t]
    \centering
    \includegraphics[width=1\linewidth]{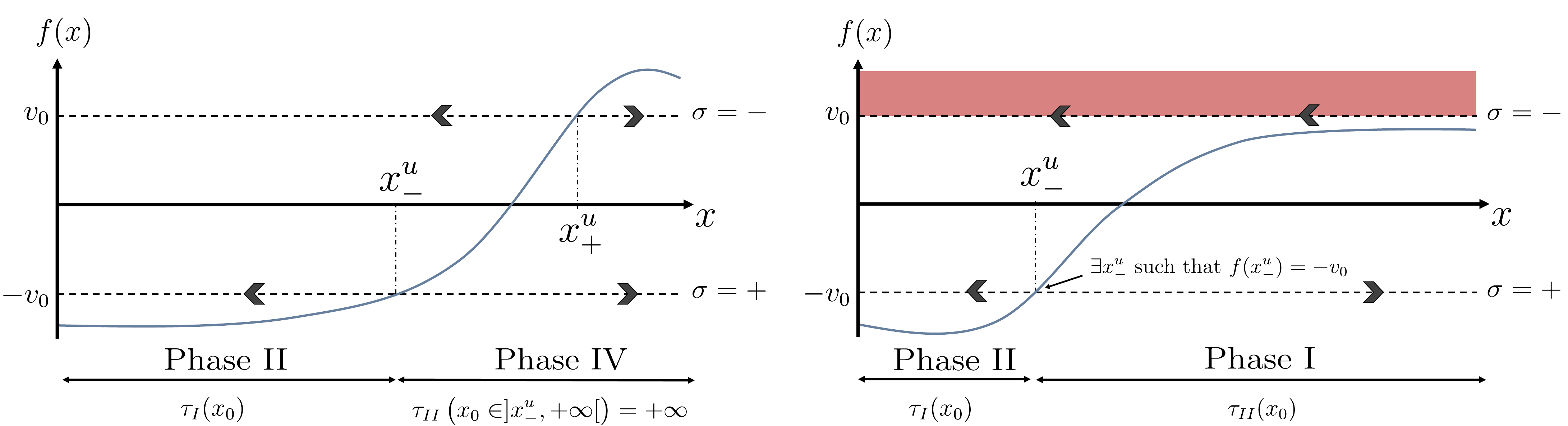}
  \caption{\textbf{Left panel:} the force has an unstable negative turning point $x_-^u$, and an unstable positive turning point $x_+^u$, with $x_+^u>x_-^u$. The force at the origin is such that $f(0)<-v_0$. On $[0,x_-^u[$, the MFPT is derived as in Phase II, while for $x_0>x_-^u$, it diverges as in Phase IV. \textbf{Right panel:} the force has an unstable negative turning point $x_-^u$. We also have $f(0)<-v_0$ and $f(x)<v_0$. On $[0,x_-^u[$ the MFPT is computed as in Phase II. However, on the right of $x_-^u$, the force is such that $|f(x)|<v_0$ and we use continuity at $x_-^u$ to calculate the MFPT. The direction of the RTP's velocity in the state $\sigma = \pm$ is indicated by the arrows on the dotted lines. If you see an arrow pointing to the right (left), it means the velocity is positive (negative) in that particular region.}
  \label{MFPTPhase5} 
\end{figure}
 \item \textit{Phase IV.} Consider a case where $-v_0< f(0) < v_0$ and there is a positive unstable turning point $x_+^u$, i.e., $f(x_+^u) = v_0$ and $f'(x_+^u)>0$. Therefore, if the position of the RTP $x(t)$ is such that $x(t) > x_+^u$ then its velocity is necessarily positive in both states $\sigma(t) = \pm 1$ (see the right panel of Fig. \ref{MFPTPhase34}). Hence, if a particle reaches or surpasses this point $x_+^u$, it will never be able to return to the origin. Clearly, such a trajectory will give an infinite contribution to the MFPT -- see the left panel of Figure \ref{MFPTPhase34} -- and in this case we simply have 
 \begin{eqnarray} \label{tau_phase4}
\tau_\gamma(x_0) = +\infty \;.
 \end{eqnarray}
 An illustration of this phase is the (inverted) double-well potential $V(x) = \alpha/2\, (|x|-1)^2$ when $-v_0<\alpha < 0$ - see section~\ref{doublewellsection}.
 \end{itemize}

These four phases are the building blocks to derive the MFPT for a general force $f(x)$. The general strategy is to study the force 
$f(x)$ on distinct intervals, each corresponding to one of the four phases discussed above. The full expression of the MFPT is then obtained by ensuring the continuity at the boundaries of these intervals. Below, to illustrate this construction, we provide the full solutions in two different situations: 
%for an RTP (i) in a double-well potential in Section \ref{doublewellsection}, and (ii) in a logarithmic potential in  Section \ref{logpotsection}.
 
 \begin{itemize}
 
 \item \textit{Combination of phases: Example 1.} Consider a force such that $f(0) < -v_0$ which displays an unstable negative turning point $x_-^u$, and, in addition, an unstable positive turning point $x_+^u$ such that $x_+^u > x_-^u$. This phase is illustrated on the left panel of Fig.~\ref{MFPTPhase5}. For $x_0 \in [0,x_-^u[$, the MFPT can be computed as a phase II case, while for $x_0 \in ]x_-^u,+\infty[$, this is a phase IV case and the MFPT diverges. An example of such a force is solved in section \ref{phase2and4DWsection}. It corresponds to the inverted double-well $V(x) = \alpha/2(|x|-1)^2$ when $\alpha < -v_0$ (see Fig. \ref{MFPTDWphases}). 

Similarly, we could also consider the case $|f(0)|<v_0$ with a stable turning point $x_-^s$, and an unstable turning point $x_-^u>x_-^s$ in addition to the positive turning point $x_+^u$. In this case, the MFPT on the interval $[0,x_-^u[$ is described by phase III, i.e., Eq.~(\ref{phase3sol}).

\item \textit{Combination of phases: Example 2.} Consider now another, more complicated, example where $f(0) < -v_0$ such that there exists an unstable negative turning point $x_-^u$, and $|f(x)|<v_0$, see the right panel of Fig.~\ref{MFPTPhase5} for an illustration. Concrete examples include for instance $f(x)=-1/(1+x)$ deriving from a log-potential $V(x) = \text{log}(1+|x|)$ that we study in Section \ref{logpotsection} (see Fig. \ref{MFPTLogphase}). This situation requires a careful analysis. Indeed we have to solve the MFPT separately for $x_0 \in [0,x_-^u[$, denoted by ${\tau_{\text{\tiny I}}}_\gamma(x_0)$ and for $x_0 \in ]x_-^u,+\infty[$, denoted by ${\tau_{\text{\tiny II}}}_\gamma(x_0)$. We have here four integration constants to fix (i.e., two for each interval). First, on $[0,x_-^u[$, as $f(0)<-v_0$, it is a phase II case, and ${\tau_{\text{\tiny I}}}_\gamma(x_0)$ is thus given by Eq.~(\ref{solutionf(0)leqmainresults}), that fixes two integration constants. A third integration constant is fixed by imposing the continuity of the MFPT at $x_-^u$. To fix the remaining constant, we introduce a reflecting wall at infinity, as done in phase I. We provide the details of the derivation in Appendix \ref{Phase5derivation} and the solution on $]x_-^u,+\infty[$ is given by
\begin{eqnarray}
    {\tau_{\text{\tiny II}}}_\gamma(x_0)={\tau_{\text{\tiny I}}}_\gamma(x_-^u)-\int_{x_-^u}^{x_0}dz\, \frac{1}{v_0^2-f(z)^2}\, \int^{+\infty}_z dy\, \left(f'(y)-2\gamma\right)\text{exp}\left[\int_y^zdu\, \frac{-2\gamma\, f(u)}{v_0^2-f(u)^2}\right]\, .\label{phase5asolution}
\end{eqnarray}
This formula is valid provided the integrals over $z$ and $y$ are well defined. If this not the case, this means that the MFPT is infinite, i.e., ${\tau_{\text{\tiny II}}}_\gamma(x_0) = + \infty$ for $x_0 > x_-^u$. For instance, if $f(x) \to C$ as $x\to\infty$ with $|C|<v_0$ a constant, then  ${\tau_{\text{\tiny II}}}_\gamma(x_0)$ is finite for $C<0$ and infinite for $C>0$. The marginal case $C=0$ is studied in detail in Section \ref{logpotsection} for the case of the logarithmic potential. This formula (\ref{phase5asolution}) would also hold for potentials of the form $V(x)= \alpha |x|^p$ with $p<1$. Note that we can also consider the case $|f(0)|<v_0$ with a stable turning point $x_-^s$, and another unstable turning point $x_-^u>x_-^s$ -- illustrated for instance by a potential of the form $V(x) = \log(1+x^2)$, as studied in \cite{Velocity_RTP}. The reasoning would be similar, and ${\tau_{\text{\tiny I}}}_\gamma(x_0)$ would be instead described by phase III, i.e., Eq.~(\ref{phase3sol}).

\end{itemize}

{Let us briefly comment on a particular class of forces, namely $f(x)$ that vanishes identically beyond a certain value $X$, i.e., $f(x)=0$ for $x > X$, such as a barrier of potential. In this case, if there is a non zero probability for an RTP to reach the point $X$ then, 
since the RTP behaves diffusively at long times, the MFPT diverges as for the free Brownian motion. As it is a rather peculiar case, we do not consider this class of force in the rest of the paper.}

Finally, we will not consider peculiar forces with a singular turning point $x_\pm$ that is neither stable nor unstable, i.e., with $f'(x_\pm)=0$. However, we have verified in some specific instances that the correct combination of the different phases discussed above yields the complete solution (for example, see the remark in Appendix \ref{phase2appendix}).

\subsection{Some applications of our results for the MFPT}
\begin{figure}[t]
    \centering
    \includegraphics[width=0.4\linewidth]{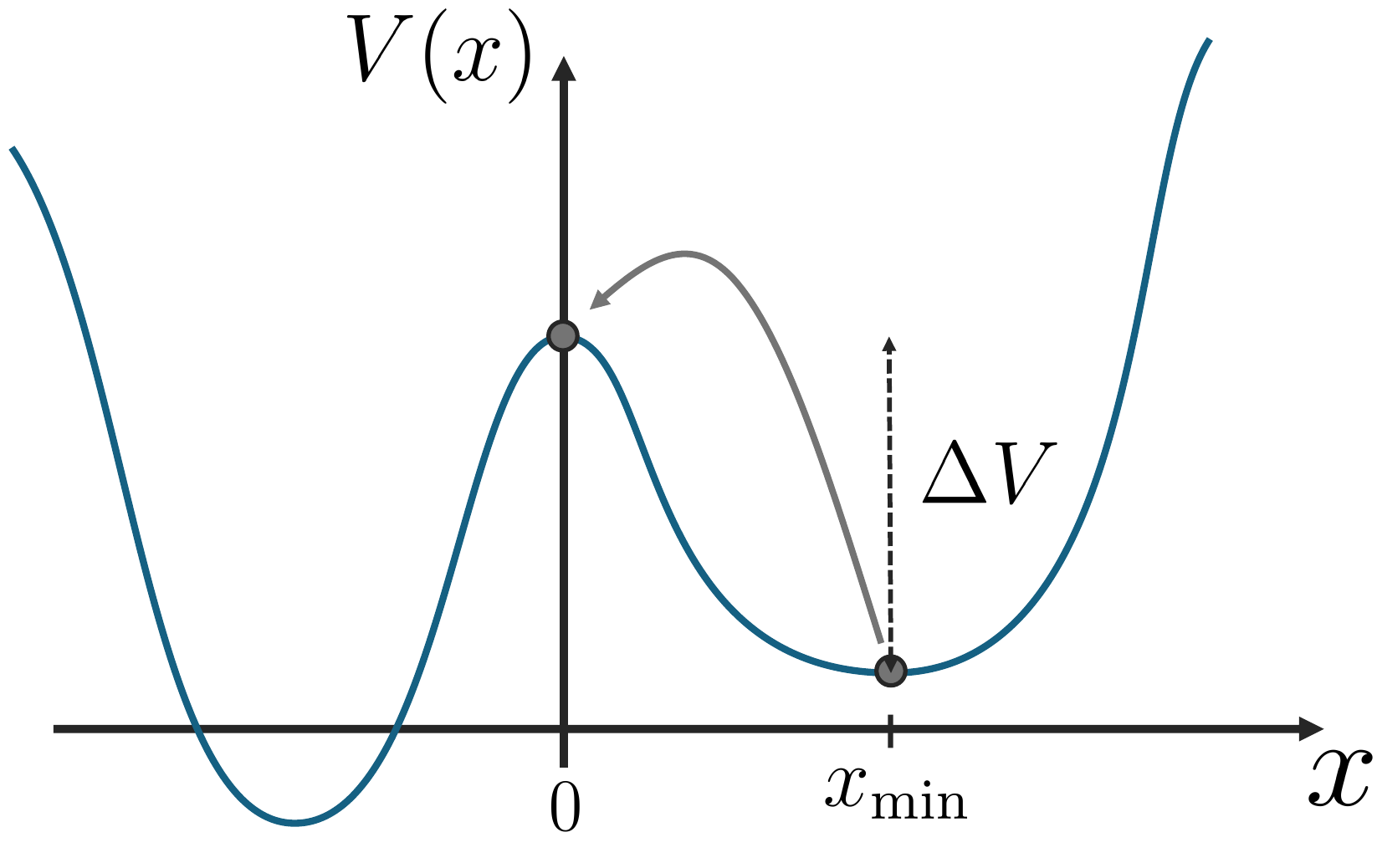}
  \caption{Consider a potential with a local minimum located at $x_{\text{min}}>0$ and a local maximum at the origin $x=0$. We want to estimate the average time needed for a particle to escape from this local minimum. This is given by the mean first-passage time to the origin. For a diffusive particle, it is simply proportional to $\exp(\Delta V/D)$ where $\Delta V$ is the barrier height and $D$ the diffusion coefficient. However, for an RTP, we show that the MFPT is approximated by $\exp(\Delta W/D)$ where $\Delta W$ is the height of an ``active external potentia'' given in Eq.~(\ref{activekramers}).}
  \label{kramersfig} 
\end{figure}

\begin{itemize}

\item \textit{Kramers' law for a one-dimensional RTP.} For a passive or diffusive particle with diffusion coefficient $D$, Kramers' law gives the main contribution to the averaged time needed for a particle to cross a barrier in the weak noise limit $D \ll \Delta V$ where $\Delta V$ is the barrier height. Let us suppose that the potential has a local minimum at $x_{\text{min}}>0$, and it can only escape through the left where a local maximum is located at the origin (see Fig. \ref{kramersfig}). The barrier height is then $\Delta V = V(0)-V(x_{\text{min}})$, and the MFPT is simply proportional to $\exp(\Delta V/D)$ (Arrhenius law). It is natural to ask: how does this Arrhenius law get modified for active particles, such as an RTP? Using our result for the MFPT, we show in section \ref{kramersApplication} that $V(x)$ has to be replaced by an effective ``active external potential'' $W(x)$ \cite{Kardar, RTPPSG}. Indeed, in the weak noise limit $\gamma \to \infty$, we obtain
\begin{eqnarray}
    \log\left(\tau_\gamma(x_0)\right) \isEquivTo{\gamma \to +\infty} \exp\left(\Delta W/D_{\text{eff}}\right)\quad \, , \quad \Delta W =W(0)-W(x_{\text{min}}) = \int_0^{x_{\text{min}}}du\, \frac{f(u)}{1-\frac{f^2(u)}{v_0^2}}\, ,
    \label{activekramers}
\end{eqnarray}
where $D_{\text{eff}} = \frac{v_0^2}{2 \gamma}$. Interestingly, one clearly sees on this expression that the effective active barrier $\Delta W$ is {\it always} greater than the passive one $\Delta V$. 

\item \textit{Relaxation/trapping time of an RTP in a harmonic well.}
When confined in a harmonic potential $V(x)=\mu x^2/2$, an RTP reaches a stationary state with finite support $[-x_e,+x_e]$ with $x_e = v_0/\mu$. In fact, it is easy to see that if the position $x(t_1) \in [-x_e,+x_e]$, it remains trapped inside this interval for all times $t \geq t_1$. Therefore, if the initial position $x_0 \geq x_e$, the first-passage time to $x_e$ is precisely the time after which the particle gets trapped in $[-x_e,+x_e]$ forever. Hence, for $x_0 \gg x_e$, the MFPT to $x_e$ is a good approximation of the relaxation time to the stationary state. In section \ref{relaxationsection} we compute the MFPT to $x_e = v_0/\mu$ for $x_0>v_0/\mu$ and we obtain 
\begin{equation}
\tau_{\rm trap}(x_0,\gamma) = \frac{1}{2\gamma} + \frac{2\gamma + \mu}{\gamma + \mu} \frac{x_0-v_0/\mu}{2v_0} \, {}_3F_2\left(\{1,1,2(1+\frac{\gamma}{\mu})\};\{2,2+\frac{\gamma}{\mu}\};-\frac{\mu\, x_0}{2v_0} + \frac{1}{2}\right)\, .
\end{equation}
In particular, for $x_0 \gg v_0/\mu$, the MFPT (and thus the relaxation time) grows as the logarithm of the initial position $\tau_{\rm trap}(x_0,\gamma)\approx \frac{1}{\mu}\log(\frac{\mu\, x_0}{v_0})$. Note also that this is a monotonically decreasing function of $\gamma$, with $\tau_{\rm trap}(x_0,\gamma) \approx 1/(2\gamma)$ as $\gamma \to 0$ and $\tau_{\rm trap}(x_0,\gamma) \to (1/\mu) \log(\frac{\mu\, x_0}{v_0})$ as $\gamma \to \infty$.

\item \textit{Optimal search strategy.} Finally, we discuss an interesting application of our results in the context of active particles 
subjected to stochastic resetting, i.e., an RTP which is reset to its initial position at exponentially distributed times \cite{resettingPRL, resettingReview, resettingRTP}. At large times, the PDF of the position of a free RTP in the presence of stochastic resetting reaches a stationary form, which is given by a double-exponential distribution \cite{resettingRTP}. This is also the case for an RTP without resetting moving in the presence of a linear potential. In section \ref{resettingsection}, we show that via a certain mapping of the parameters, the two processes  converge to the same stationary state (described by a double-exponential distribution) and we demonstrate that the resetting RTP is more efficient at finding a target than the potential-driven RTP, for all sets of parameters. This generalizes to active particles the result found 
in the diffusive case in \cite{optimalMFPTdiffusive}.

\end{itemize}

\section{MFPT of a Run-and-tumble particle in a double well potential}\label{doublewellsection}

This section is devoted to the calculation of the MFPT $\tau_\gamma(x_0)$ of an RTP moving inside a double well potential $V(x) = \alpha/2\, \left(|x|-1\right)^2$, with $\alpha \in \mathbb{R^*}$. The initial position of the RTP is $x_0\geq 0$. For $\alpha>0$ the local maximum of the double well is located at the origin and computing the MFPT to the origin amounts to answer the following question: what is the mean time for an RTP to escape from one side of the double well to the other side? We also consider $\alpha < 0$, i.e., an inverted double well, which allows us to illustrate the different phases discussed in Section \ref{Mainresultssection}. For $x>0$ the RTP is subjected to the force $f(x) = -\alpha(x-1)$. It turns out that there are four different situations depending on the value of $\alpha$ (see Fig. \ref{MFPTDWphases}). In the following subsections, we calculate the MFPT exactly for any $\alpha$.

\subsection{$-v_0<\alpha<0$ and $\alpha>v_0$ - Phase IV}

\begin{figure}[t]
    \centering
    \includegraphics[width=1\linewidth]{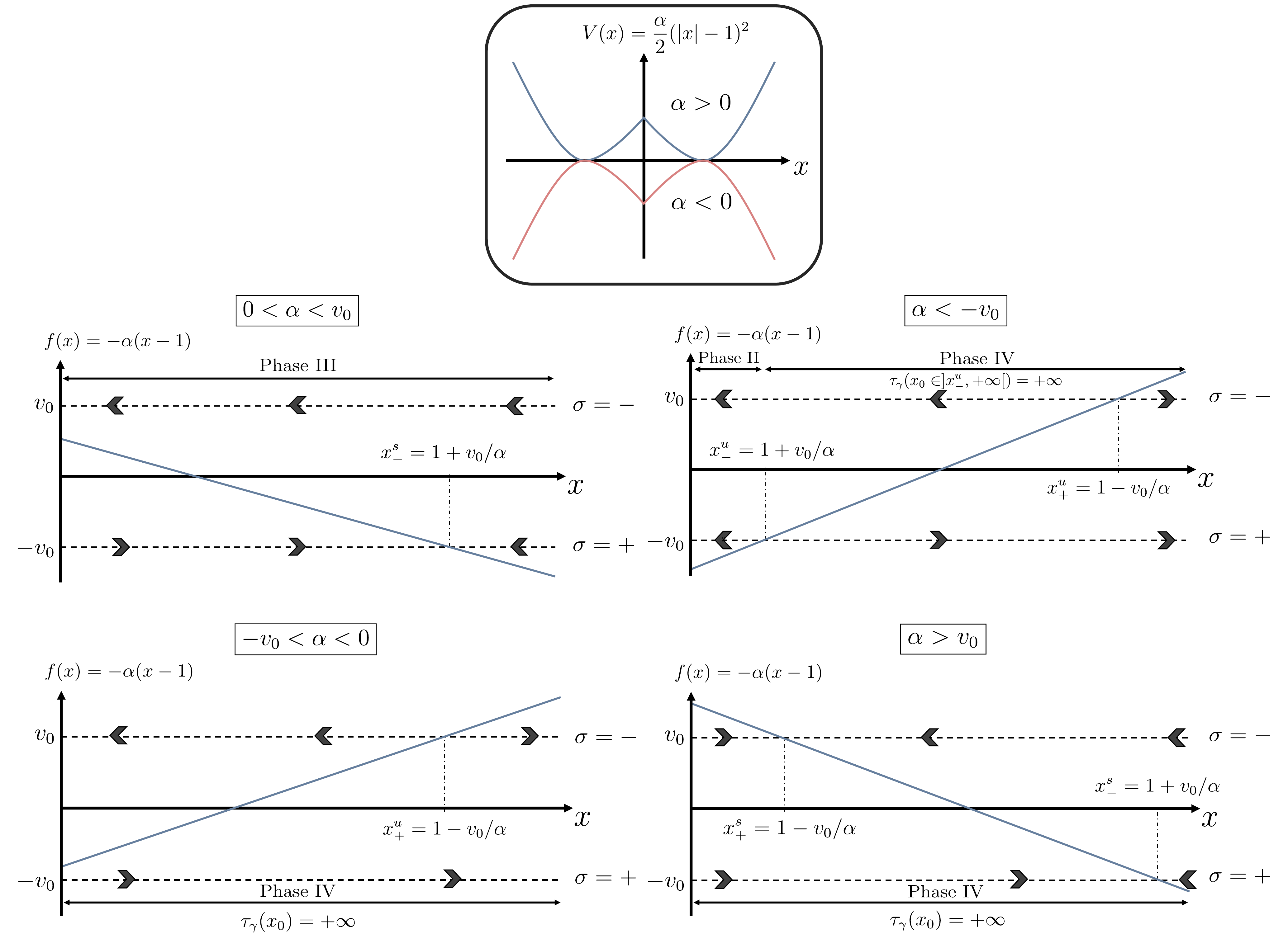}
  \caption{We show here the different behaviours of a force $f(x) = -\alpha(x-1)$ deriving from a double well potential $V(x) = \alpha/2\, \left(|x|-1\right)^2$. The little arrows represent the sign of the velocity in the two states $\sigma = \pm 1$ of the RTP. If the arrow is directed toward the right, it is positive. If it is directed toward the left, the velocity is negative. All these cases are discussed in Section~\ref{doublewellsection}.}
  \label{MFPTDWphases} 
\end{figure}

These two cases are shown in the bottom panels of Fig. \ref{MFPTDWphases}.

\begin{itemize}
    \item When $-v_0<\alpha<0$, the force is always greater than $-v_0$ and there is an unstable positive turning point $x_+^u = 1 - v_0/\alpha$. If the RTP starts its motion above $x_+^u$, it will never be able to return to the origin as the velocity of the particle is positive in both states $\pm$. If the RTP starts its motion bellow $x_+^u$, it has a non-zero probability to reach $x_+^u$ at some finite time, and hence never returns to the origin.
    \item The case $\alpha > v_0$ is similar with a stable positive turning point $x_+^s = 1 - v_0/\alpha$. However, here $f(0)>v_0$ such that no matter where the particle is initialized, it will never go back to the origin.
\end{itemize}

  Therefore in both cases $\tau_\gamma(x_0) = +\infty$.

\subsection{$0<\alpha<v_0$ - Phase III}\label{phase3DWsection}

If $0<\alpha<v_0$ the force $f(x)$ has one unique stable negative turning point $x_-^s = 1+v_0/\alpha$ such that $f(x_-^s)=-v_0$, and of course $|f(0)|=\alpha<v_0$.
The force is plotted on the top right panel of Fig. \ref{MFPTDWphases} and it is an example of Phase III (see Fig. \ref{MFPTPhase34}). In this case, the MFPT is given by Eq.~(\ref{phase3sol}) which reads here
\begin{equation}
    \begin{split}
     \tau_\gamma(x_0) &=   \frac{1}{2\gamma}+\int_{0}^{1+\frac{v_0}{\alpha}} \frac{dy}{v_0+ \alpha(y-1)} \, \text{exp}\left[\int_0^{y}du\, \frac{-2\gamma\, \alpha(u-1)}{v_0^2-\alpha^2(u-1)^2}\right]  \\
     &-\int_0^{x_0}dz\, \frac{1}{v_0^2-\alpha^2(z-1)^2} \int_{1+\frac{v_0}{\alpha}}^{z} dy\,  \left(\alpha+ 2\gamma \right)\,  \text{exp}\left[\int_y^{z}du\, \frac{2\gamma \, \alpha (u-1)}{v_0^2-\alpha^2(u-1)^2}\right]\, .
    \end{split}
    \label{phase3DWequ}
    \end{equation}
After nontrivial manipulations, it is possible to compute explicitly the integrals in Eq.~(\ref{phase3DWequ}) 
in terms of hypergeometric functions, and we give the expression in Appendix \ref{taupmexpressions} in Eq.~({\ref{MFPTDW1}}). This complicated expression becomes simpler for certain values of the parameters. For instance, for $\gamma = \alpha$, the MFPT $\tau_\gamma(x_0)$ takes a simpler form, namely
\begin{eqnarray}
    \tau_\gamma(x_0)|_{\gamma=\alpha} = \frac{1}{\alpha}\left[\frac{2\, v_0}{v_0-\alpha} - \frac{v_0}{v_0+\alpha(x_0-1)} + \text{log}\left(\frac{v_0+\alpha(x_0-1)}{v_0-\alpha}\right)\right]\, .
\end{eqnarray}
\begin{figure}[t]
    \centering
\includegraphics[width=0.8\linewidth]{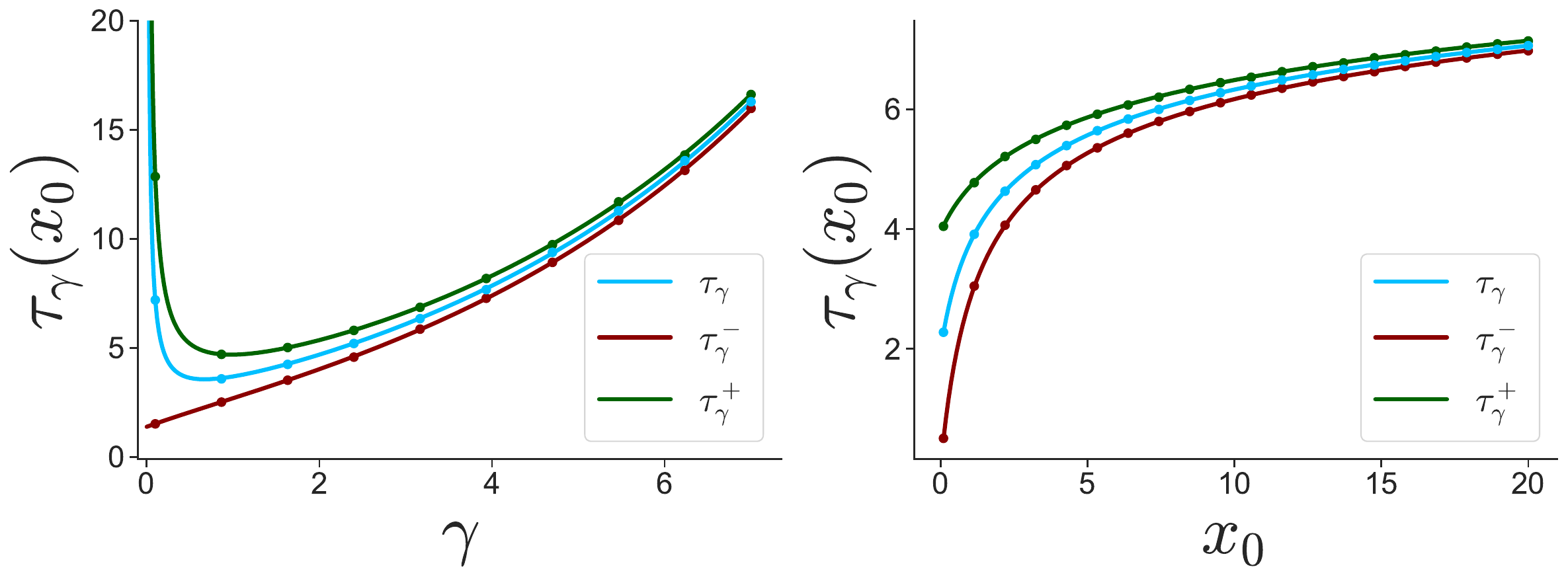}
  \caption{Plot of the MFPT for an RTP in a double well $V(x)=\alpha/2\, (|x|-1)^2$ when $0<\alpha<v_0$. The blue line corresponds to the expression for $\tau_\gamma(x_0)$ given in Eq.~(\ref{MFPTDW1}), while the red (respectively the green) line represents  $\tau_\gamma^-(x_0)$ in Eq.~(\ref{minusMFPTDW}) (respectively $\tau_\gamma^+(x_0)$ in Eq.~(\ref{plusMFPTDW})). The dots represent results of our numerical simulations. On the left panel, we show a plot of the MFPT with respect to the tumble rate $\gamma$ and fixed values of the parameters $\alpha = 1$, 
$v_0= 2$, and $x_0 = 1$. As discussed in the text, $\tau_\gamma(x_0)$ and $\tau_\gamma^+(x_0)$ exhibit a minimum at an optimal value $\gamma = \gamma_{\rm opt}$ (see also Ref. \cite{letter}). On the right panel, we show a plot of the MFPT as a function of the initial position $x_0$, with $\alpha = 1$, 
$v_0= 2$, and
$\gamma = 1.1$.}
  \label{MFPTDWsimu1} 
\end{figure}
One can also calculate both the mean first-passage times $\tau^\pm_\gamma(x_0)$ when the initial state is $\sigma(0) = \pm 1$. They can be obtained from Eqs.~(\ref{ODEtauminus}) and~(\ref{ODEtauplus}). The obtained expressions are however a bit cumbersome and we present them in the Appendix \ref{taupmexpressions}. These explicit expressions for $\tau_\gamma(x_0)$ and $\tau_\gamma(x_0)^\pm$ are very useful for numerical evaluations.
In Fig.~\ref{MFPTDWsimu1}, we present simulation results that we compare to our analytical expressions of the MFPT where we plot $\tau_\gamma(x_0)$ as well as $\tau_\gamma^{\pm}(x_0)$ both as a function of $\gamma$ (left panel) and as a function of $x_0$ (right panel). As we see, the agreement is excellent. Below, we analyse the asymptotic behaviors of $\tau_\gamma(x_0)$ as a function of $\gamma$ and $x_0$.

\vspace*{0.5cm}
\noindent\textbf{The limit $x_0 \to 0$.} In this case, one has $\tau_\gamma^-(0) =0$ (see Eq. (\ref{first_cond})), but $\tau_\gamma^+(0) > 0$. Indeed, from Eq. (\ref{MFPTDW1}) one has
\begin{eqnarray}
\lim_{x_0 \to 0} \tau_\gamma(x_0) = \tau_\gamma(0) = \frac{1}{2} \tau_\gamma^+(0) =  \frac{1}{{2 \gamma}} + \frac{{\alpha + v_0}}{{2 v_0 (\alpha + \gamma)}} \left(\frac{{2}}{{1 - \frac{{\alpha}}{{v_0}}}}\right)^{\frac{{\gamma}}{{\alpha}}} \ {}_{2}F_{1}\left(1 + \frac{{\gamma}}{{\alpha}}, 1 - \frac{{\gamma}}{{\alpha}}, 2 + \frac{{\gamma}}{{\alpha}}, \frac{{1}}{{ 2}} + \frac{{\alpha }}{{2 v_0}}\right) \;.
\end{eqnarray}

\vspace*{0.5cm}
\noindent\textbf{The limit $x_0 \to \infty$.} The large $x_0$ behavior of $\tau_\gamma(x_0)$ can be obtained from the general expression (\ref{MFPTDW1}), leading to 
\begin{eqnarray}
\tau_\gamma(x_0)\simeq \frac{1}{\alpha}\, \text{log}(x_0) \quad, \quad x_0 \to \infty \;. 
\end{eqnarray}
In fact, this is expected because for $x \gg 1$, $V(x) \sim \frac{\alpha}{2}\, x^2$ such that the double well behaves, far from the origin, as an effective harmonic potential for which the large $x_0$ behaviour of the MFPT is known to be logarithmic \cite{letter}.

As demonstrated in \cite{letter} for a potential of the form $V(x)=\alpha |x|^p$ ($p>1$), the MFPT exhibits a minimum with respect to $\gamma$ due to the trapping of the negative stable fixed point $x_-^s$. Here, when $0<\alpha<v_0$ there is also a negative stable fixed point $x_-^s$ and one can argue that in both limits $\gamma \to \infty$ and $\gamma \to 0$ the MFPT diverges, indicating the presence of a minimum.

\vspace*{0.5cm}
\noindent\textbf{The limit $\gamma \to 0$.} It is possible to show that the expression of the MFPT in Phase III (\ref{phase3sol}) behaves as  $\lim_{\gamma \to 0} \tau_\gamma(x_0) \sim \frac{1}{2\gamma}$ (for any $f(x)$ in Phase III) and hence the MFPT diverges \cite{letter}. For the double well potential, this can be checked explicitly on the exact formula (\ref{phase3DWequ}). 
Note however that $\lim_{\gamma \to 0} \tau_\gamma^-(x_0)$ is a constant which can be obtained by integrating the Langevin equation (\ref{langeRTP}) setting $\sigma(t)=-1$ for all times $t$, leading to
\begin{equation}
    \tau_{\gamma=0}^-(x_0)= -\int_{0}^{x_0} \frac{dx}{f(x)-v_0}\, .
\label{smallgammadw}
\end{equation}
This is because for all $x_0>0$, in the negative state of the RTP, the velocity is negative and directed toward the origin.

\vspace*{0.5cm}
\noindent\textbf{The limit $\gamma \to +\infty$.} This interesting limit is studied in section \ref{kramersApplication}, where we also compute the average time needed for an RTP to jump from one side of the double-well to the other, in the weak noise limit $\gamma \to \infty$ (i.e., the Kramer's formula generalized to RTP).\\

\subsection{$\alpha<-v_0$ - Combination of Phase II and Phase IV}\label{phase2and4DWsection}

This case is shown on the top right panel of Fig. \ref{MFPTDWphases}. The specificity of this case is that the dynamics has two turning points. The first one is an unstable negative turning point $x_-^u=1+v_0/\alpha$, while the other one is an unstable positive turning point $x_+^u= 1-v_0/\alpha$, with $x_-^u < x_+^u$. Here we need to distinguish two different situations $x_0 > x_-^u$ and $x_0 < x_-^u$, which we discuss separately.

\subsubsection{$x_0 > x_-^u$ - Phase IV}
When the RTP starts to the right of $x_-^u$, i.e., $x_0 > x_-^u$, the force inside the interval $]x_-^u, +\infty[$ is an example of Phase~IV (see Fig. \ref{MFPTPhase34}). For such potential, the particle has indeed a non-zero probability to reach $x_+^u$ and, in this case, will never return back to the origin. We conclude that here $\tau_\gamma(x_0) = +\infty$ for $x_0 \in ]x_-^u, +\infty[$.

\subsubsection{$x_0 \leq x_-^u$ - Phase II}\label{phase2DWsubsection}
\begin{figure}[t]
    \centering
    \includegraphics[width=0.8\linewidth]{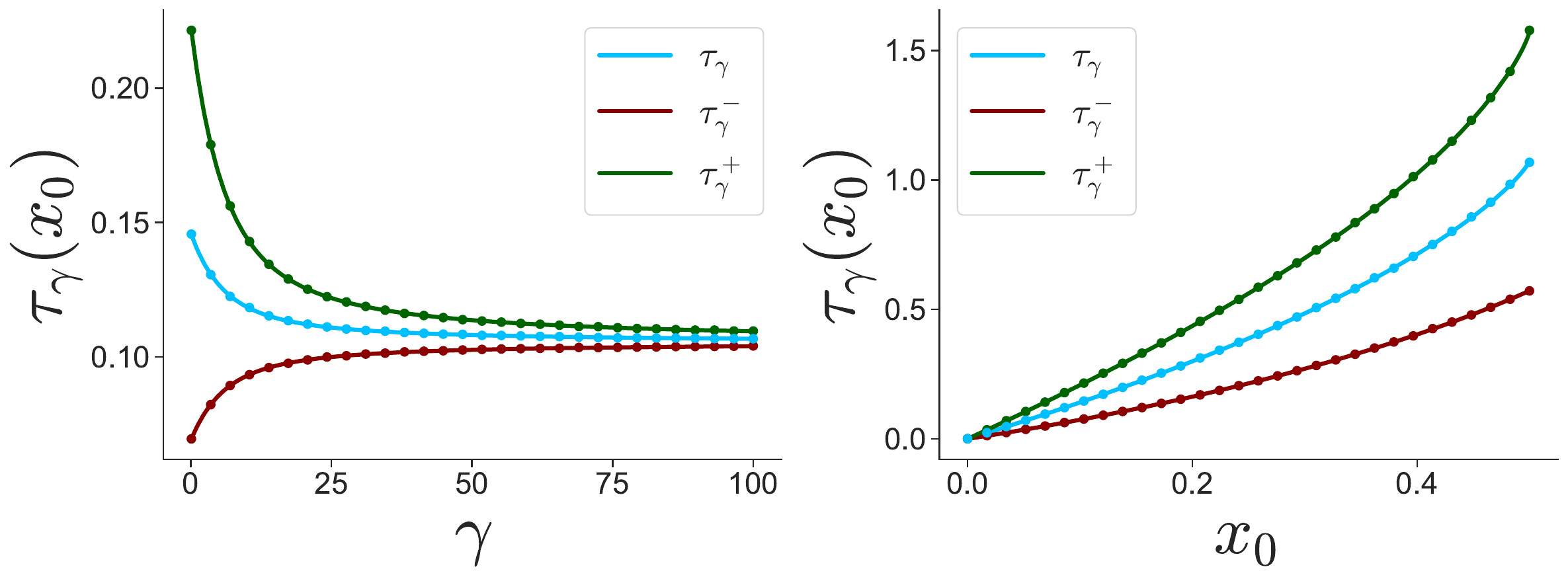}
  \caption{Plot of the MFPT for an RTP in a double well potential $V(x) = \alpha/2\, (|x|-1)^2$, when $\alpha<-v_0$. The blue line represents the MFPT $\tau_\gamma(x_0)$ given in Eq.~(\ref{TAUdwsecondformul2}), while the red and green lines correspond to $\tau_\gamma^-(x_0)$ from Eq.~(\ref{dw_E3}) and $\tau_\gamma^+(x_0)$ from Eq.~(\ref{dw_E4}), respectively. The dots on the graph represent results from numerical simulations.
The left panel displays the MFPT as a function of the tumble rate $\gamma$, while keeping the parameters fixed at $\alpha = -1$, $v_0 = 0.5$, and $x_0 = 0.1$. On the right panel, we plot the MFPT as a function of the initial position $x_0$, with $\alpha = -1$, $v_0 = 0.5$, and $\gamma = 1$.}
  \label{MFPTDWsimu2} 
\end{figure}

Here, as we have $f(0) = \alpha <-v_0$, the RTP, at the origin, has a negative velocity in both states $\sigma = \pm 1$ and thus for $x_0 \in [0,x_-^u[$, this situation is an instance of Phase II. We can thus directly apply Eq.~(\ref{solutionf(0)leqmainresults}) which reads in this case
\begin{equation}
    \tau_\gamma(x_0) =   \int_{0}^{x_0}dz\, \frac{1}{v_0^2-\alpha^2(z-1)^2}\left[\alpha\, \text{exp}\left(\int_0^{z}du\, \frac{2\gamma\, \alpha(u-1)}{v_0^2-\alpha^2(u-1)^2}\right)-\int_0^{z} dy \left(\alpha+ 2\gamma\right)\text{exp}\left(\int_y^{z}du\, \frac{2\gamma\, \alpha(u-1)}{v_0^2-\alpha^2(u-1)^2}\right) \right]\, .
    \label{TAUdwsecondformul}
\end{equation}
{The integrals in Eq.~(\ref{TAUdwsecondformul}) can be computed explicitly, once more in terms of hypergeometric functions, and the expression is given in Appendix \ref{taupmexpressions} in Eq.~(\ref{dw_E3})}. The computation of both MFPTs, $\tau_\gamma^\pm(x_0)$, can also be performed using Eqs.~(\ref{ODEtauminus}) and~(\ref{ODEtauplus}). The resulting expressions are given in Eqs. (\ref{dw_E3}) and (\ref{dw_E4}).
In Fig.~\ref{MFPTDWsimu2}, we compare our analytical results for the MFPT in Eq.~(\ref{TAUdwsecondformul2}) to numerical simulations, showing an excellent agreement. {Now, we give the asymptotic behaviors of $\tau_\gamma(x_0)$ with respect to $x_0$ and $\gamma$.}

\vspace*{0.5cm}

\noindent \textbf{The limit $x_0 \to 0$.} In contrast to the right panel of Fig.~\ref{MFPTDWsimu1}, here when $x_0\to 0$, $\tau_\gamma(x_0)$, $\tau_\gamma^+(x_0)$, and $\tau_\gamma^-(x_0)$ all tend to zero. As discussed in Section \ref{Mainresultssection}, this is a characteristic of Phase II -- see Eqs. \eqref{conditiondtau2b} and \eqref{tauminuscondition2b} -- since, when $x_0=0$, the RTP crosses the origin at $t=0^+$ with probability one. From the general formula in Eq. (\ref{solutionf(0)leqmainresults}), one finds that the leading order behavior of $\tau_\gamma(x_0)$ for $x_0 \to 0$ is given by (for $f(0)< -v_0$)
\begin{eqnarray} \label{taug_smallx0}
\tau_\gamma(x_0) = \frac{|f(0)|}{f(0)^2- v_0^2}\, x_0 + o(x_0) \;.
\end{eqnarray}
Note that this coefficient diverges when $f(0) \to -v_0$. In the case of the of the double potential, this divergence can be qualitatively 
understood since $x_-^u \to 0$ as $f(0) = \alpha \to -v_0$.

\vspace*{0.5cm}

\noindent \textbf{The limit $x_0 \to x_-^u$, and $x_0<x_-^u$.} In this limit, for a general potential belonging to Phase II, 
the behavior of $\tau_\gamma(x_0)$ is perfectly regular (for $\gamma >0$) and given by $\tau_\gamma(x_-^u)$ where $\tau_\gamma(x_0)$ is given in Eq. (\ref{solutionf(0)leqmainresults}). For the double well potential, it can be evaluated explicitly from Eq. (\ref{TAUdwsecondformul2}) setting $x_-^u = 1+v_0/\alpha$. The expression is however a bit cumbersome and we do not give it here.

\vspace*{0.5cm}

\noindent\textbf{The limit $\gamma \to 0$.} In the limit $\gamma \to 0$, it is easy to see that both $\lim_{\gamma \to 0} \tau_\gamma^\pm(x_0)$ are constants, which, as before in Eq. (\ref{smallgammadw}), can be computed by integrating the Langevin equation (\ref{langeRTP}) -- since for $0<x_0<x_-^u$, the velocity is negative in both states $\sigma = \pm 1$). This leads to 
\begin{equation}
    \tau_{\gamma=0}^{\pm}(x_0)= -\int_{0}^{x_0} \frac{dx}{f(x) \pm v_0}\,  = -\int_{0}^{x_0} \frac{dx}{\alpha(1-x) \pm v_0} = \frac{1}{\alpha} \ln \left( \frac{\alpha(1-x_0) \pm v_0}{\alpha \pm v_0}\right) \quad, \quad 0 \leq x_0 \leq x_-^u = 1 + \frac{v_0}{\alpha}  \;, \label{smallgamma}
\end{equation}
which is diverging when $\alpha \to -v_0$ as well as when $x_0 \to x_-^u$. 

\vspace*{0.5cm}
\noindent
\textbf{The limit $\gamma \to +\infty$.} For $x_0 <x_-^u$, the potential is a monotonous function of $x_0$ and the force is pulling the particle toward the origin. The large $\gamma$ limit corresponds to the diffusive limit of an RTP with a vanishing diffusive constant $D=v_0^2/(2\gamma)\to 0$ \cite{letter}. The Langevin equation is thus simply $\dot{x}(t) = f(x)$ such that the MFPT reads
\begin{equation}\label{largegammadw}
    \lim_{\gamma\to \infty}\tau_\gamma(x_0) = \lim_{\epsilon \to 0} -\int_\epsilon^{x_0}\frac{dx}{f(x)} = \lim_{\epsilon \to 0} \int_\epsilon^{x_0}\frac{dx}{\alpha(x-1)} = \frac{1}{\alpha}\log(x_0-1)\, .
\end{equation}
For the choice of parameters of the left panel of Fig.~\ref{MFPTDWsimu2}, it gives $\lim_{\gamma \to +\infty} \tau_\gamma(x_0)\approx 0.105361$ which is in agreement with the data.

\section{MFPT of a Run-and-tumble particle in a Log-potential}\label{logpotsection}

\begin{figure}[t]
    \centering
    \includegraphics[width=0.5\linewidth]{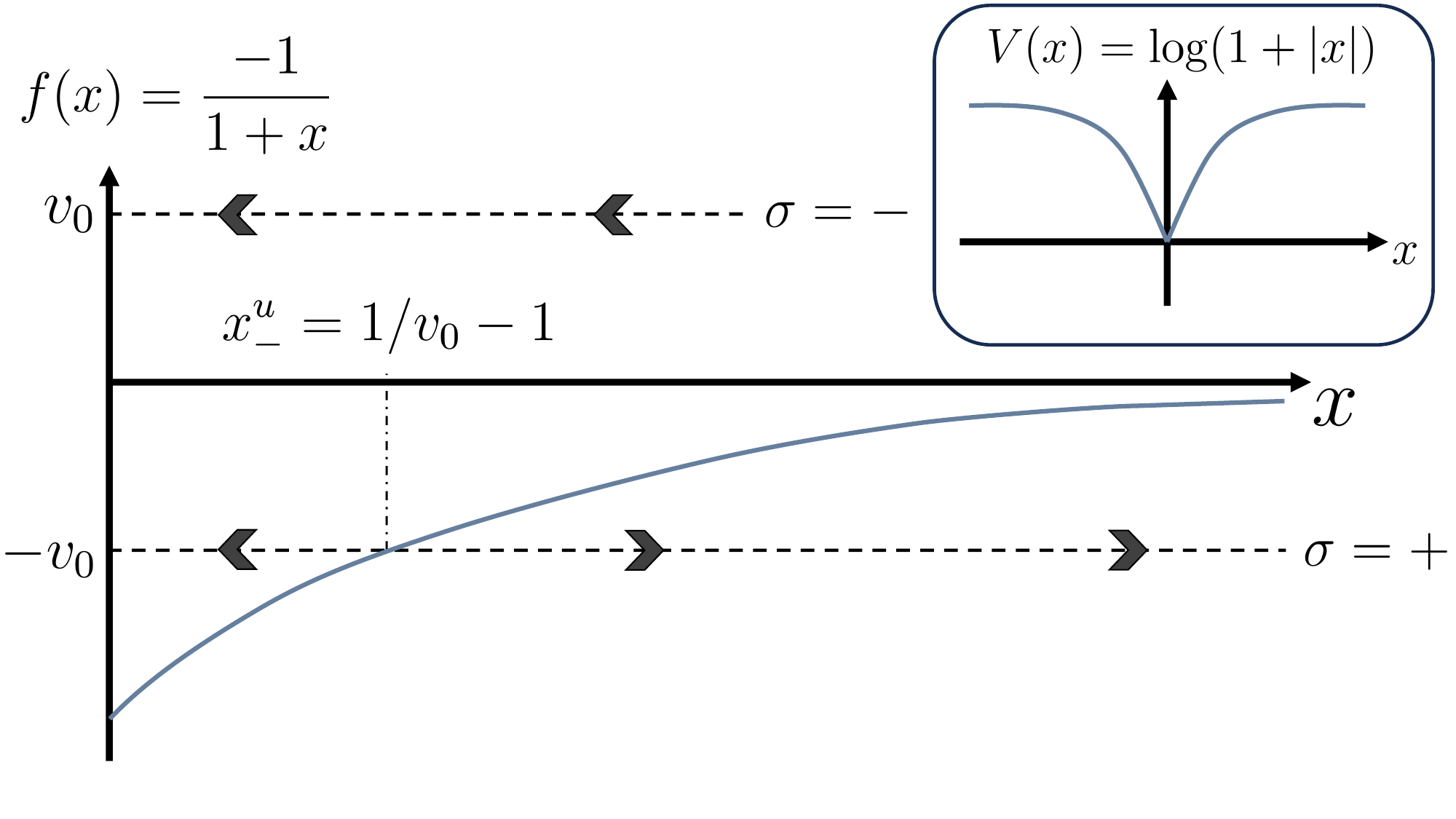}
  \caption{Schematic description of a log-potential $V(x) = \text{log}(1+|x|)$, with the associated force $f(x) = -V'(x) = - 1/(1+x)$ ($x>0$). The force has a unstable negative turning point $x_-^u=1/v_0 - 1$. One has to compute the MFPT first on $x_0 \in [0,x_-^u[$ using Eq.~(\ref{solutionf(0)leqmainresults}) and then on  $x_0 \in ]x_-^u,+\infty[$ using Eq.~(\ref{phase5asolution}).}
  \label{MFPTLogphase} 
\end{figure}
In this section, we discuss the specific case of a logarithmic potential $V(x) = \nu\, \text{log}(1+|x|)$ and, for simplicity we consider only the case $\nu = 1$. For $x>0$, the force is $f(x) = -V'(x) = - 1/(1+x)$. We consider $v_0< 1$ such that the dynamics has an unstable negative turning point $x_-^u=1/v_0 - 1$, i.e., $f(x_-^u) = -v_0$ (see Fig. \ref{MFPTLogphase}). This case is an example of the combination of two phases shown in the right panel of Fig.~\ref{MFPTPhase5}. When $x_0 \in [0,x_-^u[$ this is a Phase II case, and the MFPT on this interval, which we denote ${\tau_{\text{\tiny I}}}_\gamma(x_0)$, is given by Eq.~(\ref{solutionf(0)leqmainresults}). When $x_0 \in ]x_-^u,+\infty[$, the value of the force remains strictly between $-v_0$ and $0$ and there is no other turning point. To treat this situation, as in the case of Phase I \cite{letter,Velocity_RTP}, we introduce a reflecting barrier at $L$ (and take the limit $L \to \infty$) to obtain ${\tau_{\text{\tiny II}}}_\gamma(x_0)$. Finally, we need to impose continuity, i.e., ${\tau_{\text{\tiny I}}}_\gamma(x_-^u)={\tau_{\text{\tiny II}}}_\gamma(x_-^u)$ to fully determine the MFPT on the whole positive real line. The MFPT ${\tau_{\text{\tiny II}}}_\gamma(x_-^u)$ is then given by Eq.~(\ref{phase5asolution}).

\subsection{$x_0 \in [0,x_-^u[$ }
We consider here that the initial position of the particle is such that $x_0 \in [0,x_-^u[$. We have $f(0) < -v_0$ such that $\tau_\gamma^-(0)=\tau_\gamma^+(0)=0$. We can compute the MFPT using the formula~(\ref{solutionf(0)leqmainresults}) of Phase II to obtain
\begin{equation}\label{log_integrals_leftTP}
    {\tau_{\text{\tiny I}}}_\gamma(x_0) =   \int_{0}^{x_0}dz\frac{(1+z)^2}{v_0^2(1+z)^2-1}\left[\int_0^{z} dy \frac{1-2\gamma(1+y)^2}{(1+y)^2}\text{exp}\left(\int_y^{z}du\, \frac{-2\gamma (1+u)}{1-v_0^2(1+u)^2}\right) -\, \text{exp}\left(\int_0^{z}du\, \frac{-2\gamma (1+u)}{1-v_0^2(1+u)^2}\right)\right]\, .
\end{equation}
\begin{figure}[t]
    \centering
    \includegraphics[width=0.8\linewidth]{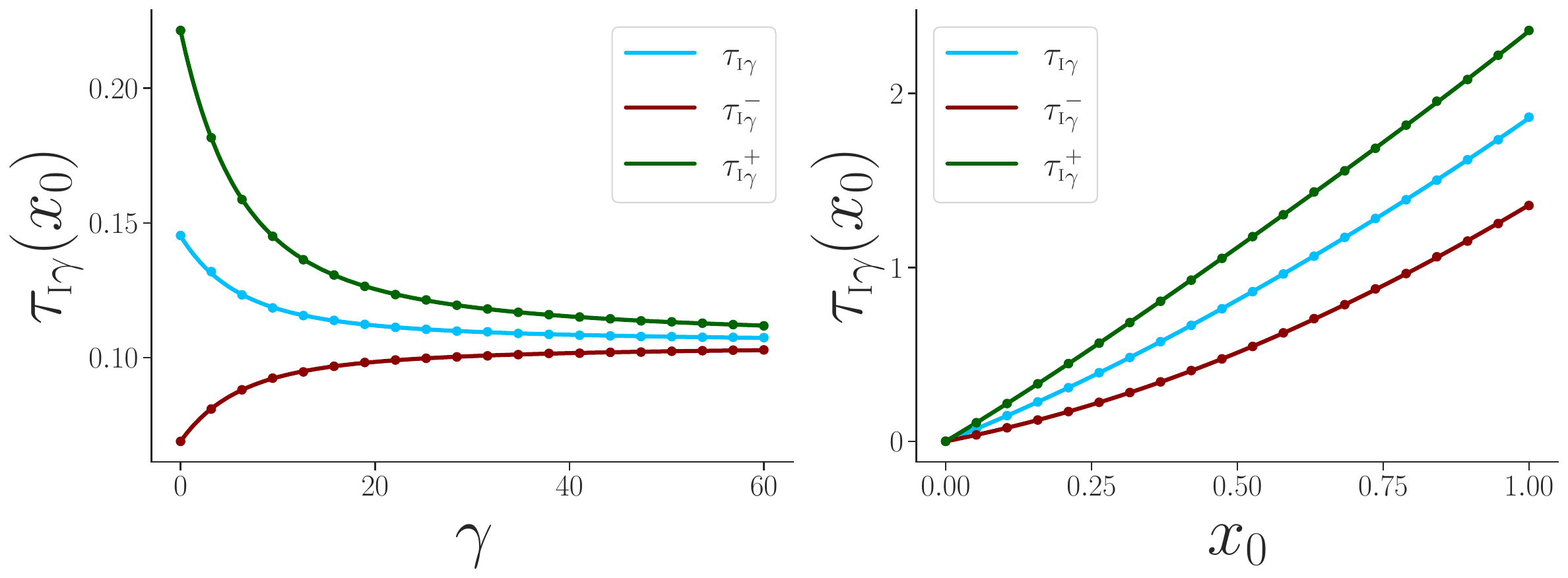}
  \caption{For $x_0 \in [0,x_-^u[$, we show results of simulation for the MFPT of an RTP (dots) inside a logarithmic potential $V(x)=\text{log}(1+|x|)$ and we compare them with our analytical solutions (lines). On the left, we plot the MFPT with respect to $\gamma$ -
$v_0= 0.5$, and $x_0 = 0.1$. On the right, we show the MFPT with respect to $x_0$ - $v_0= 0.5$, and
$\gamma = 1$.}
  \label{MFPTLogsimu1} 
\end{figure}
It is again possible to compute explicitly these integrals, and we give the result in Appendix \ref{taupmexpressions} in Eq.~({\ref{MFPTphase2LOGpot}}). Using Eqs.~(\ref{ODEtauminus}) and~(\ref{ODEtauplus}), one can also calculate the MFPTs $\tau_\pm$ and their  expressions are given in Eqs. (\ref{tauminuslog}) and (\ref{taupluslog}). In Fig.~\ref{MFPTLogsimu1} we show the results of simulation that we compare with our analytical expressions, showing an excellent agreement. Note that the results are qualitatively similar to the double well when $\alpha <-v_0$ (see Fig. \ref{MFPTDWsimu2}). This can be understood since for small $x \ll 1$, $f(x) = - 1/(1+x) \sim x-1$, which is precisely the force corresponding to a double-well potential $V(x) = \alpha/2\, (|x|-1)^2$ when $\alpha = -1$. In the following, we study ${\tau_{\text{\tiny I}}}_\gamma(x_0)$ as a function of $\gamma$ and $x_0$. In particular, we discuss interesting behaviors close to the negative unstable turning point $x_-^u$.

\vspace*{0.5cm}

\noindent \textbf{The limit $x_0 \to 0$.} In this limit, to leading order for small $x_0$, the results are similar to the one obtained in the previous section \ref{phase2and4DWsection}, see Eq. (\ref{taug_smallx0}). 

\vspace*{0.5cm}

\noindent \textbf{The limit $x_0 \to x_-^u$, and $x_0<x_-^u$.} In this limit, ${\tau_{\text{\tiny I}}}_\gamma(x_0) \to {\tau_{\text{\tiny I}}}_\gamma(x_-^u)$, which is finite (see Fig. \ref{MFPTLogsimu1}) and given by Eq.~(\ref{MFPTphase2LOGpot}) setting $x_0 = x_-^u$. The behavior of ${\tau_{\text{\tiny I}}}_\gamma(x_0) - {\tau_{\text{\tiny I}}}_\gamma(x_-^u)$ can be analysed for a generic force from the formula given in Eq.~(\ref{solutionf(0)leqmainresults}) and, interestingly, one finds that it depends on the value of the ratio $\gamma/f'(x_-^u)$. The asymptotic behavior near $x_-^u$, approaching it from below, indeed reads (see also Appendix \ref{behavior_xneg} for a general analysis of Eq. (\ref{ODE2ndTau}) near $x_-^u$)
\begin{eqnarray} \label{x0toxmu_plus}
{\tau_{\text{\tiny I}}}_\gamma(x_0) - {\tau_{\text{\tiny I}}}_\gamma(x_-^u) \simeq 
\begin{cases}
&A_- \, |x_0 - x_-^u|^{\frac{\gamma}{f'(x_-^u)}} \quad, \quad \hspace*{1.3cm}\frac{\gamma}{f'(x_-^u)} < 1 \;, \\
&B \, (x_0 - x_-^u) \log \left( \frac{1}{|x_0 - x_-^u|} \right) \quad, \quad\; \frac{\gamma}{f'(x_-^u)} = 1 \;, \\
&C \, (x_0 - x_-^u) \quad, \quad \hspace*{2.2cm} \frac{\gamma}{f'(x_-^u)} > 1 \;,
\end{cases}
\end{eqnarray}
where $A_-<0$,  $B>0$ and $C>0$ are computable constants.

\noindent \textbf{The limit $\gamma \to 0$.}
In this limit, it is easy to see that both $\lim_{\gamma \to 0} {\tau_{\text{\tiny I}}}_\gamma^\pm(x_0)$ are constants, which, as before in Eq. (\ref{smallgamma}), can be computed by integrating the Langevin equation (\ref{langeRTP}) -- since for $0<x_0<x_-^u$, the velocity is negative in both states $\sigma = \pm 1$). This leads to 
\begin{equation}
    {\tau_{\text{\tiny I}}}_{\gamma=0}^{\pm}(x_0)= -\int_{0}^{x_0} \frac{dx}{f(x) \pm v_0}\,  = -\int_{0}^{x_0} \frac{dx}{\alpha(1-x) \pm v_0} = \int_0^{x_0} dx \frac{1+x}{1 \mp v_0(1+x)} = \frac{\mp v_0 x_0 - \ln \left(1 + \frac{v_0 x_0}{v_0\mp 1} \right)}{v_0^2}\label{smallgammalog} \;.
\end{equation}
Note that ${\tau_{\text{\tiny I}}}_{\gamma=0}^{+}(x_0)$ diverges as $x \to x_-^u = 1/v_0-1$, while ${\tau_{\text{\tiny I}}}_{\gamma=0}^{-}(x_0)$ remains finite in that limit. 

\vspace*{0.5cm}

\noindent \textbf{The limit $\gamma \to \infty$.} Here, the potential is a monotonically increasing function of $x$ for $x\geq 0$. In the large $\gamma$ limit, the RTP behaves as a diffusing particle with $D=v_0^2/(2\gamma)\to 0$ \cite{letter}. Hence the dynamics is $\dot{x}(t)=f(x)$ and the limit reads
\begin{equation}\label{largegammalog}
   \lim_{\gamma\to \infty} {\tau_{\text{\tiny I}}}_{\gamma}(x_0) = \lim_{\epsilon \to 0} -\int_\epsilon^{x_0}\frac{dx}{f(x)} = \lim_{\epsilon \to 0} \int_\epsilon^{x_0}dx\, (1+x)= \frac{x_0^2}{2}+x_0\, .
\end{equation}
When $x_0 = 0.1$, we obtain $\lim_{\gamma\to \infty}\tau_\gamma(x_0) =0.105$ which is confirmed by the data in the left panel of Fig. \ref{MFPTLogsimu1}.\\
\vspace*{0.5cm}

\subsection{$x_0 \in ]x_-^u,+\infty[$}

\begin{figure}[t]
    \centering
    \includegraphics[width=0.8\linewidth]{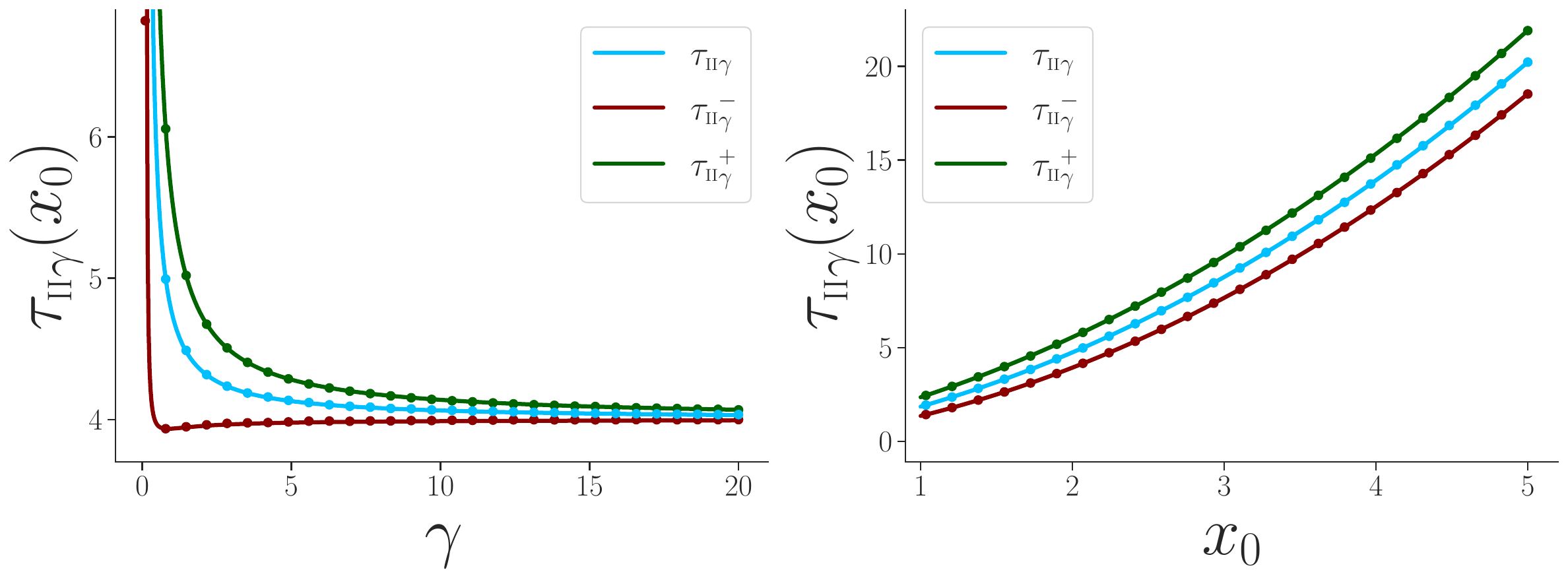}
  \caption{For $]x_-^u,+\infty[$, we show results of simulation for the MFPT of an RTP (dots) inside a logarithmic potential $V(x)=\text{log}(1+|x|)$ and we compare them with our analytical solutions (lines). On the left, we plot the MFPT with respect to $\gamma$ -
$v_0= 0.5$, and $x_0 = 2$. On the right, we show a plot of the MFPT vs $x_0$ -- with $v_0= 0.5$, and
$\gamma = 1$.}
  \label{MFPTLogsimu2} 
\end{figure}

\begin{figure}[t]
    \centering
    \includegraphics[width=0.5\linewidth]{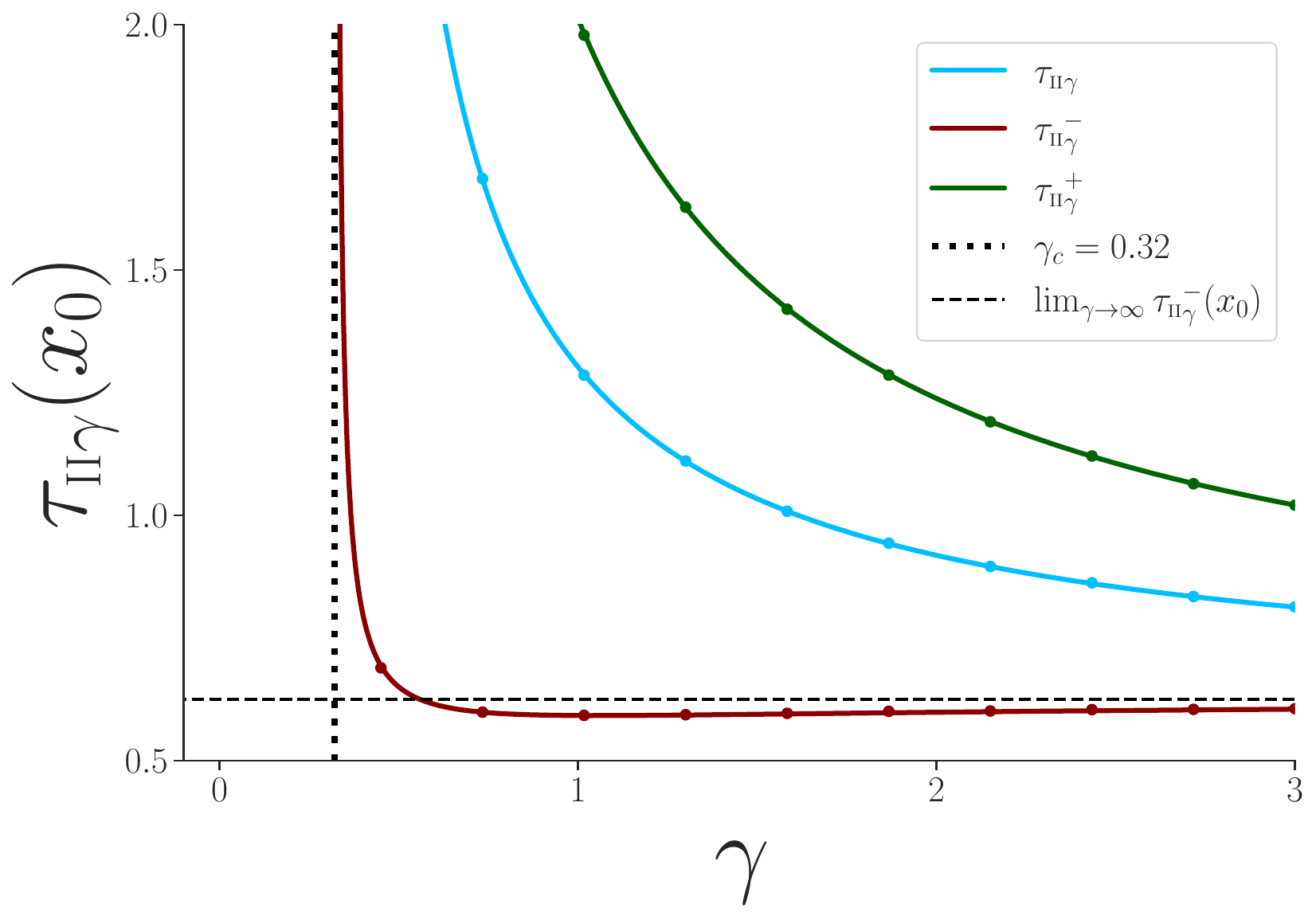}
  \caption{For the logarithmic potential, we consider a case where $v_0=0.8$ such that the negative unstable turning point is at $x_-^u=1/v_0 -1 = 0.25$, and the critical value of the tumble rate is $\gamma_c = 0.32$, below which the MFPT is infinite. We fix $x_0=0.5$ such that the RTP starts its motion on the right of the turning point. The large $\gamma$ limit reads $\lim_{\gamma\to \infty}  {\tau_{\text{\tiny II}}}_\gamma(x_0) =\lim_{\gamma\to \infty} {\tau_{\text{\tiny II}}}^-_\gamma(x_0) =\frac{x_0^2}{2}+x_0 = 0.625$. On the figure, we show the clear divergence of the MFPTs at $\gamma_c$, while ${\tau_{\text{\tiny II}}}^-_\gamma$ reaches from below the large $\gamma$ limit indicating the presence of a minimum. The solid lines correspond to our analytical prediction while the dots show numerical results from simulations.}
  \label{MFPTLogsimu_gammacritic} 
\end{figure}

The solution in the region to the right of the unstable negative turning point is given by Eq.~(\ref{phase5asolution}), i.e.,
\begin{eqnarray}
    {\tau_{\text{\tiny II}}}_\gamma(x_0)={\tau_{\text{\tiny I}}}_\gamma(x_-^u)-\int_{\frac{1}{v_0} - 1}^{x_0}dz\, \frac{(1+z)^2}{v_0^2(1+z)^2-1}\, \int_{z}^{+\infty} dy\, \left(\frac{1}{(1+y)^2}-2\gamma\right)\text{exp}\left[\int_y^zdu\, \frac{-2\gamma(1+u)}{1-v_0^2(1+u)^2}\right]\, ,\label{logregionII}
\end{eqnarray}
where we recall that $x_-^u = 1/v_0 - 1$. The first important property to notice is that the integral over $y$ in that expression is well defined if and only if $1/D_{\rm eff} = 2 \gamma/v_0^2 > 1$. Indeed, the integral in the argument of the exponential behaves, for large $y$, as $-(2\gamma)/v_0^2 \log y$, implying that the integrand behaves, for large $y$, as $y^{-2 \gamma/v_0^2}$. Therefore, ${\tau_{\text{\tiny II}}}_\gamma(x_0) = +\infty$ for $D_{\rm eff} > 1$. For fixed $v_0$, there thus exists a critical value $\gamma_c = v_0^2/2$ such that $\tau_\gamma(x_0) = + \infty$ for $\gamma < \gamma_c$ while $\tau_\gamma(x_0)$ is finite for $\gamma > \gamma_c$. This is clearly seen in our numerical simulations -- see the left panel of Fig. \ref{MFPTLogsimu2} for which $\gamma_c=1/8 = 0.125$ as well as Fig. \ref{MFPTLogsimu_gammacritic} for which $\gamma_c = 8/25=0.32$. Physically, this can be understood as follows: when $D_{\rm eff} > 1$, it is possible for the particles to diffuse fast enough such that they can escape to infinity since the force vanishes at large distance. Indeed, one can show that there is no stationary measure in this case (see also \cite{Velocity_RTP} for a similar discussion of the related potential of the form $V(x) \propto \log(1+x^2)$). Note that this divergence of the MFPT is also present for a purely passive particle in such a logarithmic potential with diffusion constant $D_{\rm eff}>1$ (see below). On the other hand, for $D_{\rm eff} < 1$, these integrals are well defined and they can again be expressed in terms of hypergeometric functions. However, the expressions are quite cumbersome and we do not give them here. 
%
%It is possible to write the explicit solution by computing the integrals in (\ref{logregionII}), or to compute them numerically. Once again, the expression of the MFPT will involve combinations of hypergeometric functions. 
%
We show in Fig.~\ref{MFPTLogsimu2} that our prediction (\ref{logregionII}) agrees perfectly with our numerical simulations. It turns out that the MFPT ${\tau_{\text{\tiny II}}}_\gamma(x_0)$ exhibits interesting behaviors as a function of both $x_0$ and $\gamma$ that we now discuss.

\begin{figure}[t]
    \centering
    \includegraphics[width=0.4\linewidth]{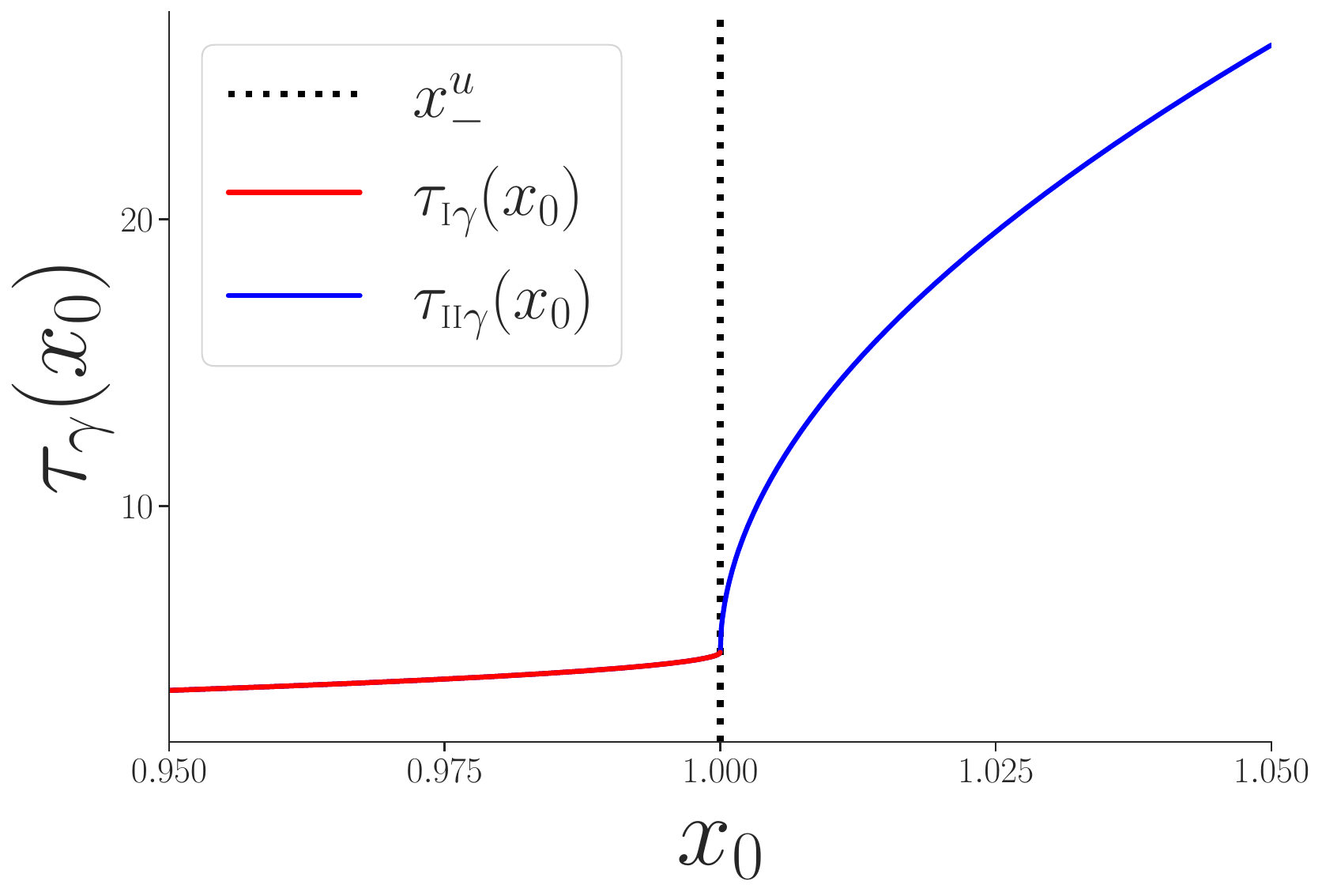}
  \caption{{\color{black}Plot of ${\tau_{\text{\tiny I}}}_\gamma(x_0)$ and ${\tau_{\text{\tiny II}}}_\gamma(x_0)$ for a log-potential when $v_0=0.5$ and $\gamma=0.13$. The negative unstable turning point is located at $x_-^u=1$. Here, $\gamma/f'(x_-^u)<1$ such that Eqs.~(\ref{x0toxmu_plus}) and~(\ref{x0toxmu_plusII}) predict a singular behavior, namely a cusp, at $x_0 = x_-^u$, which is clearly seen on the figure.}}
  \label{MFPTLogsimu_cusp} 
\end{figure}

\vspace*{0.5cm}
\noindent \textbf{The limit $x_0 \to x_-^u$, and $x_0>x_-^u$.} Let us first analyse the behavior of ${\tau_{\text{\tiny II}}}_\gamma(x_0)$ as $x_0 \to x_-^u$ from the right, i.e. $x_0>x_-^u$. The MFPT is a continuous function of $x_0$ and one has, in this limit ${\tau_{\text{\tiny II}}}_\gamma(x_0) \to {\tau_{\text{\tiny II}}}_\gamma(x_-^u) = {\tau_{\text{\tiny I}}}_\gamma(x_-^u)$. To characterize more precisely the behavior to the right of $x_-^u$, 
on can start from  the formula in Eq. (\ref{phase5asolution}), which can be analysed for a generic force $f(x)$. It turns out that the behavior of ${\tau_{\text{\tiny II}}}_\gamma(x_0)$ in this limit depends also on the value of the ratio $\gamma/f'(x_-^u)$ and is very similar to the behavior of ${\tau_{\text{\tiny I}}}_\gamma(x_0)$ in Eq. (\ref{x0toxmu_plus}). One finds indeed (see also again Appendix \ref{behavior_xneg} for a general analysis of Eq. (\ref{ODE2ndTau}) near $x_-^u$)
\begin{eqnarray} \label{x0toxmu_plusII}
{\tau_{\text{\tiny II}}}_\gamma(x_0) - {\tau_{\text{\tiny II}}}_\gamma(x_-^u) = {\tau_{\text{\tiny II}}}_\gamma(x_0) - {\tau_{\text{\tiny I}}}_\gamma(x_-^u) \simeq 
\begin{cases}
&A_+ \, |x_0 - x_-^u|^{\frac{\gamma}{f'(x_-^u)}} \quad, \quad \hspace*{1.3cm}\frac{\gamma}{f'(x_-^u)} < 1 \;, \\
&B \, (x_0 - x_-^u) \log \left( \frac{1}{|x_0 - x_-^u|} \right) \quad, \quad\; \frac{\gamma}{f'(x_-^u)} = 1 \;, \\
&C \, (x_0 - x_-^u) \quad, \quad \hspace*{2.2cm} \frac{\gamma}{f'(x_-^u)} > 1 \;,
\end{cases}
\end{eqnarray}
where $0<A_+ \neq A_-<0$ is a computable constant, while $B$ and $C$ are the same constants that appear in (\ref{x0toxmu_plus}). Note that by comparing Eqs. (\ref{x0toxmu_plus}) and (\ref{x0toxmu_plusII}) one finds that the first derivative of the MFPT with respect to $x_0$ is diverging for $\gamma/f'(x_-^u) \leq 1$ (see for instance Fig. \ref{MFPTLogsimu_cusp}) while it is finite and continuous for $\gamma/f'(x_-^u) > 1$. In fact, one can actually show that, for $\gamma/f'(x_-^u) > 1$, the second derivative generically becomes discontinuous (it can even become infinite if~$\gamma/f'(x_-^u) = 2$).    

\vspace*{0.5cm}
\noindent \textbf{The limit $x_0 \to +\infty$}. It is also interesting to investigate the large $x_0$ behavior of $\tau_\gamma(x_0)$, which can be done for general $f(x)$ from 
Eq.~(\ref{phase5asolution}). This behavior depends a priori on the large $x$ behavior of the force $f(x)$ and we have not tried to analyse it in full generality. However, this analysis can easily be done for the logarithmic potential, starting from the expression in Eq.~(\ref{logregionII}) and one finds
\begin{eqnarray} \label{largex0_log}
{\tau_{\text{\tiny II}}}_\gamma(x_0) = \frac{1}{1 - D_{\rm eff}} \left(\frac{x_0^2}{2} + x_0\right) + O(1) \quad, \quad x_0 \to \infty \;,
\end{eqnarray}
where $D_{\rm eff} = v_0^2/(2\gamma)$ while the constant term, i.e., the $O(1)$ term in Eq. (\ref{largex0_log}), can also be computed, and it is nonzero, but it has a quite complicated expression which we do not report here.

\vspace*{0.5cm}
\noindent
\textbf{The behavior of ${\tau_{\text{\tiny II}}}_\gamma(x_0)$ as a function of $\gamma$.} We have seen previously that 
${\tau_{\text{\tiny II}}}_\gamma(x_0)$ diverges for $D_{\rm eff} = v_0^2/(2 \gamma) \geq 1$. Hence, it is interesting to characterize the behavior of ${\tau_{\text{\tiny II}}}_\gamma(x_0)$ as $D_{\rm eff}$ approaches $1$ from below, e.g., 
for $D_{\rm eff} = v_0^2/(2 \gamma) < 1$ and in the limit $\gamma \to \gamma_c=v_0^2/2$. This limit can be analysed in principle for any force $f(x)$, from Eq. (\ref{phase5asolution}). However, here we restrict ourselves to the logarithmic potential $f(x) = -1/(1+x)$. 
In this limit, a careful analysis of the formula in (\ref{logregionII}) shows that the MFPT diverges as
\begin{eqnarray} \label{div_gamma}
{\tau_{\text{\tiny II}}}_\gamma(x_0) \simeq \frac{1}{\gamma - \gamma_c}\, F(v_0(1+x_0)) \quad, \quad {\rm as} \quad \,  \gamma \to \gamma_c \quad, \quad {\rm with} \quad \quad \gamma > \gamma_c \;,
\end{eqnarray}
where the function $F(x)$ is given by
\begin{eqnarray} \label{F}
F(x) = \int_1^x \frac{u^2}{\sqrt{u^2-1}}\, du = \frac{x}{4} \sqrt{x^2-1} + \frac{1}{4} \log(x + \sqrt{x^2-1}) \;, \; x \geq 1 \;.
\end{eqnarray}
It behaves as $F(x) \sim \sqrt{2(1-x)}$ for $x \to 1$ and as $F(x) \simeq x^2/4 + O(\log x)$ as $x \to \infty$. 

In the other interesting limit $\gamma \to \infty$, the MFPT ${\tau_{\text{\tiny II}}}_\gamma(x_0)$ behaves as ${\tau_{\text{\tiny I}}}_\gamma(x_0)$ given in Eq. (\ref{largegammadw}). When $x_0 = 2$, we obtain $\lim_{\gamma\to \infty}\tau_\gamma(x_0) =4$ which is confirmed by the data in the left panel of Fig. \ref{MFPTLogsimu2}. Interestingly, for intermediate values of $\gamma$, one observes (see the left panel of Fig. \ref{MFPTLogsimu2} and Fig. \ref{MFPTLogsimu_gammacritic}) that ${\tau_{\text{\tiny II}}}^-_\gamma(x_0)$ exhibits a minimum value at $\gamma = \gamma_{\rm opt}$, while ${\tau_{\text{\tiny II}}}^+_\gamma(x_0)$ and ${\tau_{\text{\tiny II}}}_\gamma(x_0)$ are monotonically decreasing function of $\gamma$. We have observed that this minimum is more pronounced as $v_0$ gets smaller and $x_0$ gets closer to $x_-^u$. It is however not clear whether this minimum disappears beyond a certain critical value of $v_0$ or $x_0$. In any case, the existence of this minimum is rather counter-intuitive and it has a different origin from the minimum value found for potentials of the form $V(x) = \alpha |x|^p$ with $p>1$ \cite{letter}. Indeed in that case, one can show that $\tau_\gamma(x_0)$ diverges in both limits $\gamma \to 0$ and $\gamma \to \infty$, suggesting the existence of an optimal rate $\gamma_{\rm opt}$. Note also that, for $p>1$, only $\tau_\gamma^+(x_0)$ and $\tau_\gamma(x_0)$ exhibit a minimum, but not $\tau_\gamma^-(x_0)$, which is quite different from what we found here (see the left panel of Fig. \ref{MFPTLogsimu2} and Fig. \ref{MFPTLogsimu_gammacritic}).  

\vspace*{0.5cm}
\noindent{\bf Remark (i):} It is interesting to compare the rich result found here for the RTP with the behavior of the MFPT for a purely passive, i.e., Brownian particle with diffusion constant $D_{\rm eff}$ in such a logarithmic potential $V(x) = \log(1+|x|)$. In this case the MFPT $\tau_D(x_0)$ for a particle starting from $x_0$ is given by~\cite{diffusivemoments, kramers}
\begin{eqnarray} \label{MFPT_passive}
\tau_D(x_0) = \frac{1}{D_{\rm eff}} \int_0^{x_0} dz \int_{z}^\infty dy \exp\left( \frac{V(z)-V(y)}{D_{\rm eff}}\right) = \frac{1}{D_{\rm eff}}  \int_0^{x_0} dz \int_{z}^\infty dy \left( \frac{1+z}{1+y}\right)^{1/D_{\rm eff}}
\end{eqnarray}
Clearly this integral is convergent if and only if $D_{\rm eff}<1$, otherwise $\tau_D(x_0) = + \infty$, as in the RTP case. Besides, for $D_{\rm eff}<1$, the double integral can trivially be performed and one finds
\begin{eqnarray} \label{MFPT_passive.2}
\tau_D(x_0) = \frac{1}{1-D_{\rm eff}} \left(x_0 + \frac{x_0^2}{2} \right) \quad , \quad {\rm for} \quad D_{\rm eff}<1 \quad {\rm and} \quad x_0 \geq 0 \;.
\end{eqnarray}
The simplicity of this result for a purely passive particle is in contrast with the wide variety of behaviors found for $\tau_\gamma(x_0)$ in the active case -- in particular in the spatial dependence of the MFPT. Note that in the limit $D_{\rm eff} \to 0$, this result (\ref{MFPT_passive.2}) coincides with the limit $\gamma \to \infty$ found for the RTP [see Eq. (\ref{largegammalog})], as it should. Interestingly, it also coincides with the large $x_0$ behavior of $\tau_\gamma(x_0)$ -- up to the next to leading order term for large $x_0$ [see Eq. (\ref{largex0_log})]. This is probably due to the fact that, the large time behavior of the RTP starting from large $x_0$ is very similar to standard Brownian diffusion with a diffusion constant $D_{\rm eff} = v_0^2/(2 \gamma)$.    

\vspace*{0.5cm}
\noindent{\bf Remark (ii):} As discussed in several places in this paper, in Ref. \cite{letter}, we have performed a detailed study of the MFPT of an RTP in the presence of a confining potential of the form $V(x) = \alpha |x|^p$ with $p \geq 1$. Specifically, for $p>1$, which falls under the Phase III case, there exists an optimal tumbling rate $\gamma_{\text{opt}}$ that minimizes $\tau_\gamma(x_0)$. A natural question to ask is whether this minimum still exists when $p<1$. For these values of $p$, the force has exactly the same properties as the one derived from the log-potential, except here $f(0) = -\infty$. The force has an unstable turning point $x_-^u$ and the MFPT is first computed for $x_0 \in [0,x_-^u[$ in Phase II using Eq.~(\ref{solutionf(0)leqmainresults_secondForm}), and then on $x_0 \in ]x_-^u,+\infty[$ using Eq.~(\ref{phase5asolution}). The behaviors of the MFPT are qualitatively similar to those observed in Fig. \ref{MFPTLogsimu1} and \ref{MFPTLogsimu2}. However, in this case the MFPT is always finite for any $\gamma >0$ -- and therefore $\gamma_c = 0$. In addition, in contrast to the logarithmic potential, the numerical evaluations of the exact formula (\ref{phase5asolution}) for $V(x) =\alpha |x|^p$ with $0<p<1$ seem to indicate that none of the MFPT $\tau_\gamma(x_0)$ and $\tau^{\pm}_\gamma(x_0)$ exhibit a minimum. This shows that the MFPT of an RTP in a potential of the form $V(x) = \alpha |x|^p$ behaves quite differently for $p<1$ and $p>1$ ($p=1$ being thus a borderline case). In fact, for passive particles, different behaviors for $p<1$ and $p>1$ were also found for other observables, such as the distribution of the time of the maximum in the stationary state \cite{time_max} or for the Kramer's escape problem, for which $p=1$ was shown to exhibit a freezing transition \cite{freezingdiffusive}. In fact, this freezing transition was recently found also for an RTP in a V-shape potential, i.e. $p=1$ \cite{SurvivalRPTlinear}.

\section{Applications}\label{application}

In this section we discuss three physical applications of our results for the MFPT.

\subsection{Average time for an RTP to jump over a high barrier - generalized Kramers' law}\label{kramersApplication}

Let us consider first a Brownian (passive) particle with diffusion coefficient $D$, moving in the presence of the double-well potential $V(x)=\frac{\alpha}{2} (|x|-1)^2$ studied in section \ref{doublewellsection}, which has a minimum at $x=1$ (assuming $x>0$). In the limit of weak noise (compared to the barrier height) $D \ll \alpha$, an interesting problem is to determine the average time needed for the particle to go from one side of the double-well (say located at $x_0>0$) to the other side of the double-well. This problem can be reformulated as follows~\cite{kramers}: what is the MFPT to the origin starting from $x_0 > 0$? In the weak noise limit, it is well known that the MFPT diverges exponentially with the barrier height: this the famous Arrhenius/Kramers' law \cite{kramers}.

To understand this, consider that without noise, the dynamics is simply given by $\dot{x}(t) = f(x)$. Regardless of the initial value $x_0$, the particle will fall into the minimum at $x=1$ and remain there indefinitely (and therefore it will never cross the barrier). When the noise is weak, the force will still attract the particle inside the minimum in a finite time. Once at the minimum, the particle has a chance to escape via thermally assisted random fluctuations of its position. The MFPT is then dominated by the mean time needed for the particle to escape from the minimum $x=1$ towards the origin $x=0$. This is a classical problem in statistical physics and this question has been studied in great detail for Brownian motion, resulting in Arrhenius/Kramers' law \cite{kramers,diffusivemoments}. More recently, this question has attracted some attention in the context of active particles \cite{Woillez, Woillez2, Hanggi_kramers}.

It is instructive to check this physical argument for a Brownian particle. For a dynamics $\dot{x}(t)= f(x) +\sqrt{2D}\eta(t)$ where $\eta(t)$ is a white noise with two-times correlations $\langle\eta(t_1)\eta(t_2)\rangle=\delta(t_2-t_1)$, the MFPT is given by the formula~(\ref{MFPT_passive}), which we recall here for convenience, namely~\cite{diffusivemoments, kramers}
\begin{eqnarray} \label{passive_tau}
\tau_D(x_0) = \frac{1}{D} \int_0^{x_0} dz \, \int_z^{\infty} dy \, \exp{\left(\frac{V(z)-V(y)}{D} \right)} \;.
\end{eqnarray}
Note that in the diffusing limit of the RTP, i.e., $D=v_0^2/(2\gamma)$ fixed with $v_0 \to \infty$ and $\gamma \to \infty$, Eq.~(\ref{phase3sol}) is indeed equivalent to Eq.~(\ref{passive_tau}) \cite{letter}. We are interested in the weak noise limit, i.e., the limit $D\to 0$ of Eq.~(\ref{passive_tau}). Performing the change of variable $y = z+u$, we have
\begin{eqnarray} 
\tau_D(x_0) = \frac{1}{D} \int_0^{x_0} dz \, \int_0^{\infty} du\,  \exp{\left(\frac{V(z)-V(z+u)}{D} \right)} \;.
\end{eqnarray}
In the limit $D \to 0$, this double integral can be evaluated by the saddle point method, leading to 
%integral is dominated by the $(z^*,u^*)$ maximizing the function $V(z)-V(z+u)$ such that 
\begin{eqnarray}
    \log(\tau_D(x_0)) \isEquivTo{D \to 0} \frac{\max_{(z,u)}(V(z)-V(z+u))}{D} \, .
\end{eqnarray}
As $V(x)$ is decreasing on $[0,1]$, the maximum is reached for $(z^*=0, u^* = 1)$ which represents the MFPT to reach the origin starting from the minimum of the well as announced. Therefore
\begin{eqnarray}
    \log(\tau_D(x_0)) \isEquivTo{D \to 0}   \frac{\Delta E}{D} \quad \, , \quad \Delta E = V(0) - V(1)\, ,
\end{eqnarray}
where $\Delta E$ is thus the height of the barrier.

In this paper, we have derived the explicit formula for the MFPT of an RTP inside a double-well $V(x)=\frac{\alpha}{2} (|x|-1)^2$ when $0<\alpha <v_0$. This is the relevant case since for $\alpha > v_0$ the particle can not reach the origin. In this case, 
the MFPT is given by Eq.~(\ref{phase3DWequ}) and we want to analyse it in the limit where $D_{\text{eff}} = v_0^2/(2\gamma) \to 0$, e.g., in the limit $\gamma \to \infty$ keeping $v_0$ fixed. In this limit, it is natural to expect that $\tau_\gamma(x_0)$ grows exponentially with $\gamma$ and we thus estimate the integrals in Eq.~(\ref{phase3DWequ}) by the saddle point method (note that the first term $1/(2\gamma)$ in Eq.~(\ref{phase3DWequ}) is subdominant compared to the other terms). Let us first analyse the first integral in Eq.~(\ref{phase3DWequ}), which leads to  
\begin{equation} \label{saddle1}
\int_0^{1+v_0/\alpha} dy\, \frac{1}{v_0 + \alpha(y-1)} \text{exp}\left[-\gamma\, \int_0^{y}du\, \frac{2\, \alpha(u-1)}{v_0^2-\alpha^2(u-1)^2}\right]  \isEquivTo{\gamma \to +\infty}  \text{exp}\left[-\gamma\, \underset{y \in [0,1+v_0/\alpha]}{\min}\left(\int_0^{y}du\, \frac{2\, \alpha(u-1)}{v_0^2-\alpha^2(u-1)^2}\right)\right]\,.
\end{equation}
It is easy to check that, for $0\leq y \leq 1+\frac{v_0}{\alpha}$, the minimum in the argument of the exponential is reached for $y=1$ such that we have
\begin{equation}
\int_0^{1+v_0/\alpha} dy\, \frac{1}{v_0 + \alpha(y-1)} \text{exp}\left[-\gamma\, \int_0^{y}du\, \frac{2\, \alpha(u-1)}{v_0^2-\alpha^2(u-1)^2}\right]  \isEquivTo{\gamma \to +\infty} \text{exp}\left[\int_0^{1}du\, \frac{2\gamma\, \alpha(1-u)}{v_0^2-\alpha^2(u-1)^2}\right]  \,.
\label{firstequivalent}
\end{equation}
The last term in Eq.~(\ref{phase3DWequ}) is a double integral which we rewrite as
\begin{eqnarray}
    \int_0^{x_0}dz\, \frac{1}{v_0^2-\alpha^2(z-1)^2} \int_{1+\frac{v_0}{\alpha}}^{z} dy\,  \left(\alpha+ 2\gamma \right)\,  e^{-\gamma \, \phi(z,y)}\quad \, , \quad \phi(z,y) = \int_z^{y}du\, \frac{2\gamma\, \alpha(u-1)}{v_0^2-\alpha^2(u-1)^2}\, .
\end{eqnarray}
We are looking for the minimum of $\phi(z,y)$. If $x_0>1$, then $\partial_z \phi(z,y) = 0$ for $z^*=1$. In that case, $z^*=1<y<1+\frac{v_0}{\alpha}$, and  $\partial_y \phi(z^*,y) = 0$ for $y^*=1$. In this case, this term is subdominant compared to the exponentially diverging log-equivalent (\ref{firstequivalent}). Instead, if $x_0<1$, separating the integral over $y$ in two regions $y>1$ and $y<1$ allows to show that the integral is of the same order as (\ref{firstequivalent}). Hence, 
\begin{eqnarray}
    \log\left(\tau_\gamma(x_0)\right) \isEquivTo{\gamma \to +\infty}\int_0^{1}du\, \frac{2\gamma\, \alpha(1-u)}{v_0^2-\alpha^2(1-u)^2} = \frac{V_{\text{eff}}(0)-V_{\text{eff}}(1)}{D_{\text{eff}}}\, ,
    \label{logequivtau}
\end{eqnarray}
where $V_{\text{eff}}(x) =- \frac{v_0^2}{2 \alpha} \log \left|1-\frac{\alpha^2}{v_0^2}(|x|-1)^2\right|$ is an effective potential. Therefore, for an RTP, in the weak noise limit, Kramers' law is modified in the sense that the averaged time to go from one side of the double-well to the other is exponentially diverging with the height of an effective barrier $\Delta V_{\text{eff}} = V_{\text{eff}}(0)-V_{\text{eff}}(1)= - \frac{v_0^2}{2 \alpha} \log \left|1-\frac{\alpha^2}{v_0^2}\right| > \Delta E$ greater than in the diffusive case. In the full diffusive limit which is retrieved when $v_0 \to \infty$, we find back the diffusive result $\log\left(\tau_\gamma(x_0)\right) \approx \frac{\Delta E}{ D_{\text{eff}}}$ with $D_{\rm eff} = v_0^2/(2 \gamma)$.

The previous analysis can actually be carried out for a general potential $V(x)$ with a single minimum located at $x_{\text{min}}$, and the top of the barrier at the origin, while $V(x)$ is an increasing function of $x$ for $x>x_{\text{min}}$ (as in Fig. \ref{kramersfig}). If the origin is accessible to an RTP located inside the minimum of the well, then from equation (\ref{phase3sol}), it is possible to show that the modified Kramers' law reads
\begin{eqnarray}
    \log\left(\tau_\gamma(x_0)\right) \isEquivTo{\gamma \to +\infty}\int_0^{x_{\text{min}}}du\, \frac{2\gamma\, f(u)}{v_0^2-f^2(u)}\, ,
\end{eqnarray}
where $f(x)$ is the force associated to the potential, i.e., $f(x) = -V'(x)$. One can re-write the expression as
\begin{eqnarray} \label{gen_kramer}
    \log\left(\tau_\gamma(x_0)\right) \isEquivTo{\gamma \to + \infty}\frac{1}{D_{\text{eff}}}\int_0^{x_{\text{min}}}du\, \frac{f(u)}{1-\frac{f^2(u)}{v_0^2}}= \frac{W(0)-W(x_{\text{min}})}{D_{\text{eff}}}\, ,
\end{eqnarray}
and $W(x) = \int^{x}du\, f(u)(1-\frac{f^2(u)}{v_0^2})^{-1}$ is called the ``active external potential'' \cite{Kardar, RTPPSG}. Note that in the double-well potential above we used the notation $W(x)=V_{\rm eff}(x)$. In the diffusive limit, we indeed obtain $\log\left(\tau_\gamma(x_0)\right) \approx (V(0)-V(x_{\text{min}}))/D_{\text{eff}}$ where the numerator is the barrier height. Note that in principle our exact formula (\ref{phase3sol}) allows to compute the pre-exponential corrections to the modified Kramer's law (\ref{gen_kramer}), as it can be done in the passive case, see e.g. \cite{kramers}. However we leave this rather technical analysis for further studies.

Finally, it is interesting to compare this formula (\ref{gen_kramer}) with the result of Ref. \cite{naftaliSS} who studied the MFPT in the weak noise limit. In that limit, the authors established a relation between the MFPT in a confining potential and the stationary distribution of the RTP $P_{\text{st}}(x)$ in the same potential, namely~\cite{naftaliSS}
\begin{eqnarray}
    \lim_{\gamma \to \infty} \log\left(\tau_\gamma(x_0)\right) \sim -\lim_{\gamma \to \infty}\log\left(P_{st}(X)\right)\, ,
    \label{MFPTdistrib}
\end{eqnarray}
where $X$ is the position of the absorbing state (here $X=0$). Using the explicit expression of $P_{st}(x)$ for a generic potential (e.g., from \cite{RTPSS, RTPSS2, RTPSS3, RTPSS4, Sevilla}) one has for $V(x)=\frac{\alpha}{2} (|x|-1)^2$ 
\begin{eqnarray}
    \frac{1}{P_{\text{st}}(0)} = (v_0^2 - \alpha^2)\, \int_{-1-\frac{v_0}{\alpha}}^{1+\frac{v_0}{\alpha}}dy\, \frac{1}{v_0^2-\alpha^2(y-1)^2}\, e^{-2\gamma\int_0^y dz\, \frac{\alpha (z-1)}{v_0^2 - \alpha^2(z-1)^2}}\, .
\end{eqnarray}
At large $\gamma$, this integral can be evaluated by a saddle-point method. It turns out the saddle point is located at $y^*=1$, leading to
\begin{eqnarray}
    \log\left(\frac{1}{P_{\text{st}}(0)}\right) \isEquivTo{\gamma \to +\infty}\int_0^{1}du\, \frac{2\gamma\, \alpha(1-u)}{v_0^2-\alpha^2(1-u)^2}\, .
\end{eqnarray}
By using the relation in (\ref{MFPTdistrib}), this result indeed coincides with our prediction in (\ref{logequivtau}).

\subsection{Mean trapping time of an RTP inside a harmonic trap}\label{relaxationsection}
When confined in a harmonic trap described by the potential $V(x)=\mu\, x^2/2$, an RTP reaches a stationary state at large time. What is remarkable is that the support of the distribution is finite and the particle is trapped in the interval $[-v_0/\mu,v_0/\mu]$ \cite{RTPSS}. This is because $\pm v_0/\mu$ are turning points of the force $f(x) = -V'(x) = -\mu \, x$. At $v_0/\mu$, the positive state of the RTP has a zero velocity while the negative state has a negative velocity. This means that once an RTP that initiates its motion at $x_0>v_0/\mu$ reaches for the first time $[0,v_0/\mu]$, it stays inside this interval forever. Therefore, in order for the particle to relax in the stationary state, it must first reach $v_0/\mu$. Consequently, the mean first-passage time to $v_0/\mu$, which we call the mean ``trapping time'' serves as a lower bound for the relaxation time of an RTP inside a harmonic trap to the stationary state. 

\begin{figure}[t]
    \centering
    \includegraphics[width=0.8\linewidth]{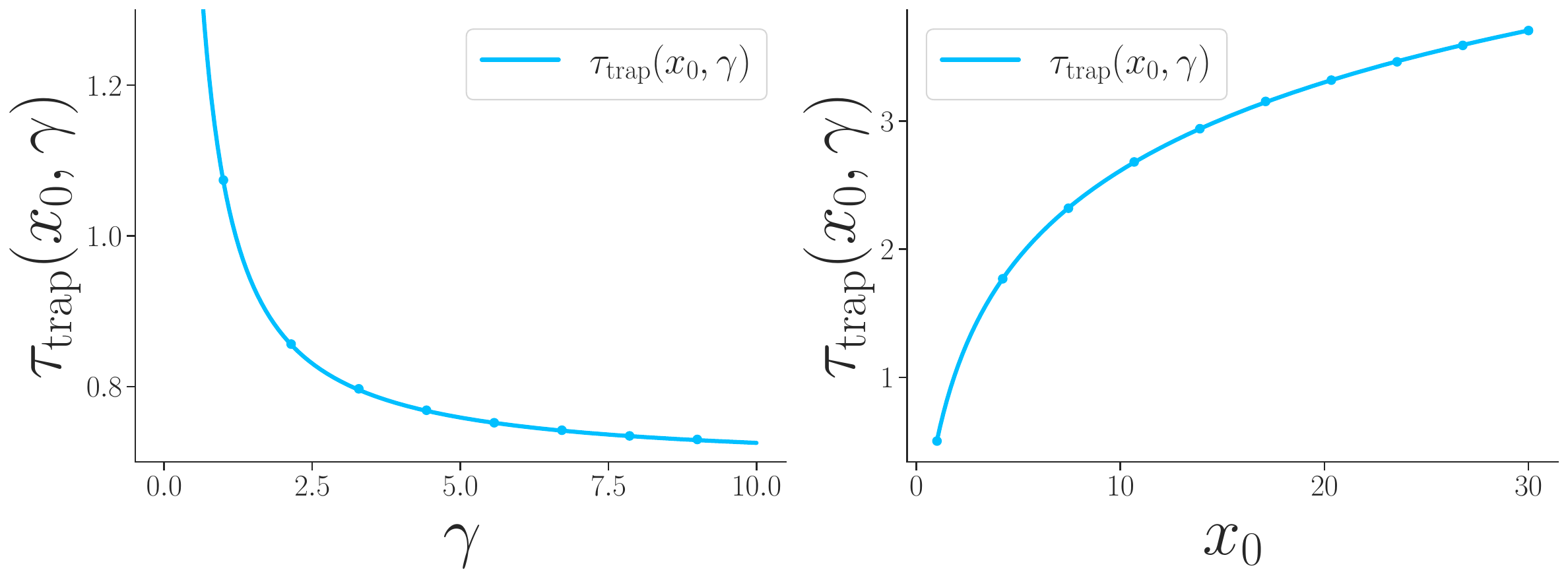}
  \caption{Plots of the mean ``trapping time'' of an RTP confined in a harmonic trap, i.e., the mean time needed for the particle to reach the right edge of the support of the stationary distribution located at $v_0/\mu$, and starting at $x_0>v_0/\mu$. Here, $v_0=1$ and $\mu=1$ such that the right edge is located at $v_0/\mu=1$. On both panels, the dots show numerical results, while the solid lines are analytical results Eq.~(\ref{hypergeomtrapping}). On the left plot, we plot the MFPT with respect to $\gamma$ fixing $x_0=2$, while on the right, we plot the MFPT with respect to $x_0$ fixing $\gamma=1$.}
  \label{MFPTtrappingfigure} 
\end{figure}

Although, up to now, we have only computed the MFPT to the origin $x=0$, it is possible -- as mentioned in the introduction -- to instead compute the MFPT to an arbitrary point by shifting the force $f(x)$. Here, we shift the potential such that the right edge of the support of the steady state is at the origin. For this purpose, the shifted potential is $V(x)=\mu\, (x+v_0/\mu)^2/2$ and $f(x) = -\mu \, x -v_0$. The limit $x_-^s\to 0$ inside Eq. (\ref{phase3sol}) is well defined and the MFPT is given by
%\begin{equation}
%    \begin{split}
%     \tau_{\rm trap}(x_0,\gamma)  &=   \frac{1}{2\gamma}+\int_0^{x_0}dz\, \frac{1}{v_0^2-f(z)^2} \int_{0}^{z} dy \left(f'(y)- 2\gamma \right) \text{exp}\left[\int_y^{z}du\, \frac{-2\gamma f(u)}{v_0^2-f(u)^2}\right]\, .
%    \end{split}
%    \end{equation}
%
{\begin{equation}\label{trapintegrals}
    \begin{split}
     \tau_{\rm trap}(x_0,\gamma)  &=   \frac{1}{2\gamma}+\int_0^{x_0-\frac{v_0}{\mu}}dz\, \frac{1}{v_0^2-f(z)^2} \int_{0}^{z} dy \left(f'(y)- 2\gamma \right) \text{exp}\left[\int_y^{z}du\, \frac{-2\gamma f(u)}{v_0^2-f(u)^2}\right]\, .
    \end{split}
    \end{equation}
    }
Performing the integrals explicitly leads to
\begin{equation}
\tau_{\rm trap}(x_0,\gamma)= \frac{1}{2\gamma} + \frac{(2\gamma + \mu)}{\gamma + \mu} \frac{(x_0-v_0/\mu)}{2v_0} \, {}_3F_2\left(\{1,1,2(1+\frac{\gamma}{\mu})\};\{2,2+\frac{\gamma}{\mu}\};-\frac{\mu\, x_0}{2v_0}+ \frac{1}{2}\right)\, .
\label{hypergeomtrapping}
\end{equation}
The relaxation time of the RTP is equal to the sum of the MFPT to $v_0/\mu$ and the relaxation time of an RTP starting from $v_0/\mu$ which is of order $1/\mu$ \cite{RTPSS}. Hence, for large $x_0$, $\tau_{\rm trap}(x_0,\gamma)\approx \frac{1}{\mu}\log(\frac{\mu\, x_0}{v_0})$ dominates the relaxation time. In Fig. \ref{MFPTtrappingfigure}, we show a perfect agreement between simulations and our theoretical prediction (\ref{hypergeomtrapping}).

\subsection{Optimal search strategy: resetting RTP vs potential driven RTP}\label{resettingsection}

The last application that we want to discuss here is the characterization of the efficiency of the search strategy of a {\it free} RTP in the presence of stochastic resetting. For this purpose, it is useful to recall what stochastic resetting is and what are its main effects on the simpler model of (passive) Brownian diffusion \cite{resettingPRL,resettingJPA} -- for a recent review on stochastic resetting see \cite{resettingReview}. Let us thus consider a single particle on an infinite line starting at the initial position $x_0$ at time $t=0$. The position of the particle at time $t$ evolves, during an infinitesimal amount of time $d t$ according to~\cite{resettingPRL,resettingJPA}
\begin{eqnarray} \label{rBM}
x(t+dt) = 
\begin{cases}
& x_0 \;, \quad \quad \quad \quad \quad \,{\rm with \quad  proba.} \quad r \,dt \;, \\
& x(t) + \eta(t) \, dt \; \quad {\rm with \quad proba.} \quad 1- r \,dt \;,
\end{cases}
\end{eqnarray}
where $r$ is the resetting rate and $\eta(t)$ is a Gaussian white noise with zero mean $\langle \eta(t)\rangle = 0$ and delta correlations $\langle \eta(t) \eta(t')\rangle = 2 D\delta(t-t')$, with $D$ the diffusion constant. In the limit $dt \to 0$, this dynamics (\ref{rBM}) defines the {\it resetting Brownian motion} (rBM) \cite{resettingPRL,resettingJPA} which has generated a lot attention during the last few years \cite{resettingReview}. Interestingly, it was shown that, in the large time limit, the distribution of the position of the particle $p(x,t)$ at time $t$ converges to a stationary distribution given by \cite{resettingPRL,resettingJPA} 
\begin{eqnarray} \label{pst}
\lim_{t \to \infty}p(x,t) = p_{\rm st}(x) = \frac{\alpha_0}{2} \exp{[-\alpha_0|x-x_0|]} \quad, \quad {\rm where} \quad \alpha_0 = \sqrt{r/D} \;. 
\end{eqnarray}
Therefore, although the rBM is a non-stationary process (one can indeed show that the rules of the dynamics (\ref{rBM}) violate detailed balance), the stationary distribution (\ref{pst}) can nevertheless be expressed as an effective Boltzmann weight $p_{\rm st}(x) \propto \exp{[-U_{\rm eff}(x)]}$ with the effective potential
\begin{eqnarray} \label{Veff}
U_{\rm eff}(x) = \alpha_0 |x-x_0| \;. 
\end{eqnarray}
Therefore, if we consider the following equilibrium Langevin dynamics
\begin{eqnarray} \label{langevin}
\frac{dx}{dt} = - D \partial_x U_{\rm eff}(x) + \eta(t) \;,
\end{eqnarray}
where $\eta(t)$ is the same Gaussian white noise as in (\ref{rBM}), then the stationary state of Eq. (\ref{langevin}) is characterized by the same stationary distibution $p_{\rm st}(x) \propto \exp[-U_{\rm eff}(x)]$ -- although of course the two dynamics (\ref{rBM}) and (\ref{langevin}) are quite different. 

An other remarkable property of the rBM is that the resetting parameter $r$ can be tuned to minimize the MFPT to the origin. Indeed, for any finite $r>0$ the MFPT is finite (while it is infinite for the standard Brownian motion corresponding to $r=0$). In addition, there exists an optimal value of the resetting rate $r_{1}$ that minimizes this MFPT \cite{resettingPRL,resettingJPA}. In Ref. \cite{optimalMFPTdiffusive}, the authors asked the following question: since the non-equilibrium resetting dynamics (\ref{rBM}) and the equilibrium Langevin process (\ref{langevin}) lead to the same stationary state, can one compare the MFPT to the origin of these two processes? Interestingly, they showed that 
the optimal ``nonequilibrium'' MFPT (i.e., with resetting with $r=r_1$) is always smaller than the ``equilibrium'' MFPT (evaluated at its optimal value of $r=r_2 \neq r_{1}$). Loosely speaking, ``nonequilibrium offers a better search strategy than equilibrium'' \cite{optimalMFPTdiffusive}. In this section, using our results derived for the MFPT of an RTP in a confining potential, we address this question of the efficiency of the search strategy offered by a free RTP subjected to resetting, in the same spirit as in Ref. \cite{optimalMFPTdiffusive}.

%In \cite{optimalMFPTdiffusive}, authors compared the MFPT to the origin for (i) a diffusing particle driven by resetting at $X_r$ with initial position $x_0=X_r$ and (ii) a diffusing particle in an external potential $V(x) = \sqrt{r/D} |x-x_0|$. This potential was chosen since the steady state of a diffusing particle under resetting is also precisely $P(x) \sim \exp\left[-\sqrt{r/D} |x-x_0|\right]$. The MFPT was computed in both cases, and the optimal ``nonequilibrium" MFPT (resetting) is shown to be always smaller than the ``equilibrium" MFPT (potential driven). In other works, the nonequilibrium is better than equilibrium.

We thus consider a free RTP evolving under Eq. (\ref{langeRTP}) with $f(x) = 0$ where we add resetting to the dynamics -- as in Eq. (\ref{rBM}). This comprises simultaneously resetting both the position and the velocity. With rate $r$ the particle thus resets to its initial position $x_0$, while the velocity $\sigma$ is set $\pm 1$ with probability $1/2$, i.e. the velocity is randomized. This resetting protocol is referred to as position resetting and velocity randomization \cite{resettingRTP}. In this case, the system also reaches a steady state characterized by a steady state distribution of the position of the particle $P_{{\rm st},r}(x)$ given by~\cite{resettingRTP} 
\begin{equation}\label{Pandlambdar}
P_{{\rm st},r}(x) = \frac{\lambda_r}{2}\, \exp\left[-\lambda_r|x-x_0|\right] \quad , \quad \lambda_r =\left(\frac{r(r+2\gamma)}{v_0^2}\right)^{\frac{1}{2}}\, .
\end{equation}
On the other hand, an RTP in a linear potential $V(x)=\alpha|x-x_0|$, if $\alpha<v_0$, reaches also a steady state described by (see e.g., \cite{RTPSS})
\begin{equation} \label{Pst_lin}
P_{{\rm st}}(x) = \frac{\gamma\, \alpha}{v_0^2-\alpha^2}\,  \exp\left[-\frac{2\gamma \alpha}{v_0^2-\alpha^2}|x-x_0|\right]\, .
\end{equation}
Therefore, the two steady states (\ref{Pandlambdar}) and (\ref{Pst_lin}) coincide if we choose $\lambda_r = {2\gamma \alpha}/({v_0^2-\alpha^2)}$, i.e.
\begin{equation}\label{mapping}
\alpha = v_0\, \sqrt{\frac{r}{2\gamma +r}}\, .
\end{equation}
Hence, in the spirit of the work \cite{optimalMFPTdiffusive} described above, we can compare the optimal MFPT of the potential driven particle to the resetting driven one.

The MFPT of the resetting RTP was derived in \cite{resettingRTP} and reads
\begin{equation}
T_r(x_0) = \frac{1}{r}\left[-1+e^{\frac{\sqrt{1+u}}{u}z} \frac{u}{1+u-\sqrt{1+u}}\right]\quad \, , \quad u=\frac{2\gamma}{r}\quad \, , \quad z=\frac{2\gamma x_0}{v_0} \, .\label{tauresettinguz}
\end{equation}
To calculate the MFPT of an RTP within a linear potential centered at the particle's initial position, it is convenient to consider the potential $V(x)=\alpha|x-x_1|$ and then take the limit as $x_1$ approaches $x_0$. When $\alpha<v_0$, the force is in phase I, and the MFPT is given by Eq.~(\ref{phase1solLinf}). The force is given by $f(x)=-V'(x)= -\alpha\, \text{sign}(x-x_1)$, so it is crucial to carefully split the integrals in Eq.~(\ref{phase1solLinf}) into two separate cases to account for the regions where $x>x_1$ and $x<x_1$. Upon calculating these integrals, one finds (see also \cite{SurvivalRPTlinear, SurvivalRPTlinearMORI})
\begin{eqnarray}
\tau_\gamma(x_0) &&=  -\frac{x_0}{\alpha} + \frac{v_0}{2\alpha \gamma}+\frac{v_0(v_0+\alpha)}{\alpha^2 \gamma}\left(e^{\frac{2\alpha \gamma}{v_0^2 - \alpha^2}x_0}-1\right)\\ 
&&= \frac{1}{r}\frac{(1+u)}{u}\left[2\, e^{\frac{\sqrt{1+u}}{u}z}\left(1+\sqrt{\frac{1}{1+u}}\right)-\sqrt{\frac{1}{1+u}}(z+1)-2\right]\, \, , \quad u=\frac{2\gamma}{r}\quad \, , \quad z=\frac{2\gamma x_0}{v_0} \, ,\label{tauRTPpotentialuz}
\end{eqnarray}
where we have used the relation (\ref{mapping}) in the second equality.

\begin{figure}[t]
    \centering
    \includegraphics[width=0.5\linewidth]{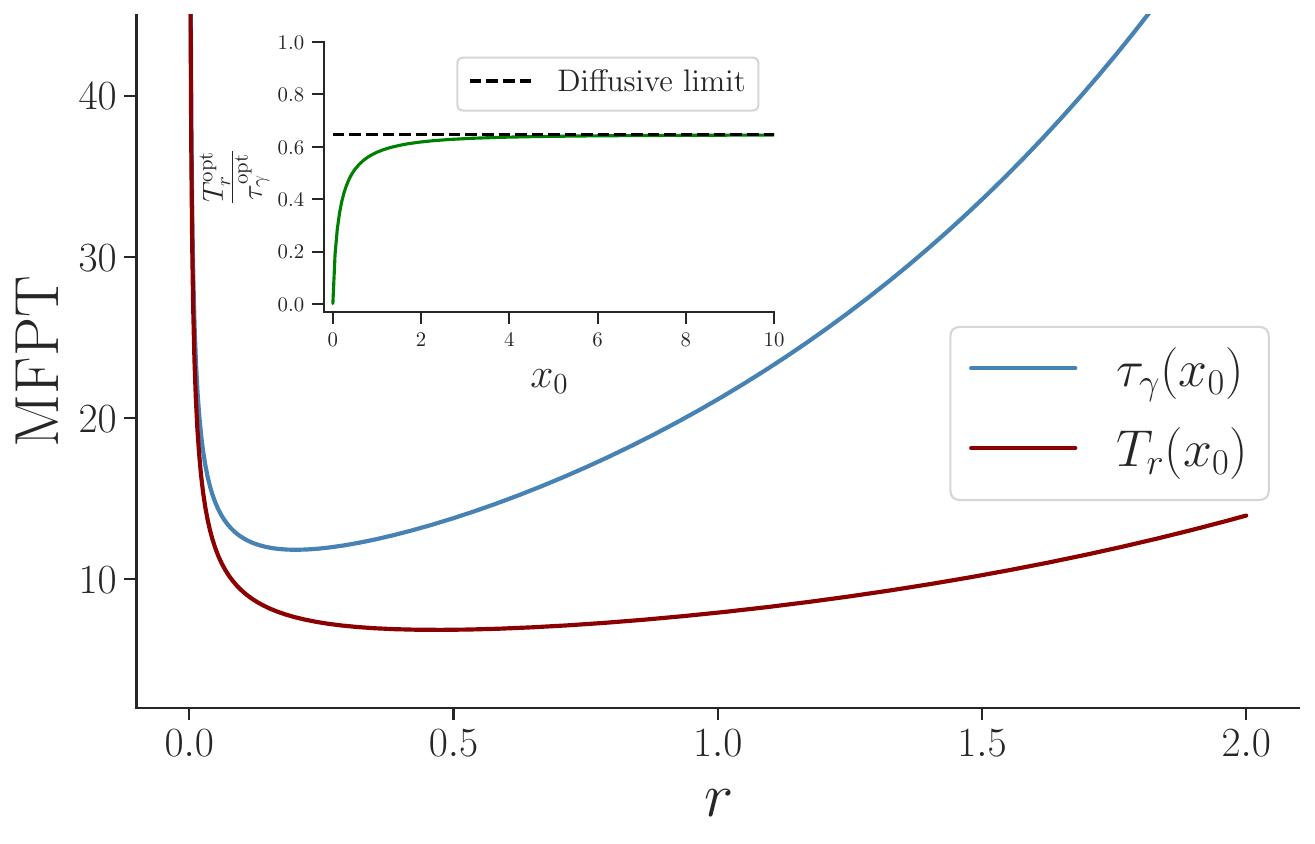}
  \caption{For $\gamma = 1$, $x_0=1$ and $v_0=1$, we plot $T_r(x_0)$ and $\tau_\gamma(x_0)$ versus $r$. Both curves exhibit a minimum at an optimal tumbling rate, with $T_r(x_0)$ always smaller than $\tau_\gamma(x_0)$. The inset shows that the ratio $T_r^{\text{opt}}/\tau_\gamma^{\text{opt}}$ converges to $0.646728\ldots$ for large value of $x_0$ matching the ratio of the optimal MFPTs for a diffusive particle.} 
  \label{resettingRTP_vs_RTP} 
\end{figure}

We now introduce the difference between the mean first-passage times and study its sign to determine which strategy is more efficient in finding the target at the origin
\begin{eqnarray}
    \delta(u) &&= r\, \left(T_r(x_0) - \tau_\gamma(x_0)\right)\\
    &&= \frac{2+u +\sqrt{1+u}(1+z)}{u}-\frac{e^{\frac{\sqrt{1+u}}{u}z} (2+u)}{(1+u)-\sqrt{1+u}}\, .
\end{eqnarray}
Note that we have multiplied the difference of the MFPTs by a factor $r$ due to the scaling form in Eqs.~(\ref{tauresettinguz}) and (\ref{tauRTPpotentialuz}) -- since $r>0$ this does not affect the analysis of the sign of $\delta(u)$. Using the fact that $e^{\frac{\sqrt{1+u}}{u}z} \geq 1 +\frac{\sqrt{1+u}}{u}z$, it is possible to verify that $\delta(u)$ is always negative for fixed $z>0$. This implies that the resetting RTP is always more efficient in reaching the origin than the potential-driven RTP, for any value of $r$. Hence as in the passive case, resetting offers a better search strategy ! 

For $x_0>0$, both $T_r(x_0)$ and $\tau_\gamma(x_0)$ have a minimum with respect to $r$. In the large $x_0$ limit, with fixed $\gamma$ and $v_0$, the particles that reach the origin have experienced a large number of tumbles, and have taken a long time, so their behavior is essentially diffusive. Therefore, we expect the ratio $T_r^{\mathrm{opt}}/\tau^{\mathrm{opt}}_\gamma$ to take the diffusive value $ 0.646728\ldots$ found in \cite{optimalMFPTdiffusive} when $x_0 \to \infty$. In fact, for diffusive particles, the minimum of the MFPTs is reached at a specific rate $r^* \propto x_0^{-2}$. Using this scaling, one can solve the minimisation equations $\partial_r T_r(x_0)=0$ and $\partial_r \tau_\gamma(x_0)=0$ and identify the optimal rate in both cases. Plugging this back in the original expression, we obtain in the large $x_0$ limit
\begin{eqnarray}
    &&T_r^{\text{opt}}(x_0) \approx \left(\frac{e^{C_1 - 1}}{C_1^2}\right)\, \frac{x_0^2}{D_{\text{eff}}}\, ,\text{ where} \quad 2(1-e^{C_1})-C_1 =0 \, , \quad C_1= 1.54414\ldots\, , \\
    &&\tau_\gamma^{\text{opt}}(x_0) \approx \left(\frac{2(e^{C_2}-1)-C_2)}{C_2^2}\right)\, \frac{x_0^2}{D_{\text{eff}}}\, ,\text{ where} \quad 4+2e^{C_2}(C_2-2)+C_2 =0 \, , \quad C_2= 2.38762\ldots\, ,
\end{eqnarray}
with $D_{\text{eff}}=\frac{v_0^2}{2\gamma}$. These coincide with the results derived in \cite{optimalMFPTdiffusive} and as expected, it gives $T_r^{\mathrm{opt}}/\tau^{\mathrm{opt}}_\gamma\approx  0.646728\ldots$. In Fig. \ref{resettingRTP_vs_RTP}, we show a plot of $T_r(x_0)$ and $\tau_\gamma(x_0)$ when $\gamma=1$; $x_0=1$ and $v_0=1$. We also show the ratio $T_r^{\mathrm{opt}}/\tau^{\mathrm{opt}}_\gamma$ with respect to $x_0$ when $\gamma=1$ and $v_0=1$.

\section{Conclusion}\label{conclusionsection}

The mean first-passage time is an important quantity to probe the first-passage properties of a stochastic process. Generally, its calculation is challenging, especially for active particles, for which few analytical results are available. In this paper, we focused on a one-dimensional run-and-tumble particle moving within an arbitrary potential. We considered only the case of a particle subjected to a telegraphic noise, excluding thermal noise. Depending on the shape of the potential, we have derived explicit expressions in different phases corresponding to different shapes of potentials. The method developed in this paper provides a systematic method to compute the MFPT for a very wide range of potentials. We have illustrated it with the calculation of the MFPT for an RTP in the presence of a logarithmic potential and a double-well potential, which both exhibit quite rich behaviors. 
In addition, we have proposed three applications to demonstrate the physical relevance of our results. Among these, we derived Kramers' law for a one-dimensional RTP. Another application is the study of optimal search strategies, showing that a resetting RTP is more efficient at finding a target than a potential-driven RTP.

The present work can be extended in several interesting direction. For instance, 
although run times were traditionally believed to follow an exponential distribution (see e.g., \cite{Naturecoli}), more recent measurements suggest that a power-law distribution may be more accurate in some cases \cite{natureruntime}. In this paper, we considered run times that are exponentially distributed which is commonly assumed in RTP models and is considered a reasonable approximation based on experimental data \cite{PRLcoli}. However, it would be interesting to compute the MFPT for other distribution of run times. Another interesting generalization would be to consider a space-dependent tumble rate $\gamma(x)$ \cite{inhomoRTP,inhomoRTP2}. Following our derivations, it should be possible to write an explicit expression for this specific case (e.g. see the appendices of \cite{RTPPSG}). Finally, it would also be interesting to study the MFPT of confined RTP in higher dimensions, for which very few exact results exist.

\acknowledgments
We would like to thank the Isaac Newton Institute for Mathematical Sciences, Cambridge, for support and hospitality during the programme {\it Stochastic systems for anomalous diffusion}, 
where part of this work was done. This work was supported by EPSRC grant no EP/K032208/1. We also acknowledge support from ANR Grant No. ANR-23-CE30-0020-01 EDIPS.

\appendix
\section*{Appendices}
\section{Derivation of the backward Fokker-Planck equations for the survival probabilities $Q^\pm(x,t)$}\label{survivalappendix}

\begin{figure}[t]
    \centering
    \includegraphics[width=0.6\linewidth]{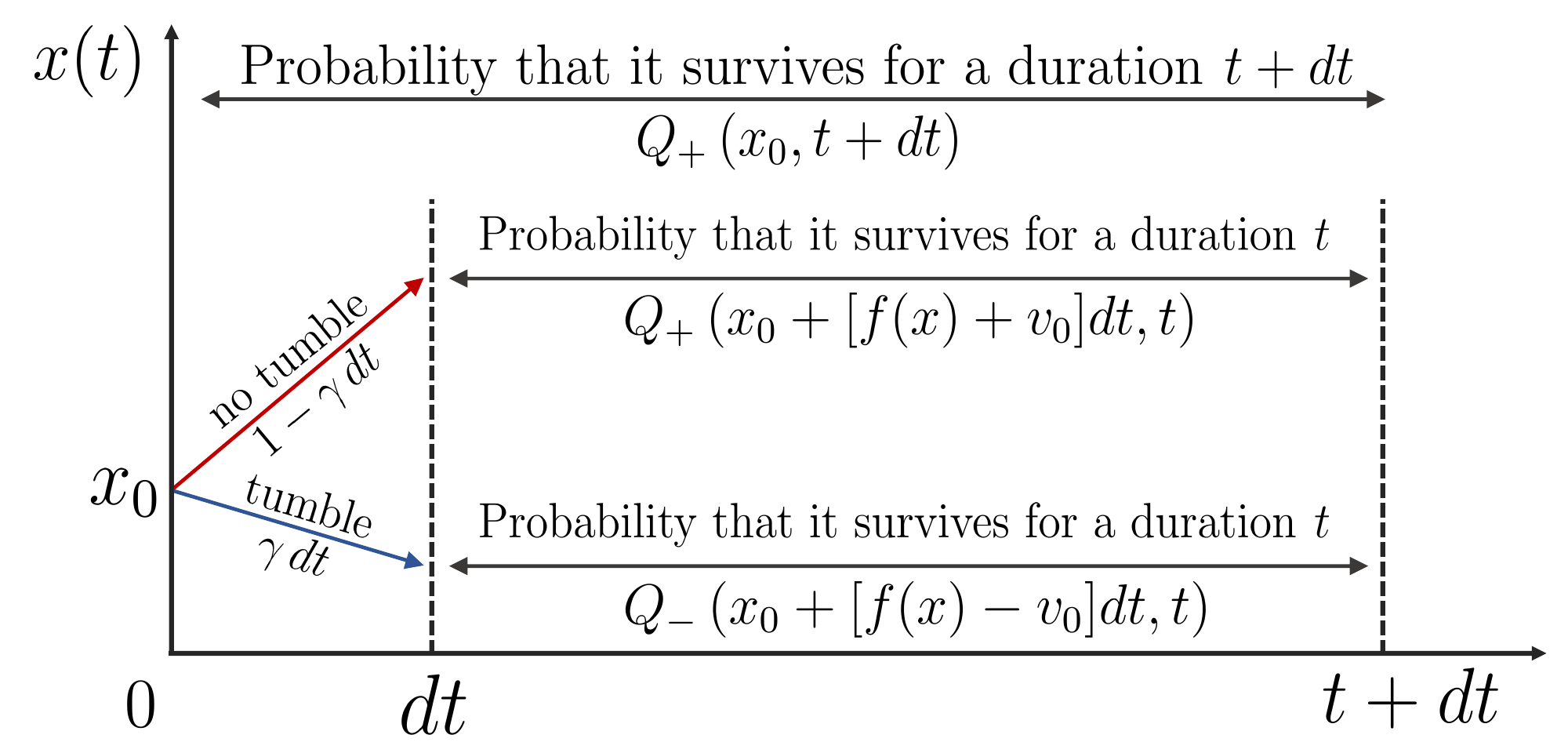}
  \caption{Illustration for the derivation of the backward Fokker-Plank equation for an RTP starting in state $+$ at position $x_0$. The probability for the RTP to survives for a duration $t+dt$ can be written as follows: between time $0$ and time $dt$, the RTP tumbles with probability $\gamma\, dt$ while it stays in the same state with probability $1-\gamma\, dt$. In both cases, it must still survive for a duration $t$.}
  \label{BackwardFP} 
\end{figure}

Consider a run-and-tumble particle (RTP) in the positive region of the 1D line whose position is denoted by $x(t)$, with $x(0) \in [0,+\infty[$. The equation of motion reads 
\begin{equation}
    \dot{x}(t)=f(x)+v_0\, \sigma(t)\, ,
\label{RTPmotion}
\end{equation}
where $\sigma(t)$ represents a telegraphic noise that alternates between two possible values, +1 or -1, at exponentially distributed random times. The RTP is subjected to a force $f(x)$ that derives from a potential such that $f(x) = -V'(x)$. We denote by $Q^+(x,t)$ and $Q^-(x,t)$ the survival probabilities of RTP's with respectively $\sigma(0) = +1$, or $\sigma(0) = -1$. It is the probability that the particle stayed in the positive region up to time $t$ having started its motion at position~$x(0)$. Let us derive the backward Fokker-Planck equation for $Q^+$, and $Q^-$. For this purpose, one can first write the disretised version of Eq. (\ref{RTPmotion}). If at time $t$, the particle is in state $\sigma(t) = +1$, then one has
\begin{equation}
x(t+dt) = \begin{cases}
x(t) + \left[f(x) + v_0\right]dt &\text{, \; with \, probability } 1 - \gamma\, dt \text{, and } \sigma(t+dt) = +1 \\
x(t)&, \; \;\, \text{with \, probability } \gamma\, dt \text{, and } \sigma(t+dt) = -1 \;.
\end{cases}
\end{equation}
On the opposite, if at time $t$, the particle is in state $\sigma(t) = -1$, one has
\begin{equation}
x(t+dt) = \begin{cases}
x(t) + \left[f(x) - v_0\right]dt &\text{, \; with probability } 1 - \gamma\, dt \text{, and } \sigma(t+dt) = -1 \\
x(t)&, \; \text{with probability } \gamma\, dt \text{, and } \sigma(t+dt) = +1 \;.
\end{cases}
\end{equation}
Now suppose that the RTP starts its motion at time $t=0$ at position $x(0)=x_0$. We want to write the probability $Q^\pm(x,t+dt)$ that it survives for a duration $t+dt$. If the initial state is $\sigma(0) = +1$, we have two possibilities as explained in Fig. \ref{BackwardFP}. First, with probability $(1-\gamma\, dt)$, there is no change of state such that $\sigma(dt)=+1$. From time $t=0$ to time $t=dt$, the particle has speed $\dot{x}=f(x)+v_0$ so that at time $dt$, its position is $x_0+\left[f(x) +v_0\right]dt$. On the other hand, with probability $\gamma\, dt$, the particle tumbles and $\sigma(dt)=-1$ while its new position is $x_0+\left[f(x) -v_0\right]dt$. In both cases, after a duration $dt$, the particle still needs to survive for a time $t$. Hence, we have 
\begin{equation} \label{FP_1}
    Q^+(x_0,t+dt)= (1-\gamma\, dt)\, Q^+(x_0+\left[f(x_0) +v_0\right]dt,t) + \gamma \, dt\,  Q^-(x_0+\left[f(x) -v_0\right]dt,t)\, .
\end{equation}
One can Taylor expand the first term at first order in $dt$, which gives
\begin{equation}
    Q^+(x_0+\left[f(x_0) +v_0\right]dt,t) = Q^+(x_0,t) + \left[f(x_0) +v_0\right]dt \, \partial_{x_0} Q^+(x_0,t)\, .
\end{equation}
Injecting it back in (\ref{FP_1}), and keeping only terms of order $dt$, we obtain
\begin{equation}
    Q^+(x_0,t+dt)= Q^+(x_0,t) +  \left[f(x_0) +v_0\right]dt \, \partial_{x_0} Q^+(x_0,t)  - \gamma \, dt Q^+(x_0,t)+ \gamma \, dt Q^-(x_0,t) \, .
\end{equation}
Finally, taking the limit $dt \to 0$ leads to
\begin{equation}
    \partial_t Q^+(x_0,t) = \left[f(x_0) + v_0\right]\partial_{x_0} Q^+(x_0,t) -\gamma\, Q^+(x_0,t) + \gamma\, Q^-(x_0,t)\, .
\label{BFPplus}
\end{equation}
An analogous reasoning gives us the equation for $Q^-(x_0,t)$, namely
\begin{equation}
    \partial_t Q^-(x_0,t) = \left[f(x_0) - v_0\right]\partial_{x_0} Q^-(x_0,t) -\gamma\, Q^-(x_0,t) + \gamma\, Q^+(x_0,t)\, .
\end{equation}
We have therefore derived the announced backward Fokker-Plank equations~(\ref{survivalequation1}) and (\ref{survivalequation2}) given in the text.

\section{Higher order moments of the distribution of the first-passage time}\label{momentsappendix}

In this Appendix, we derive a recursive relation for the moments of the distribution of the first-passage time $F_{\rm fp}(x_0,t) = - \partial_t Q(x_0,t)$ where we recall $Q(x_0,t) = \frac{1}{2} (Q^+(x_0,t)+Q^-(x_0,t))$. We thus define $\langle T^{n}_\pm \rangle = -\int_0^{+\infty}dt\, t^n\, \partial_t Q^\pm(x_0,t)$ such that $\langle T^{n} \rangle = \frac{1}{2}(\langle T^{n}_+ \rangle+\langle T^{n}_- \rangle)$. Here, we assume all moments $\langle T^{n} \rangle$ to be well-defined. We also have the identity $\langle T^{}_\pm \rangle \equiv T_\gamma^\pm(x_0)$. Following from the derivation of equations~(\ref{coupledtaupm1}) and (\ref{coupledtaupm2}), it is easy to generalised the two coupled equations to the $n^{\text{th}}$ moments yielding
\begin{eqnarray}\label{coupledtaupmN1}
    && \left(f(x_0)+ v_0\right) \partial_{x_0}\langle T^{n}_+ \rangle - \gamma \langle T^{n}_+ \rangle + \gamma \langle T^{n}_- \rangle= -n\, \langle T^{n-1}_+ \rangle \, ,\\
&&\left(f(x_0) - v_0\right) \partial_{x_0}\langle T^{n}_- \rangle + \gamma \langle T^{n}_+ \rangle  -\gamma \langle T^{n}_- \rangle  =- n \langle T^{n-1}_- \rangle\, .
\label{coupledtaupmN2}
\end{eqnarray}
It is possible to re-write the system in terms of the differential operator $\mathcal{L}_\pm$ 
\begin{eqnarray}\label{coupledtaupmN1bis}
    && \mathcal{L}_+ \langle T^{n}_+ \rangle  =\left[\left(f(x_0)+ v_0\right) \partial_{x_0} - \gamma  \right] \langle T^{n}_+ \rangle = -n\, \langle T^{n-1}_+ \rangle - \gamma \langle T^{n}_- \rangle\, ,\\
&&\mathcal{L}_- \langle T^{n}_- \rangle  =\left[\left(f(x_0) - v_0\right) \partial_{x_0}- \gamma \right] \langle T^{n}_- \rangle  = -n \langle T^{n-1}_- \rangle -\gamma \langle T^{n}_+ \rangle  \, .
\label{coupledtaupmN2bis}
\end{eqnarray}
Applying $\mathcal{L}_-$ on eq.~(\ref{coupledtaupmN1bis}) and $\mathcal{L}_+$ on eq.~(\ref{coupledtaupmN2bis}) gives the second order differential equations
\begin{eqnarray}\label{coupledorder2A}
  \hspace*{-0.5cm}  && \left(f(x_0)^2-v_0^2\right)\partial_{x_0}^2\langle T^{n}_+ \rangle  + \left[(f(x_0)-v_0)f'(x_0) -2\gamma f(x_0)\right] \partial_{x_0}\langle T^{n}_+ \rangle  = -n\left(f(x_0)-v_0\right)\partial_{x_0}\langle T^{n-1}_+ \rangle+2n\gamma \langle T^{n-1} \rangle\, ,\\
\hspace*{-0.5cm} && \left(f(x_0)^2-v_0^2\right)\partial_{x_0}^2\langle T^{n}_- \rangle  + \left[(f(x_0)+v_0)f'(x_0) -2\gamma f(x_0)\right] \partial_{x_0}\langle T^{n}_- \rangle  = -n\left(f(x_0)+v_0\right)\partial_{x_0}\langle T^{n-1}_- \rangle+2n\gamma \langle T^{n-1} \rangle\, .
\label{coupledorder2B}
\end{eqnarray}
Summing the two second order equations gives
\begin{eqnarray}
    && 2\left(f(x_0)^2-v_0^2\right)\partial_{x_0}^2\langle T^{n} \rangle  + 2f(x_0)\left[f'(x_0) -2\gamma \right] \partial_{x_0}\langle T^{n} \rangle  -v_0 f'(x_0)\partial_{x_0} (\langle T^{n}_+ \rangle-\langle T^{n}_- \rangle)\nonumber \\
    &&= -2 n f(x_0)\partial_{x_0}\langle T^{n-1} \rangle + 4n\gamma \langle T^{n-1} \rangle + n v_0\,  \partial_{x_0} (\langle T^{n-1}_+ \rangle-\langle T^{n-1}_- \rangle)\, .
\label{couplesum}
\end{eqnarray}
Similarly, by summing up Eqs. (\ref{coupledtaupmN1}) and (\ref{coupledtaupmN2}) gives us the following useful relation
\begin{eqnarray}
    v_0\, \partial_{x_0}(\langle T^{n}_+ \rangle-\langle T^{n}_- \rangle)=-2 n \langle T^{n-1} \rangle - 2 f(x_0) \partial_{x_0}\langle T^{n} \rangle\, .
\end{eqnarray}
Plugging this inside (\ref{couplesum}) leads to the final result announced in the main text in equation (\ref{highermoments}) for $n\geq 2$
\begin{eqnarray}
    && \left(v_0^2-f(x_0)^2\right)\partial_{x_0}^2\langle T^{n} \rangle  + 2f(x_0)\left[\gamma-f'(x_0) \right] \partial_{x_0}\langle T^{n} \rangle \nonumber \\
    &&= n\, (f'(x_0)-2\gamma)\langle T^{n-1} \rangle+2n f(x_0)\partial_{x_0}\langle T^{n-1} \rangle+ n^2 \langle T^{n-2} \rangle\, .
\label{couplesum2}
\end{eqnarray}
In principle, Eq.~(\ref{couplesum2}) is of first order and could be solved formally. However, to express the explicit solution we need two conditions that may depend on the value of $n$.
Note that in the diffusive limit, i.e., when $v_0\to \infty$ and $\gamma \to \infty$ with $D=\frac{v_0^2}{2 \gamma}$ fixed 
we retrieve the recursive relation \cite{diffusivemoments} 
\begin{eqnarray} \label{singular}
    D\, \partial_{x_0}^2 \langle T^{n} \rangle  + f(x_0)\partial_{x_0} \langle T^{n} \rangle  =-n\langle T^{n-1} \rangle  \,,
\end{eqnarray}
which is sometimes called in the literature ``Pontryagin equation''~\cite{diffusivemoments}. 

\section{Behavior of $\tau_\gamma(x)$ close to a negative turning point $x_-$}\label{behavior_xneg}

To understand the behavior of $\tau_\gamma(x_0)$ and $\tau^{\pm}_\gamma(x_0)$ near a negative fixed point $x_-$ such that $f(x_-)=-v_0$, it is useful to linearize Eqs. \eqref{ODE2ndTau}-\eqref{ODEtauplus}. They read, setting $x_0 = x_- + u$ and $u>0$
\begin{eqnarray} \label{tau_lin}
u \tau''_\gamma(u) + \left(1 - \frac{\gamma}{f'(x_-)}\right)\tau'_\gamma(u) = \frac{f'(x_-)-2 \gamma}{2 v_0} \;,
\end{eqnarray}
as well as
\begin{eqnarray}
&& \tau_\gamma^-(u) = - \frac{1}{2 \gamma} - \frac{f'(x_-)}{2\gamma} u \tau'_\gamma(u) + \tau_\gamma(u) \;, \label{taum_lin}  \\
&& \tau_\gamma^+(u) = + \frac{1}{2 \gamma} + \frac{f'(x_-)}{2\gamma} u \tau'_\gamma(u) + \tau_\gamma(u) \;. \label{taup_lin}
\end{eqnarray}
For this type of differential equation (\ref{tau_lin}) the special point $u=0$ is called a ``regular singular'' point see e.g. \cite{regularsingular}.

The solution of Eq. (\ref{tau_lin}) reads
\begin{eqnarray} \label{tau_lin_sol}
\tau_\gamma(u) = c_1 + c_2 u^{\frac{\gamma}{f'(x_-)}} + \frac{f'(x_-)(f'(x_-)-2\gamma)}{2v_0(f'(x_-)-\gamma)}\, u \;.    
\end{eqnarray}
Since $\tau_\gamma(u)$ cannot diverge as $u \to 0$, one sees that if $f'(x_-)<0$ (i.e., at a stable fixed point) then the constant $c_2$ must vanish and $\tau_\gamma(u)$ is a regular function (at least at leading order) close to $u=0$. However, if $f'(x_-)>0$ (at an unstable fixed point), then $c_2$ can be nonzero and in this case $\tau_\gamma(u)$ is a nonanalytic function of $u$. This is indeed what we found in the case of the logarithmic potential $f(x) = -1/(1+x)$. Interestingly, for an unstable fixed point, using Eqs. (\ref{taum_lin}) and (\ref{taup_lin}), one can show that $\tau_\gamma^-(u)$ is a regular function, even if $c_2 \neq 0$, while $\tau_\gamma^+(u)$ has the same nonanalytic behavior as $\tau_\gamma(u)$ in Eq. (\ref{tau_lin_sol}) with different coefficients $c_1 \to c_1^+$ and $c_2 \to c_2^+$. We have checked that our explicit solution of the logarithmic potential is fully consistent with this general analysis, leading to (\ref{tau_lin_sol}). 

Note that in principle a more detailed analysis of the full Eq.~(\ref{ODE2ndTau}) around the regular singular point $x=x_-$ can be done, beyond the leading order leading to (\ref{tau_lin_sol}). This can be achieved using the general theory of regular singular points for second order differential equations \cite{regularsingular}.

\section{Derivation of the MFPT in Phase II}\label{phase2appendix}
In this section, we derive the MFPT of an RTP in Phase II (see (\ref{solutionf(0)leqmainresults})). In this phase, $f(x)<-v_0$, and in particular $f(0)<-v_0$ such that the two states of the RTP have a negative velocity at the origin leading to the conditions
\begin{eqnarray}
     &&\tau_\gamma^+(x_0=0) = 0\, , \label{conditiondtau2b_appendix}\\
     &&\tau_\gamma^-(x_0=0) = 0\, . \label{tauminuscondition2b_appendix}
\end{eqnarray}
With these conditions, we can solve the differential equation satisfied by the MFPT $\tau_\gamma(x_0)$~\cite{letter}
\begin{equation}
\left[v_0^2 -f(x_0)^2\right] \partial^2_{x_0} \tau_\gamma + 2f(x_0)\left[\gamma-f'(x_0) \right]\partial_{x_0} \tau_\gamma = f'(x_0)- 2\gamma\, .
\end{equation}We can also express $\tau_\gamma^\pm$ in terms of $\tau_\gamma$, namely~\cite{letter}
\begin{equation}
\begin{split}
&\tau_\gamma^-(x_0) = \frac{1}{2\gamma}\frac{f(x_0)}{v_0} - \frac{1}{2\gamma v_0}\left[v_0^2-f(x_0)^2\right]\partial_{x_0} \tau_\gamma + \tau_\gamma\, ,\\
&\tau_\gamma^+(x_0) = - \frac{1}{2\gamma}\frac{f(x_0)}{v_0} +  \frac{1}{2\gamma v_0}\left[v_0^2-f(x_0)^2\right]\partial_{x_0} \tau_\gamma + \tau_\gamma \, .
\end{split}
\label{tauminustauplusappendices}
\end{equation} 
Let us introduce $w(x_0)=\partial_{x_0}\tau_\gamma$ such that the equation satisfied by $w(x)$ is now of first order\begin{equation}
\left[v_0^2 -f(x_0)^2\right] \partial_{x_0}w(x_0) + 2f(x_0)\left[\gamma-f'(x_0) \right]w(x_0) = f'(x_0)- 2\gamma\, .
\label{eq1storder}
\end{equation}First, let us solve the homogeneous equation \begin{equation}
\left[v_0^2 -f(x_0)^2\right] \partial_{x_0}w_H(x_0) + 2f(x_0)\left[\gamma-f'(x_0) \right] w_H(x_0) = 0\, ,
\end{equation}
whose solution is 
\begin{equation}
    w_H(x_0)= \frac{A}{v_0^2-f(x_0)^2}G(x_0)\, ,
\end{equation} with $G(x_0) = \text{exp}\left[\int_0^{x_0}dx \frac{-2\gamma f(x)}{v_0^2-f(x)^2}\right]$, and $A$ is a yet unknown integration constant. 

Using the method of variation of constants one can show that the full solution is 
\begin{equation}
    w(x_0) = \frac{G(x_0)}{v_0^2-f(x_0)^2}\left[\int_0^{x_0} dy \frac{f'(y)- 2\gamma}{G(y)} + A\right]\, .
\label{wsolwithA}
\end{equation}
By integrating Eq. (\ref{wsolwithA}) we get \begin{equation}
    \tau_\gamma(x_0) = \int_{0}^{x_0}dz\, \frac{G(z)}{v_0^2-f(z)^2}\int_0^{z} dy \frac{f'(y)- 2\gamma}{G(y)} + A\, \int_{0}^{x_0}dz\, \frac{G(z)}{v_0^2-f(z)^2} + B\, .
\end{equation}
Let us now compute $\tau_\gamma^-(x_0)$ using Eq. (\ref{tauminustauplusappendices}). One has
\begin{equation}
\begin{split}
        \tau_\gamma^-(x_0) &= \frac{1}{2\gamma}\frac{f(x_0)}{v_0} - \frac{1}{2\gamma v_0}\left[v_0^2-f(x_0)^2\right]\frac{G(x_0)}{v_0^2-f(x_0)^2}\left[\int_0^{x_0} dy \frac{f'(y)- 2\gamma}{G(y)} + A\right]\,  \\
        &+\int_{0}^{x_0}dz\, \frac{G(z)}{v_0^2-f(z)^2}\int_0^{z} dy \frac{f'(y)- 2\gamma}{G(y)} + A\, \int_{0}^{x_0}dz\, \frac{G(z)}{v_0^2-f(z)^2} + B\, ,
\end{split}
\end{equation}
where $B$ is an integration constant, yet to be determined. Taking the limit $x_0 \to 0$, using $\tau_\gamma^-(x_0)=0$, we obtain \begin{equation}
    0 = \frac{f(0)}{2\gamma v_0} -\frac{A}{2\gamma v_0} + B\, .
\end{equation}
Hence, \begin{equation}
    B = \frac{1}{2\gamma v_0}\left(A-f(0)\right)\, .
\end{equation}
The MFPT thus simplifies to \begin{equation}
    \tau_\gamma(x_0) = -\frac{f(0)}{2 \gamma v_0} +  \int_{0}^{x_0}dz\, \frac{G(z)}{v_0^2-f(z)^2}\int_0^{z} dy \frac{f'(y)- 2\gamma}{G(y)} + A\, \left[\int_{0}^{x_0}dz\, \frac{G(z)}{v_0^2-f(z)^2} + \frac{1}{2\gamma v_0}\right]\, .
\end{equation}
Now, we fix $A$, using that $\tau_\gamma^+(0) = 0$ [see Eq. (\ref{conditiondtau2b_appendix})], which leads to
\begin{equation}
    0 = -\frac{f(0)}{\gamma v_0} + \frac{A}{ \gamma v_0}\, ,
\end{equation}
so that $A = f(0)$.
Hence we arrive at the formula (\ref{solutionf(0)leqmainresults}) valid in the Phase II
\begin{equation}
    \tau_\gamma(x_0) =   \int_{0}^{x_0}dz\, \frac{G(z)}{v_0^2-f(z)^2}\left[\int_0^{z} dy \frac{f'(y)- 2\gamma}{G(y)} + f(0)\right]\, .
\label{solutionf(0)leq}
\end{equation}
It is also possible to get rid of the pre-factor $f(0)$ which is useful for instance for numerical integrations when $f(0)=-\infty$. Using integation by parts, one can show that Eq.~(\ref{solutionf(0)leqmainresults}) can be re-written as

\begin{equation}
    \tau_\gamma(x_0) =  \int_{0}^{x_0}dz\, \frac{f(z)}{v_0^2-f(z)^2} - \int_{0}^{x_0}dz\, \frac{2 \gamma\, v_0^2}{v_0^2-f(z)^2}\int_0^{z} dy\, \frac{1}{v_0^2-f(y)^2}\, \text{exp}\left(\int_y^{z}du\, \frac{-2\gamma f(u)}{v_0^2-f(u)^2}\right)\, .
\label{solutionf(0)leqmainresults_secondForm}
\end{equation}

\begin{figure}[t]
    \centering
    \includegraphics[width=0.5\linewidth]{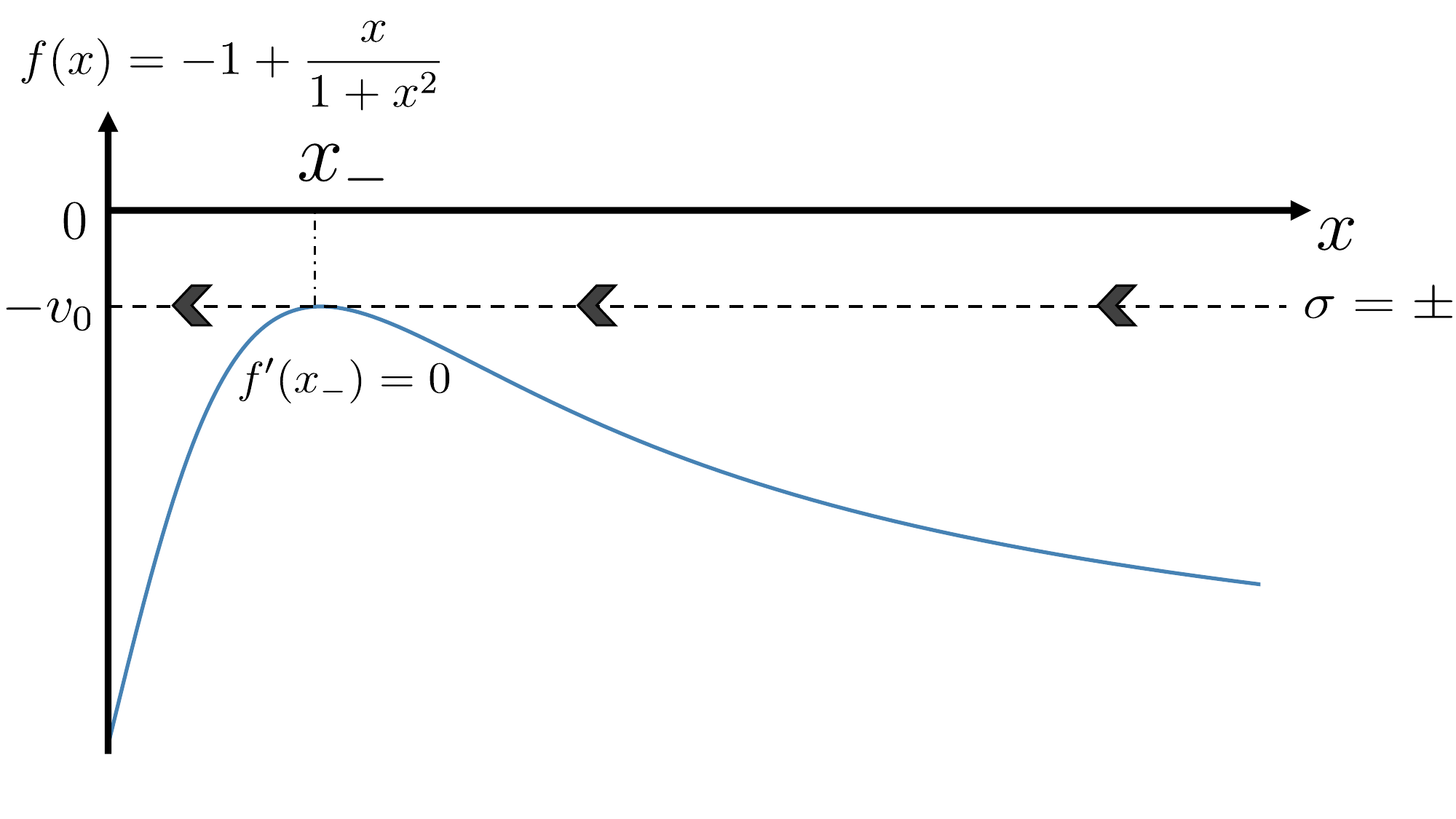}
  \caption{Plot of $f(x) = -1 + \frac{x}{1+x^2}$. When $v_0 = 1/2$, the force has a turning point $x_-$ such that $f(x_-)= -v_0$. It is neither stable nor unstable as $f'(x_-)=0$.}
  \label{Force_null_derivative} 
\end{figure}

\noindent
{\bf Remark.} Throughout this derivation we have supposed that $f(x)<-v_0$. However, Eq.~(\ref{solutionf(0)leq}) is also applicable for a force with a turning point $x_-$ such that $f(x_-) = -v_0$, but $f'(x_-)=0$ (in this case $x_-$ is called an ``irregular singular point''). For instance, consider the function $f(x) = -1 + {x}/{(1+x^2)}$ which has a maximum at $-1/2$. For $v_0 = 1/2$ we have solved numerically Eq.~(\ref{solutionf(0)leq}) and verified that our analytical solution agrees with the simulation results. In Fig. \ref{Force_null_derivative}, we show a plot of $f(x) = -1 + {x}/{(1+x^2)}$.

\section{Derivation of the MFPT for a combination of phase II and phase I}\label{Phase5derivation}

We consider here the case where the force has an unstable negative turning point $x_-^u$, such that $f(x_-^u) = -v_0$ and $f'(x_-^u) > 0$. In addition $f(0)<-v_0$, and $f(x)<v_0$ as it is the case for the log-potential studied in Section \ref{logpotsection}. First, we have to compute the MFPT ${\tau_{\text{\tiny I}}}_\gamma(x_0)$ on $[0,x_-^u[$. As we have $f(0)<-v_0$, the solution is directly given by the formula of phase II given in Eq.~(\ref{solutionf(0)leqmainresults}). For $x_0 \in ]x_-^u,+\infty[$, we have $|f(x)|<v_0$ such that we fix the first integration constant by introducing a reflective barrier at $x=L$, and then we send $L \to \infty$ as in Phase I leading to (for details see \cite{letterSM}),
\begin{equation}
    {\tau_{\text{\tiny II}}}_\gamma^+(x_0)= \int_{0}^{x_0}dz\, \frac{2\gamma}{v_0+f(z)}\, \int_{z}^{L} \frac{dy}{v_0-f(y)} \frac{G(z)}{G(y)} + B\, ,
    \label{reflectivetau+appendix}
\end{equation}
where we recall $G(z) = \text{exp}\left[\int_0^{z}dx \frac{-2\gamma f(x)}{v_0^2-f(x)^2}\right]$. From
\begin{equation}
    \left[v_0 + f(x_0)\right] \partial_{x_0}{\tau_{\text{\tiny II}}}_\gamma(x_0) - \gamma {\tau_{\text{\tiny II}}}_\gamma^+(x_0)+ \gamma {\tau_{\text{\tiny II}}}_\gamma^-(x_0)= -1 \, ,
\end{equation}
and using that $ {\tau_{\text{\tiny II}}}_\gamma(x_0) = 1/2 \left({\tau_{\text{\tiny II}}}_\gamma^+(x_0) +  {\tau_{\text{\tiny II}}}_\gamma(x_0)^-\right)$, we can write ${\tau_{\text{\tiny II}}}_\gamma(x_0)$ as
\begin{eqnarray}
     {\tau_{\text{\tiny II}}}_\gamma(x_0) = {\tau_{\text{\tiny II}}}_\gamma^+(x_0) -\frac{1}{2\, \gamma}\left[1+\left(v_0+f(x_0)\right)\partial_{x_0}{\tau_{\text{\tiny II}}}_\gamma^+(x_0)\right]\, .
\end{eqnarray}
This gives
\begin{equation}
     {\tau_{\text{\tiny II}}}_\gamma(x_0) =\int_{0}^{x_0}dz\, \frac{2\gamma}{v_0+f(z)}\, \int_{z}^{L} \frac{dy}{v_0-f(y)} \frac{G(z)}{G(y)} + B - \frac{1}{2 \gamma} -\int_{x_0}^L \frac{dy}{v_0 -f(y)}\frac{G(x_0)}{G(y)}\, .
\end{equation}
Finally, we fix the constant $B$ by continuity, by imposing ${\tau_{\text{\tiny I}}}_\gamma(x_-^u) ={\tau_{\text{\tiny II}}}_\gamma (x_-^u)$, and we obtain

\begin{equation}
   B =  {\tau_{\text{\tiny I}}}_\gamma(x_-^u) + \int_{0}^{x_-^u}dz\, \frac{2\gamma}{v_0+f(z)}\, \int_{z}^{L} \frac{dy}{v_0-f(y)} \frac{G(z)}{G(y)}  + \frac{1}{2 \gamma} + \int_{x_-^u}^L \frac{dy}{v_0 -f(y)}\frac{G(x_-^u)}{G(y)}\, .
\end{equation}
For $\epsilon\ll 1$, we have $f(x_-^u + \epsilon) = -v_0 + \epsilon f'(x_-^u)$, and it is possible to show (as done in \cite{letterSM}) that $G(x_-^u)\sim |\epsilon|^{\frac{\gamma}{f'(x_-^u)}}$. As $x_-^u$ is an unstable fixed point (i.e., $f'(x_-^u)>0$), the exponent is positive, and we conclude that $G(x_-^u)\to 0$ as $\epsilon\to 0$. Therefore, 
\begin{equation}
   B =  {\tau_{\text{\tiny I}}}_\gamma(x_-^u) + \int_{0}^{x_-^u}dz\, \frac{2\gamma}{v_0+f(z)}\, \int_{z}^{L} \frac{dy}{v_0-f(y)} \frac{G(z)}{G(y)}  + \frac{1}{2 \gamma} \, .
\end{equation}
Hence, taking the limit $L\to \infty$, the MFPT reads
\begin{eqnarray}
    {\tau_{\text{\tiny II}}}_\gamma(x_0)={\tau_{\text{\tiny I}}}_\gamma(x_-^u) - \int_{x_0}^{+\infty}\frac{dy}{v_0-f(y)}\, \frac{G(x_0)}{G(y)} + \int_{x_-^u}^{x_0}dz\, \frac{2 \gamma}{v_0 + f(z)}\, \int_z^{+\infty}\frac{dy}{v_0-f(y)}\frac{G(z)}{G(y)}\, ,
\end{eqnarray}
and we can re-write it as
\begin{eqnarray}
    {\tau_{\text{\tiny II}}}_\gamma(x_0)={\tau_{\text{\tiny I}}}_\gamma(x_-^u)-\int_{x_-^u}^{x_0}dz\, \frac{1}{v_0^2-f(z)^2}\, \int_{z}^{+\infty} dy\, \left(f'(y)-2\gamma\right)\text{exp}\left[\int_y^zdu\, \frac{-2\gamma\, f(u)}{v_0^2-f(u)^2}\right]\, .
\end{eqnarray}
To prove this, as both formulae coincide at $x_0=x_-^u$ (since $G(x_0=x_-^u) = 0$), it is enough to show that their derivatives are the same (see Eqs. (138)-(143) of \cite{letterSM}). This gives the formula (\ref{phase5asolution}) announced in the main text.

\section{Explicit expressions for $\tau_\gamma$ and $\tau_\gamma^\pm$ for different potentials}\label{taupmexpressions}

In this appendix, we give the explicit formulae derived for $\tau_\gamma$  and $\tau_\gamma^\pm$ for an RTP moving in a double-well potential, and a log-potential. The $\pm$ indicates the initial value $\pm 1$ of the telegraphic noise $\sigma(t)$ in (\ref{langeRTP}). The expressions of $\tau_\gamma^\pm$ are deduced from the ones of $\tau_\gamma$ (where the initial value of $\sigma(t)$ is chosen at random with probability $1/2$) using Eqs. (\ref{ODEtauminus}) and (\ref{ODEtauplus}).

\subsection{Double-well - $0<\alpha<v_0$}
For a double-well potential $V(x) = \alpha/2\, \left(|x|-1\right)^2$ studied in Section \ref{doublewellsection}, when $0<\alpha<v_0$, the MFPT is given by Eq.~(\ref{phase3DWequ}). The integrals can be computed explicitly after some manipulations, and it leads to 
\begin{eqnarray}
  &&\tau_\gamma(x_0) =  \frac{1}{{2 \gamma}} + \frac{{\alpha + v_0}}{{2 v_0 (\alpha + \gamma)}} \left(\frac{{2}}{{1 - \frac{{\alpha}}{{v_0}}}}\right)^{\frac{{\gamma}}{{\alpha}}} \ {}_{2}F_{1}\left(1 + \frac{{\gamma}}{{\alpha}}, 1 - \frac{{\gamma}}{{\alpha}}, 2 + \frac{{\gamma}}{{\alpha}}, \frac{{1}}{{ 2}} + \frac{{\alpha }}{{2 v_0}}\right)  \nonumber \\
  && + \frac{{\sqrt{\pi}}}{{v_0}} \frac{{\Gamma\left(1 + \frac{{\gamma}}{{\alpha}}\right)}}{{\Gamma\left(\frac{{1}}{{2}} + \frac{{\gamma}}{{\alpha}}\right)}} \left[\ {}_{2}F_{1}\left(\frac{{1}}{{2}}, 1 + \frac{{\gamma}}{{\alpha}}, \frac{{3}}{{2}}, \frac{{\alpha^{2}}}{{v_0^{2}}}\right) - (1 - x_0) \ {}_{2}F_{1}\left(\frac{{1}}{{2}}, 1 + \frac{{\gamma}}{{\alpha}}, \frac{{3}}{{2}}, \frac{{\alpha^{2} (1-x_0)^{2}}}{{v_0^{2}}}\right)\right] \nonumber \\
  && + \frac{{\alpha + 2 \gamma}}{{2 v_0^{2}}}\left[ \ {}_{3}F_{2}\left(\left\{1, 1, \frac{{3}}{{2}} + \frac{{\gamma}}{{\alpha}}\right\};\left\{ \frac{{3}}{{2}}, 2 \right\}; \frac{{\alpha^{2}}}{{v_0^{2}}}\right)- (1 - x_0)^{2} \ {}_{3}F_{2}\left(\left\{1, 1, \frac{{3}}{{2}} + \frac{{\gamma}}{{\alpha}}\right\};\left\{ \frac{{3}}{{2}}, 2 \right\}; \frac{{\alpha^{2} (1 - x_0)^{2}}}{{v_0^{2}}}\right)\right]\, .
  \label{MFPTDW1}
\end{eqnarray}
From Eqs. (\ref{ODEtauminus}) and (\ref{ODEtauplus}) we deduce 
\begin{eqnarray}
    &&\tau_\gamma^-(x_0) = \frac{1}{2\gamma}+\frac{\alpha(1-x_0)}{2\gamma v_0} + \frac{\alpha+v_0}{2(\alpha+\gamma)v_0}\left(\frac{2v_0}{v_0-\alpha}\right)^{\frac{\gamma}{\alpha}}\ {}_{2}F_{1}\left(1+\frac{\gamma}{\alpha},1-\frac{\gamma}{\alpha},2+\frac{\gamma}{\alpha},\frac{\alpha+v_0}{2v_0}\right)\nonumber \\
    &&- \frac{\alpha+2\gamma}{2\gamma v_0}\left(1-\frac{\alpha^2(1-x_0)^2}{v_0^2}\right)(1-x_0)\ {}_{2}F_{1}\left(1,\frac{3}{2}+\frac{\gamma}{\alpha},\frac{3}{2},\frac{\alpha^2(1-x_0)^2}{v_0^2}\right)  \nonumber \\
    && \nonumber \\
    &&-\frac{\sqrt{\pi}}{2\alpha} \frac{\Gamma\left(\frac{\gamma}{\alpha}\right)}{\Gamma\left(\frac{1}{2}+\frac{\gamma}{\alpha}\right)} \left[\left(1-\frac{\alpha^2(1-x_0)^2}{v_0^2}\right)^{-\frac{\gamma}{\alpha}} - \frac{2\gamma}{v_0}\left(\ {}_{2}F_{1}\left(\frac{1}{2},1+\frac{\gamma}{\alpha},\frac{3}{2},\frac{\alpha^2}{v_0^2}\right) - (1-x_0)\ {}_{2}F_{1}\left(\frac{1}{2},1+\frac{\gamma}{\alpha},\frac{3}{2},\frac{\alpha^2(1-x_0)^2}{v_0^2}\right)\right)\right]\nonumber \\
    &&+ \frac{\alpha+2\gamma}{2v_0^2}\left[\ {}_{3}F_{2}\left(\left\{1,1,\frac{3}{2}+\frac{\gamma}{\alpha}\right\};\left\{ \frac{3}{2},2 \right\} ;\frac{\alpha^2}{v_0^2}\right) - (1-x_0)^2 \ {}_{3}F_{2}\left(\left\{1,1,\frac{3}{2}+\frac{\gamma}{\alpha}\right\};\left\{ \frac{3}{2},2 \right\} ;\frac{\alpha^2(1-x_0)^2}{v_0^2}\right)\right]\, ,
    \label{minusMFPTDW}
\end{eqnarray}
and also
\begin{eqnarray}
    &&\tau_\gamma^+(x_0) = \frac{1}{2\gamma}-\frac{\alpha(1-x_0)}{2\gamma v_0} + \frac{\alpha+v_0}{2(\alpha+\gamma)v_0}\left(\frac{2v_0}{v_0 - \alpha}\right)^{\frac{\gamma}{\alpha}} \ {}_{2}F_{1}\left(1+\frac{\gamma}{\alpha},1-\frac{\gamma}{\alpha},2+\frac{\gamma}{\alpha},\frac{\alpha+v_0}{2v_0}\right)\nonumber \\
    &&+ \frac{\alpha+2\gamma}{2\gamma v_0}\left(1-\frac{\alpha^2(1-x_0)^2}{v_0^2}\right)(1-x_0)\ {}_{2}F_{1}\left(1,\frac{3}{2}+\frac{\gamma}{\alpha},\frac{3}{2},\frac{\alpha^2(1-x_0)^2}{v_0^2}\right)  \nonumber \\
    && \nonumber \\
    &&+\frac{\sqrt{\pi}}{2\alpha} \frac{\Gamma\left(\frac{\gamma}{\alpha}\right)}{\Gamma\left(\frac{1}{2}+\frac{\gamma}{\alpha}\right)} \left[\left(1-\frac{\alpha^2(1-x_0)^2}{v_0^2}\right)^{-\frac{\gamma}{\alpha}} + \frac{2\gamma}{v_0}\left(\ {}_{2}F_{1}\left(\frac{1}{2},1+\frac{\gamma}{\alpha},\frac{3}{2},\frac{\alpha^2}{v_0^2}\right) - (1-x_0)\ {}_{2}F_{1}\left(\frac{1}{2},1+\frac{\gamma}{\alpha},\frac{3}{2},\frac{\alpha^2(1-x_0)^2}{v_0^2}\right)\right)\right]\nonumber \\
    &&+ \frac{\alpha+2\gamma}{2v_0^2}\left[\ {}_{3}F_{2}\left(\left\{1,1,\frac{3}{2}+\frac{\gamma}{\alpha}\right\};\left\{ \frac{3}{2},2 \right\} ;\frac{\alpha^2}{v_0^2}\right) - (1-x_0)^2 \ {}_{3}F_{2}\left(\left\{1,1,\frac{3}{2}+\frac{\gamma}{\alpha}\right\};\left\{ \frac{3}{2},2 \right\} ;\frac{\alpha^2(1-x_0)^2}{v_0^2}\right)\right]\, .
    \label{plusMFPTDW}
\end{eqnarray}

\subsection{Double-well - $\alpha<-v_0$ - $x_0 \leq x_-^u$}

Again, for the double-well $V(x) = \alpha/2\, \left(|x|-1\right)^2$, it is possible to derive the MFPT when $\alpha<-v_0$ leading to the expression (\ref{TAUdwsecondformul}) for $x_0 \leq x_-^u = 1+ v_0/\alpha$.
The explicit integrations give
\begin{eqnarray}
  &&\tau_\gamma(x_0) =  
-\frac{{(\alpha + 2\gamma)}}{{v_0^2}} \, {}_{2}F_{1}\left(\frac{1}{2}, -\frac{\gamma}{\alpha}, \frac{3}{2}, \frac{{\alpha^2}}{{v_0^2}}\right) \,  \left[{}_{2}F_{1}\left(\frac{1}{2}, 1 + \frac{\gamma}{\alpha}, \frac{3}{2}, \frac{{\alpha^2}}{{v_0^2}}\right) - (1 - x_0) \,  {}_{2}F_{1}\left(\frac{1}{2}, 1 + \frac{\gamma}{\alpha}, \frac{3}{2}, \frac{{\alpha^2(1 - x_0)^2}}{{v_0^2}}\right)\right] \nonumber \\
&&+ 
\frac{{\alpha}}{{v_0^2}} \, \left[{}_{2}F_{1}\left(1, \frac{1}{2} - \frac{\gamma}{\alpha}, \frac{3}{2}, \frac{{\alpha^2}}{{v_0^2}}\right) - (1 - \frac{{\alpha^2}}{{v_0^2}})^{\frac{{\gamma}}{{\alpha}}} \,  (1 - x_0) \,  {}_{2}F_{1}\left(\frac{1}{2}, 1 + \frac{\gamma}{\alpha}, \frac{3}{2}, \frac{{\alpha^2(1 - x_0)^2}}{{v_0^2}}\right)\right] + \nonumber \\
&&\frac{{(\alpha + 2\gamma)}}{{2v_0^2}} \, \left[{}_{3}F_{2}\left(\left\{1, 1, \frac{3}{2} + \frac{\gamma}{\alpha}\right\}; \left\{\frac{3}{2}, 2\right\}; \frac{{\alpha^2}}{{v_0^2}}\right) - (1 - x_0)^2 \, {}_{3}F_{2}\left(\left\{1, 1, \frac{3}{2} + \frac{\gamma}{\alpha}\right\}; \left\{\frac{3}{2}, 2\right\}; \frac{{\alpha^2(1 - x_0)^2}}{{v_0^2}}\right)\right]\, . \label{TAUdwsecondformul2}
\end{eqnarray}
Then, we give the expressions of the MFPT conditioned on the initial sign of the telegraphic noise 
\begin{eqnarray}
    &&\tau_\gamma^-(x_0) = \frac{\alpha}{{2 \gamma v_0^2}} \left[v_0 (1 - x_0) + 2 \gamma \,  _2F_1\left(1, \frac{1}{2} - \frac{\gamma}{\alpha}, \frac{3}{2}, \frac{\alpha^2}{v_0^2}\right)\right] \nonumber \\
    &&+ \frac{\alpha + 2 \gamma}{{2 \gamma v_0}} \left(1 - \frac{\alpha^2 (1 - x_0)^2}{v_0^2}\right)^{-\frac{\gamma}{\alpha}} \left[\, _2F_1\left(\frac{1}{2}, -\frac{\gamma}{\alpha}, \frac{3}{2}, \frac{\alpha^2}{v_0^2}\right) - (1 - x_0)\,  _2F_1\left(\frac{1}{2}, -\frac{\gamma}{\alpha}, \frac{3}{2}, \frac{\alpha^2 (1 - x_0)^2}{v_0^2}\right)\right] \nonumber \\
    &&- \frac{(\alpha + 2 \gamma)}{{v_0^2}} \,  _2F_1\left(\frac{1}{2}, -\frac{\gamma}{\alpha}, \frac{3}{2}, \frac{\alpha^2}{v_0^2}\right) \left[\, _2F_1\left(\frac{1}{2}, 1+\frac{\gamma}{\alpha}, \frac{3}{2}, \frac{\alpha^2}{v_0^2}\right) - (1 - x_0) \, _2F_1\left(\frac{1}{2}, 1+\frac{\gamma}{\alpha}, \frac{3}{2}, \frac{\alpha^2 (1 - x_0)^2}{v_0^2}\right)\right] + \nonumber \\
    &&- \left(\frac{v_0^2 - \alpha^2}{v_0^2 - \alpha^2 (-1 + x_0)^2}\right)^{\frac{\gamma}{\alpha}} \left[\frac{\alpha}{{2 \gamma v_0}} + \frac{\alpha (1 - x_0)}{v_0^2} \, _2F_1\left(1, \frac{1}{2} - \frac{\gamma}{\alpha}, \frac{3}{2}, \frac{\alpha^2 (1 - x_0)^2}{v_0^2}\right)\right] \nonumber \\
    &&+ \frac{{(\alpha + 2\gamma)}}{{2v_0^2}} \, \left[{}_{3}F_{2}\left(\left\{1, 1, \frac{3}{2} + \frac{\gamma}{\alpha}\right\}; \left\{\frac{3}{2}, 2\right\}; \frac{{\alpha^2}}{{v_0^2}}\right) - (1 - x_0)^2 \, {}_{3}F_{2}\left(\left\{1, 1, \frac{3}{2} + \frac{\gamma}{\alpha}\right\}; \left\{\frac{3}{2}, 2\right\}; \frac{{\alpha^2(1 - x_0)^2}}{{v_0^2}}\right)\right],
    \label{dw_E3}
\end{eqnarray}
and
\begin{eqnarray}
    &&\tau_\gamma^+(x_0) = \frac{\alpha}{{2 \gamma v_0^2}} \left[ 2 \gamma \,  _2F_1\left(1, \frac{1}{2} - \frac{\gamma}{\alpha}, \frac{3}{2}, \frac{\alpha^2}{v_0^2}\right)-v_0 (1 - x_0) \right] \nonumber \\
    &&- \frac{\alpha + 2 \gamma}{{2 \gamma v_0}} \left(1 - \frac{\alpha^2 (1 - x_0)^2}{v_0^2}\right)^{-\frac{\gamma}{\alpha}} \left[\, _2F_1\left(\frac{1}{2}, -\frac{\gamma}{\alpha}, \frac{3}{2}, \frac{\alpha^2}{v_0^2}\right) - (1 - x_0)\,  _2F_1\left(\frac{1}{2}, -\frac{\gamma}{\alpha}, \frac{3}{2}, \frac{\alpha^2 (1 - x_0)^2}{v_0^2}\right)\right] \nonumber \\
    &&- \frac{(\alpha + 2 \gamma)}{{v_0^2}} \,  _2F_1\left(\frac{1}{2}, -\frac{\gamma}{\alpha}, \frac{3}{2}, \frac{\alpha^2}{v_0^2}\right) \left[\, _2F_1\left(\frac{1}{2}, 1+\frac{\gamma}{\alpha}, \frac{3}{2}, \frac{\alpha^2}{v_0^2}\right) - (1 - x_0) \, _2F_1\left(\frac{1}{2}, 1+\frac{\gamma}{\alpha}, \frac{3}{2}, \frac{\alpha^2 (1 - x_0)^2}{v_0^2}\right)\right] + \nonumber \\
    &&+ \left(\frac{v_0^2 - \alpha^2}{v_0^2 - \alpha^2 (-1 + x_0)^2}\right)^{\frac{\gamma}{\alpha}} \left[\frac{\alpha}{{2 \gamma v_0}} - \frac{\alpha (1 - x_0)}{v_0^2} \, _2F_1\left(1, \frac{1}{2} - \frac{\gamma}{\alpha}, \frac{3}{2}, \frac{\alpha^2 (1 - x_0)^2}{v_0^2}\right)\right] \nonumber \\
    &&+ \frac{{(\alpha + 2\gamma)}}{{2v_0^2}} \, \left[{}_{3}F_{2}\left(\left\{1, 1, \frac{3}{2} + \frac{\gamma}{\alpha}\right\}; \left\{\frac{3}{2}, 2\right\}; \frac{{\alpha^2}}{{v_0^2}}\right) - (1 - x_0)^2 \, {}_{3}F_{2}\left(\left\{1, 1, \frac{3}{2} + \frac{\gamma}{\alpha}\right\}; \left\{\frac{3}{2}, 2\right\}; \frac{{\alpha^2(1 - x_0)^2}}{{v_0^2}}\right)\right]\, .
    \label{dw_E4}
\end{eqnarray}

\subsection{Log-potential}
For a potential $V(x) = \text{log}(1+|x|)$, the dynamics has an unstable negative turning point $x_-^u = 1/v_0-1$. For $x_0 \in [0,x_-^u[$, the MFPT is given by Eq.~(\ref{log_integrals_leftTP}). Performing the integrals yields
\begin{eqnarray}
    &&{\tau_{\text{\tiny I}}}_\gamma(x_0) = -\frac{{v_0^2}}{{2 \gamma (2 \gamma - v_0^2)}} \left[ (1 - 2 \gamma) - 2 \gamma^2 x_0 (2 + x_0)/v_0^2 - (1 - v_0^2) \, _2F_1\left(1, \frac{1}{2} - \frac{\gamma}{v_0^2}, \frac{1}{2}, v_0^2\right) \right] \nonumber \\
    &&+ \frac{{v_0^2}}{{2 \gamma (2 \gamma - v_0^2)}} (1 - v_0^2)^{-\frac{\gamma}{v_0^2}} \left[1 - 2 \gamma - (1 - v_0^2) \, _2F_1\left(1, \frac{1}{2} - \frac{\gamma}{v_0^2}, \frac{1}{2}, v_0^2\right)]\right] \nonumber \\
    &&\times \left[(1 + x_0) \left(1 - v_0^2 (1 + x_0)^2 \right)^{\frac{\gamma}{v_0^2}} + \, _2F_1\left(\frac{1}{2}, -\frac{\gamma}{v_0^2}, \frac{3}{2}, v_0^2\right) - (1 + x_0) \, _2F_1\left(\frac{1}{2}, -\frac{\gamma}{v_0^2}, \frac{3}{2}, v_0^2 (1 + x_0)^2\right)\right] \nonumber \\
    &&+ \frac{{v_0^2}}{{2 (2 \gamma - v_0^2)}} \left[ \, _3F_2\left(\{1, 1, \frac{1}{2} - \frac{\gamma}{v_0^2}\}; \left\{\frac{1}{2}, 2\right\}; v_0^2\right) - (1 + x_0)^2 \, _3F_2\left(\{1, 1, \frac{1}{2} - \frac{\gamma}{v_0^2}\}; \left\{\frac{1}{2}, 2\right\}; v_0^2 (1 + x_0)^2\right) \right]\, .
\label{MFPTphase2LOGpot}
\end{eqnarray}
Using Eqs. (\ref{ODEtauminus}) and (\ref{ODEtauplus}) we find
\begin{eqnarray}
    &&{\tau_{\text{\tiny I}}}^-_\gamma(x_0)= \frac{{(1 - v_0^2)^{-\frac{{\gamma}}{{v_0^2}}}}}{{2 \gamma (2 \gamma - v_0^2)}} v_0 \left[2 \gamma - 1 + \left(1 - v_0^2\right) \, _2F_1\left(1, \frac{1}{2} - \frac{{\gamma}}{{v_0^2}}, \frac{1}{2}, v_0^2\right)\right] \nonumber\\
    &&\times \left\{v_0 (1 + x_0) \, _2F_1\left(\frac{1}{2}, -\frac{{\gamma}}{{v_0^2}}, \frac{3}{2}, v_0^2 (1 + x_0)^2\right) + \left(1 - v_0 (1 + x_0)\right) \left(1 - v_0^2 (1 + x_0)^2\right)^{\frac{{\gamma}}{{v_0^2}}} - v_0 \, _2F_1\left(\frac{1}{2}, -\frac{{\gamma}}{{v_0^2}}, \frac{3}{2}, v_0^2\right)\right\}  \nonumber \\
    &&+ \frac{1}{{2 \gamma (2 \gamma - v_0^2) (1 + x_0)}} \left[(2 \gamma - 1) v_0^2 (1 + x_0) + 2 \gamma^2 x_0 (1 + x_0) (2 + x_0) + v_0 (1 - 2 \gamma (1 + x_0)^2) \right. \nonumber\\
    && \left. + v_0^2 (1 - v_0^2) (1 + x_0) \, _2F_1\left(1, \frac{1}{2} - \frac{{\gamma}}{{v_0^2}}, \frac{1}{2}, v_0^2\right) - v_0 (1 - v_0^2 (1 + x_0)^2) \, _2F_1\left(1, \frac{1}{2} - \frac{{\gamma}}{{v_0^2}}, \frac{1}{2}, v_0^2 (1 + x_0)^2\right)\right] \nonumber \\
    &&+ \frac{{v_0^2}}{{2 (2 \gamma - v_0^2)}} \left[_3F_2\left(\left\{1, 1, \frac{1}{2} - \frac{{\gamma}}{{v_0^2}}\right\}; \left\{\frac{1}{2}, 2\right\}; v_0^2\right) - (1 + x_0)^2 \, _3F_2\left(\left\{1, 1, \frac{1}{2} - \frac{{\gamma}}{{v_0^2}}\right\}; \left\{\frac{1}{2}, 2\right\}; v_0^2 (1 + x_0)^2\right)\right]\, ,
    \label{tauminuslog}
\end{eqnarray}
and
\begin{eqnarray}
    &&{\tau_{\text{\tiny I}}}^+_\gamma(x_0) = \frac{{(1 - v_0^2)^{-\frac{{\gamma}}{{v_0^2}}}}}{{2 \gamma (2 \gamma - v_0^2)}} v_0 \left[2 \gamma - 1 + \left(1 - v_0^2\right) \, _2F_1\left(1, \frac{1}{2} - \frac{{\gamma}}{{v_0^2}}, \frac{1}{2}, v_0^2\right)\right] \nonumber\\
    &&\times \left\{v_0 (1 + x_0) \, _2F_1\left(\frac{1}{2}, -\frac{{\gamma}}{{v_0^2}}, \frac{3}{2}, v_0^2 (1 + x_0)^2\right) - \left(1 + v_0 (1 + x_0)\right) \left(1 - v_0^2 (1 + x_0)^2\right)^{\frac{{\gamma}}{{v_0^2}}} - v_0 \, _2F_1\left(\frac{1}{2}, -\frac{{\gamma}}{{v_0^2}}, \frac{3}{2}, v_0^2\right)\right\}  \nonumber \\
    &&+ \frac{1}{{2 \gamma (2 \gamma - v_0^2) (1 + x_0)}} \left[(2 \gamma - 1) v_0^2 (1 + x_0) + 2 \gamma^2 x_0 (1 + x_0) (2 + x_0) - v_0 (1 - 2 \gamma (1 + x_0)^2) \right. \nonumber\\
    && \left. + v_0^2 (1 - v_0^2) (1 + x_0) \, _2F_1\left(1, \frac{1}{2} - \frac{{\gamma}}{{v_0^2}}, \frac{1}{2}, v_0^2\right) + v_0 (1 - v_0^2 (1 + x_0)^2) \, _2F_1\left(1, \frac{1}{2} - \frac{{\gamma}}{{v_0^2}}, \frac{1}{2}, v_0^2 (1 + x_0)^2\right)\right] \nonumber \\
    &&+ \frac{{v_0^2}}{{2 (2 \gamma - v_0^2)}} \left[_3F_2\left(\left\{1, 1, \frac{1}{2} - \frac{{\gamma}}{{v_0^2}}\right\}; \left\{\frac{1}{2}, 2\right\}; v_0^2\right) - (1 + x_0)^2 \, _3F_2\left(\left\{1, 1, \frac{1}{2} - \frac{{\gamma}}{{v_0^2}}\right\}; \left\{\frac{1}{2}, 2\right\}; v_0^2 (1 + x_0)^2\right)\right]\, .
    \label{taupluslog}
\end{eqnarray}


\begin{thebibliography}{26}

\bibitem{colorednoise} P. H\"anggi, P. Jung,  {\em Colored Noise in Dynamical Systems}, Adv. Chem. Phys. {\bf 89}, 239 (1995).

\bibitem{activeintro1} M. C. Marchetti, J. F. Joanny, S. Ramaswamy, T. B. Liverpool, J. Prost,  M. Rao, R. Aditi Simha, {\em Hydrodynamics of soft active matter}, Rev. Mod. Phys. {\bf 85}, 1143 (2013). 

\bibitem{activeintro2} C. Bechinger, R. Di Leonardo, H. L\"owen, C. Reichhardt, G. Volpe, G. Volpe, {\em Active particles in complex and crowded environments}, Rev. Mod. Phys. {\bf 88}, 4 045006 (2016). 

\bibitem{activeintro3} S. Ramaswamy, {\em Active matter}, J. Stat. Mech. 054002 (2017). 

\bibitem{activeintro4} E. Fodor, C. Marchetti, {\em The statistical physics of active matter: From self-catalytic colloids to living cells}, Physica A {\bf 504}, 106 (2018). 

\bibitem{activeintro5} P. Romanczuk, M. B\"ar, W. Ebeling, B. Lindner, L. Schimansky-Geier, {\em Active Brownian particles}, Eur. Phys. J. Spec. Top. {\bf 202}, 1 (2012).

\bibitem{Berg2004} H. C. Berg, {\em E. Coli in Motion}, (Springer Verlag, Heidelberg, Germany) (2004).

\bibitem{Tailleur_RTP}
J. Tailleur, M. E. Cates, {\em Statistical mechanics of interacting run-and-tumble bacteria}, Phys. Rev. Lett. {\bf 100}, 218103 (2008).

\bibitem{Cates2012} M. E. Cates, {\em Diffusive transport without detailed balance: Does microbiology need statistical physics?}, Rep. Prog. Phys. {\bf 75}, 042601 (2012). 

\bibitem{Kac1974}
M. Kac, {\em A stochastic model related to the telegrapher's equation}, Rocky Mountain J. Math. {\bf 4}, 497 (1974).

\bibitem{Ors1990}
E. Orsingher, {\em Random motions governed by third-order equations}, Stoch. Process. Their Appl. {\bf 34}, 49 (1990)

\bibitem{PRWWeiss} G. H. Weiss, {\em Some applications of persistent random walks and the telegrapher's equation}, Physica A {\bf 311}, 381 (2002).


\bibitem{RTPSS} A. Dhar, A. Kundu, S. N. Majumdar, S. Sabhapandit, G. Schehr, {\em Run-and-tumble particle in one-dimensional confining potentials}, Phys. Rev. E {\bf 99}, 032132 (2019).

\bibitem{Sevilla}
F. J. Sevilla, A. V. Arzola, E. P. Cital, {\em Stationary superstatistics distributions of trapped run-and-tumble particles}, Phys. Rev. E {\bf 99}, 012145 (2019).

\bibitem{Velocity_RTP} P. Le Doussal, S. N. Majumdar, G. Schehr, {\em Velocity and diffusion constant of an active particle in a one-dimensional force field}, EPL {\bf 130}, 40002 (2020).








\bibitem{Lee2013} C. F. Lee, {\em Active particles under confinement: aggregation at the wall and gradient formation inside a channel}, New J. Phys. {\bf 15} 055007 (2013).

\bibitem{Yang2014} X. Yang, M. L. Manning, M. C. Marchetti, {\em Aggregation and segregation of confined active particles}, Soft Matter {\bf 10}, 6477 (2014).

\bibitem{Uspal2015} W. E. Uspal, M. N. Popescu, S. Dietrich, M. Tasinkevych, {\em Self-propulsion of a catalytically active particle near a planar wall: from reflection to sliding and hovering}, Soft Matter {\bf 11}, 434 (2015).

\bibitem{Duzgun2018} A. Duzgun, J. V. Selinger, {\em Active Brownian particles near straight or curved walls: Pressure and boundary layers}, Phys. Rev. E {\bf 97}, 032606 (2018).

\bibitem{AngelaniHardWalls} L. Angelani, {\em Confined run-and-tumble swimmers in one
dimension}, J. Phys. A: Math. Theor. {\bf 50}, 325601 (2017).

\bibitem{hardWallsJoanny} C. Sandford, A. Y. Grosberg, J.-F. Joanny, {\em Pressure and flow of exponentially self-correlated active particles},
Phys. Rev. E {\bf 96}, 052605 (2017).

\bibitem{hardWallsCaprini} L. Caprini, U. M. B. Marconi, {\em Active particles under confinement and effective force generation among surfaces}, Soft matter {\bf 14}, 9044 (2018).



\bibitem{redner}
S. Redner, {\it A guide to first-passage processes}, Cambridge University Press, (2001).

\bibitem{Metzler_book}
R. Metzler, S. Redner, G. Oshanin, {\it First-passage phenomena and their applications},
Vol. 35, World Scientific, (2014).

\bibitem{EVS1} A. J. Bray, S. N. Majumdar, G. Schehr, {\it Persistence and first-passage properties in nonequilibrium systems}, Adv. Phys. {\bf 62}, 225 (2013).




\bibitem{MalakarRTP} K. Malakar, V. Jemseena, A. Kundu, K. Vijay Kumar, S. Sabhapandit, S. N. Majumdar, S. Redner, A. Dhar, {\em 
Steady state, relaxation and first-passage properties of a run-and-tumble particle in one-dimension},
Phys. Rev. E {\bf 103}, 032607 (2021) %{\red and references therein}.







\bibitem{Singh2020} P. Singh, S. Sabhapandit, A. Kundu, {\em Run-and-tumble particle in inhomogeneous media in one dimension}, J. Stat. Mech. 083207 (2020).

\bibitem{Singh2022} P. Singh, S. Santra, A. Kundu, {\em Extremal statistics of a one-dimensional run and tumble particle with an absorbing wall}, J. Phys. A: Math. Theor. {\bf 55} 465004 (2022).

\bibitem{MFPT1DABP} S. A. Iyaniwura, Z. Peng, {\em Asymptotic analysis and simulation of mean first-passage time for active Brownian particles in 1-D}, SIAM J. Appl. Math. \textbf{84}, 1079 (2024).



\bibitem{Debruyne} B. De Bruyne, S. N. Majumdar, G. Schehr, {\em Survival probability of a run-and-tumble particle in the presence of a drift}, J. Stat. Mech. 043211 (2021).

\bibitem{RTPsurvivalMori} F. Mori, P. Le Doussal, S. N. Majumdar, G. Schehr, {\em Universal Survival Probability for a d-Dimensional Run-and-Tumble Particle}, Phys. Rev. Lett. {\bf 124}, 090603 (2020).

\bibitem{TVB12}
V. Tejedor, R. Voituriez, O. B{\' e}nichou, {\em Optimizing persistent random searches}, Phys. Rev. Lett. {\bf 108}, 088103 (2012).

\bibitem{RBV16}
J.-F. Rupprecht, O. B{\'e}nichou, R. Voituriez, {\em Optimal search strategies of run-and-tumble walks}, Phys. Rev. E {\bf 94}, 012117 (2016).




\bibitem{BressloffStickyBoundaries} P. C. Bressloff, {\em Encounter-based model of a run-and-tumble particle II: absorption at sticky boundaries}, J. Stat. Mech. 043208 (2023).

\bibitem{AngelaniGenericBC} L. Angelani, {\em One-dimensional run-and-tumble motions with generic boundary conditions}, J. Phys. A: Math. Theor. {\bf 56}, 455003 (2023).

\bibitem{EPJEAngelani} L. Angelani, {\em Optimal escapes in active matter}, Eur. Phys. J. E {\bf 47}, 9 (2024).

\bibitem{RTPpartiallyAbsorbingTarget} E. Jeon, B. Go, Y. W. Kim, {\em Searching for a partially absorbing target by a run-and-tumble particle in a confined space}, Phys. Rev. E \textbf{109}, 014103 (2024).

\bibitem{letter} M. Gu\'eneau, S. N. Majumdar, G. Schehr, {\em Optimal mean first-passage time of a run-and-tumble particle in a class of one-dimensional confining potential}, EPL {\bf 145} 61002 (2024).

\bibitem{ExitProbaShort} M. Gu{\'e}neau, L. Touzo, {\em Relating absorbing and hard wall boundary conditions for active particles}, J. Phys. A: Math. Theor. {\bf 57}, 225005 (2024).

\bibitem{Kardar}
Y. B. Dor, E. Woillez, Y. Kafri, M. Kardar, A. P. Solon, {\it Ramifications of disorder on active particles in one dimension}, Phys. Rev. E {\bf 100}, 052610 (2019).

\bibitem{naftaliSS} N. R. Smith, O. Farago, {\em Nonequilibrium steady state for harmonically confined active particles}, Phys. Rev. E {\bf 106}, 054118 (2022).


\bibitem{SurvivalRPTlinear} S. K. Nath, S. Sabhapandit, {\em Survival probability and position distribution of a run and tumble particle in $U(x)=\alpha|x|$ potential with an absorbing boundary}, J. Stat. Mech. 093205 (2024).

\bibitem{SurvivalRPTlinearMORI} F. Mori, S. N. Majumdar, G. Schehr,
{\em Time to reach the maximum for a stationary stochastic process},
Phys. Rev. E {\bf 106}, 054110 (2022).

\bibitem{resettingRTP} M. R. Evans, S. N. Majumdar, {\em Run and tumble particle under resetting: a renewal approach}, J. Phys. A: Math. Theor. {\bf 51} 475003 (2018). 

\bibitem{resettingPRL} M. R. Evans, S. N. Majumdar, {\em Diffusion with stochastic resetting}, Phys. Rev. Lett. {\bf 106}, 160601 (2011).

\bibitem{resettingReview} M. R. Evans, S. N. Majumdar, G. Schehr, {\em Stochastic resetting and applications}, J. Phys. A: Math. Theor. {\bf 53} 193001 (2020).





\bibitem{kramers} P. Hänggi, P. Talkner, M. Borkovec, {\em Reaction-rate theory: fifty years after Kramers}, Rev. Mod. Phys. {\bf 62}, 251 (1990).

\bibitem{RTPSS4} C. Van den Broeck, P. H{\"a}nggi, {\em Activation rates for nonlinear stochastic flows driven by non-Gaussian noise}, Phys. Rev. A~{\bf 30}, 2730 (1984).

\bibitem{coloredKramers} A. J. Bray, A. J. McKane, 
{\em Instanton Calculation of the Escape Rate for Activation over a Potential Barrier Driven by Colored Noise}, Phys. Rev. Lett. {\bf 62}, 493 (1989).

\bibitem{delago}
A. Militaru, M. Innerbichler, M. Frimmer, F. Tebbenjohanns, L. Novotny, C. Dellago, {\it Escape dynamics of active particles in multistable potentials}, Nature Comm. {\bf 12}, 2446 (2021).


\bibitem{Woillez}  E. Woillez, Y. Kafri, V. Lecomte, {\em Nonlocal stationary probability distributions and escape rates for an active Ornstein–Uhlenbeck particle}, J. Stat. Mech. 063204 (2020).

\bibitem{Woillez2} E. Woillez, Y. Zhao, Y. Kafri, V. Lecomte, J. Tailleur, {\em Activated Escape of a Self-Propelled Particle from a Metastable State}, Phys. Rev. Lett. {\bf 122}, 258001 (2019).

\bibitem{Hanggi_kramers} A. Geiseler, P. H\"anggi, G. Schmid, {\em Kramers escape of a self-propelled particle}, Eur. Phys. J. B {\bf 89}, 175 (2016).



\bibitem{Sharma} A. Sharma, R. Wittmann, J. M. Brader, {\em Escape rate of active particles in the effective equilibrium approach}, Phys. Rev. E {\bf 95}, 012115 (2017). 

\bibitem{WexlerKramers} D. Wexler, N. Gov, K. O. Rasmussen, G. Bel,  {\em Dynamics and escape of active particles in a harmonic trap}, Phys. Rev. Res.~{\bf 2}, 013003 (2020). 


\bibitem{kramersCaprini} L. Caprini, F. Cecconi, U. Marini Bettolo Marconi, {\em Correlated escape of active particles across a potential barrier}, The Journal of Chemical Physics, {\bf 155}(23) (2021).

\bibitem{RTPPSG} P. Le Doussal, S. N. Majumdar, G. Schehr, 
{\em Velocity and diffusion constant of an active particle in a one dimensional force field},
EPL {\bf 130}, 40002 (2020).

\bibitem{diffusivemoments} D. A. Gorokhov, G. Blatter, {\em Diffusion and creep of a particle in a random potential}, Phys. Rev. B {\bf 58}(1), 213 (1998).

\bibitem{letterSM} M. Gu\'eneau, S. N. Majumdar, G. Schehr, {\em Optimal mean first-passage time of a run-and-tumble particle in a class of one-dimensional confining potential}, Supp. Mat.  arXiv:2311.06923.

\bibitem{Grad}
I. S. Gradshteyn, I. M. Ryzhik, {\it Table of integrals, series, and products}, Academic press, (2014).




\bibitem{optimalMFPTdiffusive} M. R. Evans, S. N. Majumdar, K. Mallick, {\em Optimal diffusive search: nonequilibrium resetting versus equilibrium dynamics}, J. Phys. A: Math. Theor. {\bf 46}, 185001 (2013). 


\bibitem{time_max} F. Mori, S.~N. Majumdar, G. Schehr, {\em Distribution of the time of the maximum for stationary processes}, Europhys. Lett. {\bf 135}, 30003 (2021)

\bibitem{freezingdiffusive} S. Sabhapandit, S. N. Majumdar, {\em Freezing Transition in the
Barrier Crossing Rate of a Diffusing Particle},
Phys. Rev. Lett.  {\bf 125}, 200601 (2020)


\bibitem{RTPSS3} R. Lefever, W. Horsthemke, K. Kitahara, I. Inaba, {\em Phase Diagrams of Noise Induced Transitions: Exact Results for a Class of External Coloured Noise}, Prog. Theor. Phys. {\bf 64}, 1233 (1980).




\bibitem{RTPSS2} V. I. Klyatskin, {\em Dynamic systems with parameter fluctuations of the telegraphic-process type},  Radiofizika {\bf 20}, 562 (1977).




%\bibitem{RTPSS1} V I. Klyatskin, Radiophys. Quantum El. {\bf 20}, 382 (1978).




























\bibitem{resettingJPA}
M. R. Evans, S. N. Majumdar, {\it Diffusion with optimal resetting}, J. Phys. A: Math. Theor. {\bf 44}, 435001 (2011).










\bibitem{Naturecoli} H. C. Berg, D. A. Brown, {\em Chemotaxis in Escherichia coli analysed by three-dimensional tracking}, Nature  {\bf 239}, 500 (1972).



\bibitem{natureruntime} E. Korobkova, T. Emonet, J.M. Vilar, T.S Shimizu, P. Cluzel, {\em From molecular noise to behavioural variability in a single bacterium}, Nature {\bf 428}(6982):574-8 (2004)

\bibitem{PRLcoli} C. Kurzthaler, Y. Zhao, N. Zhou, J. Schwarz-Linek, C. Devailly, J. Arlt, J.-D. Huang, W. C. K. Poon, T. Franosch, J. Tailleur, V.~A Martinez, {\em Characterization and control of the run-and-tumble dynamics of Escherichia Coli}, Phys. Rev. Lett. {\bf 132}, 038302 (2024).





\bibitem{inhomoRTP} P. Singh, S. Sabhapandit, A. Kundu, {\em Run-and-tumble particle in inhomogeneous media in one dimension}, J. Stat. Mech, 083207 (2020).

\bibitem{inhomoRTP2} C. Monthus, {\em Large deviations at various levels for run-and-tumble processes with space-dependent velocities and space-dependent switching rates}, J. Stat. Mech. 083212 (2021).

\bibitem{regularsingular} \url{http://scipp.ucsc.edu/~haber/ph116C/singularpoint_12b.pdf}


\end{thebibliography}
\end{document}